\documentclass{article}
\usepackage{amssymb}
\usepackage{graphicx}
\usepackage{amsmath}

\input{tcilatex}

\begin{document}

\begin{center}
{\Huge From Ginzburg-Landau to Hilbert-Einstein via Yamabe}

\bigskip\ \ \ \ \ \ 

\ \ \ \ \ \textbf{Arkady L.Kholodenko}$^{a}$\textbf{, Ethan E.Ballard }$^{b}$%
\ $\footnote{\textit{E-mail addresses}: string@clemson.edu,
eballar@clemson.edu}$

\ \ 

$\ \ \ \ \ \ \ ^{a}$ \ \ \textit{375 H.L.Hunter Laboratories, Clemson
University Clemson, SC 29634-0973, USA. }

$\ \ \ \ \ \ \ ^{b}$ \ \ \textit{Department of Materials Science and
Engineering, Clemson University, Clemson, SC 29634,USA}
\end{center}

\bigskip

\textbf{Abstract}

\bigskip

In this paper, based on some mathematical results obtained by Yamabe, \
Osgood, Phillips and Sarnak, we demonstrate that in dimensions three and
higher the famous Ginzburg-Landau equations used in theory of phase
transitions can be obtained (without any approximations) by minimization of
the Riemannian-type Hilbert-Einstein action functional for pure gravity in
the presence of cosmoloigical term. We use this observation in order to
bring to completion the work by Lifshitz (done in 1941) on group-theoretical
refinements of the Landau theory of phase transitions. In addition, this
observation allows us to develop a systematic extension to higher dimensions
of known string-theoretic path integral methods developed for calculation of
observables in two dimensional conformal field theories.

\bigskip

\bigskip

\bigskip

\section{Introduction}

\subsection{\protect\bigskip \textbf{\ Landau theory versus other
theoretical methods for predicting crystal structure}}

An article [\textbf{1}] in \textit{Nature }written in 1988 referred to the
inability, at that time, to predict the crystal structure of simple
crystalline solids from their chemical composition as "a continuing scandal"
in solid state physics. \ Recently, Oganov and Glass developed a method for
prediction of the most stable structures along with some low energy
metastable states for a given compound without hints supplied by
experimental data for this compound [\textbf{2}]. Their method combines 
\textit{ab initio} electronic structure calculations with an evolunionary
algorithm. \ Only the chemical composition is needed as an input. The method
allows prediction of crystal strucutures at any P-T conditions.

In view of thise remarkable achievements, it is of interest to study to what
extent much earlier phenominological results by Landau [\textbf{3}] can help
in solving the same problem of structure prediction. \ Landau's results were
subsequently improved by Lifshitz in 1941 [\textbf{4}] and used later by
Ginzburg and Landau in 1950 [\textbf{5}] in connection with their study of
superconductivity, as is well known. The Landau theory can be used only in a
rather narrow domain of temperatures near criticality. Being thermodynamical
in nature, it contains parameters whose actual numerical value is to a large
degree arbitrary. \ In principle, they can be deduced from experimental
data. \ The original motivation for such a theory came from experimental
study of order-disorder phase transitions in alloys. The first paragraph of
Chapter 14 of Vol. 5 of the famous Landau-Lifshitz course in theoretical
physics [\textbf{6}] begins with description of the concept of order (to be
defined below) in a typical CuZn alloy. For pure Cu, the lattice is face
centered cubic (fcc), while for Zn, it is hexagonal closed packed (hcp) [%
\textbf{7}]. For an alloy composed of these two metals, the system must
\textquotedblright decide\textquotedblright\ what kind of lattice to have.
The decision is made based on thermodynamical considerations. \
Specifically, it is believed that if the system has reached an equilibrium
at a given temperature and pressure, its free energy should be minimal.
Since a system as simple as CuZn exhibits a very complicated phase diagram,
depicted in Fig.1, such considerations are not too informative. \footnote{%
A large number of such binary alloys, their phase diagrams and their
symmetry descriptions are listed in Ref.[\textbf{8}]. As compared to the
textbook by Landau and Lifshitz [6], where actually only a small portion of
CuZn phase diagram is discussed (e.g. $\beta $ and $\beta ^{\prime }$ phases
in Fig.1). Ref.[\textbf{8}] indicates that most binary alloys, including
CuZn, exibit unexpectedly complex phase diagrams similar to that given in
Fig.1. Subsection 6.2. contains additional information helpful for its
understanding.}

\medskip

\begin{figure}[tbp]
\begin{center}
\includegraphics*[width=5.00 in]{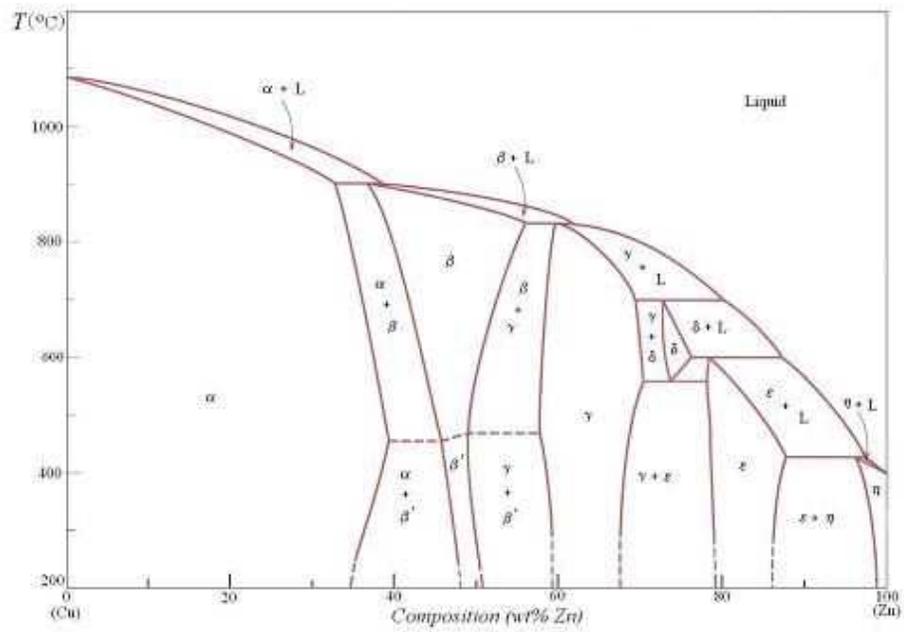}
\end{center}
\caption{Phase diagram for CuZn alloy}
\end{figure}

Empirically, it is known that about 3/4 of all elements in nature are
metals. When mixed, about 3/4 of these crystallize into one of the three
most frequently encountered lattices with almost equal probability [\textbf{8%
}]. These are: fcc, Fig.2a), bcc, Fig.3 a), and hcp, Fig.2 b). In addition,
Fig.1 exhibits more complex lattices, e.g. $\gamma $ and $\delta $, whose
group-theoretic description is given in Table 1.

\ \ \ \ \ \ \ \ \ \ \ \ \ \ \ \ \ \ \ \ \ \ \ \ \ \ \ \ \ \ \ \ \ \ \ \ \ \
\ \ \ 

\ \ \ \ \ \ \ \ \ \ \ \ \ \ \ \ \ \ \ \ \ \ \ \ \ \ \ \ \ \ \ \ \ \ \ \ \ \
\ \ \ \ \ Table 1.\ \ \ 

\ \ \ \ \ \ \ \ \ \ \ \ \ \ \ \ \ \ \ \ \ \ \ \ \ \ \ \ \ \ \ \ \ \ \ \ \ \
\ \ \ \ \ \ \ \ \ \ \ \ \ \ \ \ \ \ \ 

\begin{tabular}{|c|c|c|c|}
\hline
Phase & Composition(at \% Zn) & Symbol and \# & Prototype \\ \hline
$\alpha $ & $0-38.27$ & $Fm\bar{3}m$ \ \ \ \ \ \ \ \ \ \ \ \ \ \ \ \ \ \ $\
225$ & $fcc$ \\ \hline
$\beta $ & $36.1-55.8$ & $Im\bar{3}m$ \ \ \ \ \ \ \ \ \ \ \ \ \ \ \ \ \ \ \
\ $229$ & $bcc$ \\ \hline
$\beta ^{^{\prime }}$ & $44.8-50$ & $Pm\bar{3}m$ \ \ \ \ \ \ \ \ \ \ \ \ \ \
\ \ \ \ \ $221$ & \textbf{Z}$^{3}$ \\ \hline
$\gamma $ & $57.0-70.0$ & $I\bar{4}3m$ \ \ \ \ \ \ \ \ \ \ \ \ \ \ \ \ \ \ \
\ \ $217$ & $bcc$ \\ \hline
$\delta $ & $72.45-76.0$ & $P\bar{6}$ \ \ \ \ \ \ \ \ \ \ \ \ \ \ \ \ \ \ \
\ \ \ \ \ \ $174$ & $hcp$ \\ \hline
$\varepsilon $ & $78.0-88.0$ & $P6_{3}/mmc$ \ \ \ \ \ \ \ \ \ \ \ \ \ \ \ $%
194$ & $hcp$ \\ \hline
$\eta $ & $97.17-100$ & $P6_{3}/mmc$ \ \ \ \ \ \ \ \ \ \ \ \ \ \ \ $194$ & $%
hcp$ \\ \hline
\end{tabular}

\bigskip

A brief summary of crystallographic terminology is provided in Appendix A.
The CuZn system was used in Ref.[\textbf{6}] as a good example of the second
order phase transition. A phase transition for which the order parameter
changes continuously (with concentration) till it becomes zero at the
transition point is of \textit{second order} (or \textit{continuous}). \ In
reality, however, such a transition takes place in a rather narrow
concentration range (roughly, between 30 and 60 \% of Zn in Cu), as can be
seen in Fig.1. Nevertheless, this example was used by Landau to introduce
the concept of an order parameter. To this purpose, using Fig.1, we fix the
temperature so that the horizontal line will cross the phase boundary within
the concentration range just mentioned. Then, on the l.h.s.with respect to
this crossing we obtain the $\beta $-phase depicted in Fig.3a) where both Cu
and Zn can occupy each lattice site with equal probability (disordered
state). To the r.h.s.of this crossing, the occupation probabilities for Cu
and Zn become different (ordered state). Following Landau [\textbf{3}], we
introduce the \textit{nonnegative order parameter }$\varphi $ measuring the
degree of such ordering. Clearly, it should be proportional to the
difference between the occupation probability 1/2 in the disordered state
(either for Cu or Zn) and that for the ordered state which should be less
than 1/2. As Fig.1 and Table 1 indicate, the symmetry of the low temperature 
$\beta ^{\prime }$-phase is that of two interpenetrating cubic lattices (as
depicted in Fig.3b)), while the symmetry of the high temperature $\beta $%
-phase is that of

%\FRAME{ftbpFU}{3.2024in}{2.7959in}{0pt}{\Qcb{a) Basic
%fundamental domain for the fcc lattice along with its primitive cell; b) the
%same for the hcp lattice and its primitive cell}}{}{smallfigure2-25.jpg}{%
%\special{language "Scientific Word";type "GRAPHIC";maintain-aspect-ratio
%TRUE;display "USEDEF";valid_file "F";width 3.2024in;height 2.7959in;depth
%0pt;original-width 3.4895in;original-height 3.0415in;cropleft "0";croptop
%"1";cropright "1";cropbottom "0";filename
%'smallfigure2-25.JPG';file-properties "NPEU";}}

\begin{figure}[ptb]
\begin{center}
\includegraphics[width=3.2024in]{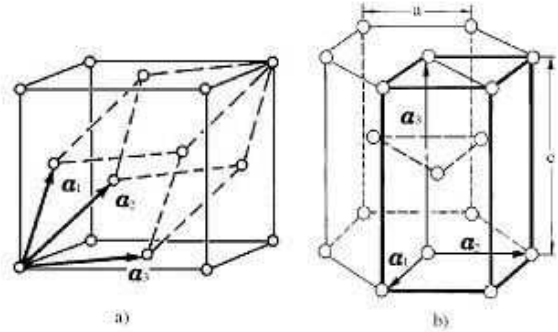}
\end{center}
\caption{a) Basic fundamental domain for the fcc lattice along with its
primitive cell; b) the same for the hcp lattice and its primitive cell}
\end{figure}

%\FRAME{ftbpFU}{2.8599in}{%
%2.7864in}{0pt}{\Qcb{The simplest example of superlattice formation: a bcc $%
%\protect\beta -$type lattice a) is converted into $\protect\beta ^{\prime }-$%
%type lattice made of two interpenetrating \textbf{Z}$^{3}-$type lattices. To
%facilitate understanding, a two dimensional square lattice \textbf{Z}$^{2}$
%depicted in a$^{\prime }$) is splitting into two square sublattices in $%
%b^{\prime }$)}}{}{figure3.jpg}{\special{language "Scientific Word";type
%"GRAPHIC";maintain-aspect-ratio TRUE;display "USEDEF";valid_file "F";width
%2.8599in;height 2.7864in;depth 0pt;original-width 10.1149in;original-height
%9.8545in;cropleft "0";croptop "1";cropright "1";cropbottom "0";filename
%'figure3.jpg';file-properties "NPEU";}}

\begin{figure}[tbp]
\begin{center}
\includegraphics[width=2.8599in]{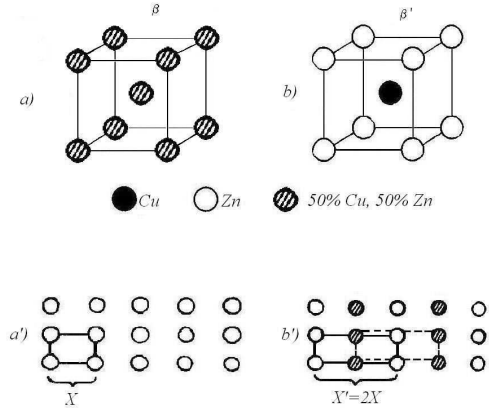}
\end{center}
\caption{The simplest example of superlattice formation: a bcc $\protect%
\beta -$type lattice a) is converted into $\protect\beta ^{\prime }-$type
lattice made of two interpenetrating \textbf{Z}$^{3}-$type lattices. To
facilitate understanding, a two dimensional square lattice \textbf{Z}$^{2}$
depicted in a$^{\prime }$) is splitting into two square sublattices in $%
b^{\prime }$)}
\end{figure}
the bcc lattice and is higher. This fact is in accord with the general
observation [\textbf{6}] that the higher symmetry phase usually occurs at
higher temperatures. Having said this, one has to take into account that the
order parameter $\varphi $ was defined without reference to temperature so
that the order-disorder transition thus described is only
concentration-dependent. This means that infinitesimal changes in
concentration can cause changes in symmetry.

To include both the temperature and concentration dependence, the notion of
an order parameter should be generalized. This can be accomplished if,
instead of fixing the temperature, we fix the concentration in the
concentration range specified above. Then, a vertical line in Fig.1 also
will cross the phase boundary so that the higher symmetry phase $\beta $
indeed occurs at higher temperatures. In view of this, to describe a phase
diagram locally in the sense just described, Landau [\textbf{3}] proposed
the following \ phenomenological expansion for the free energy functional $%
\mathcal{F}\{\varphi \}$ valid in the vicinity of the transition temperature 
$T_{c}:$ 
\begin{equation}
\mathcal{F}\{\varphi \}=\mathcal{F}_{0}+a\varphi +A\varphi ^{2}+C\varphi
^{3}+B\varphi ^{4}+...\text{ \ \ .}  \tag{1.1}
\end{equation}%
It is expected, that for $T>T_{c}$ we are left only with $\mathcal{F}%
\{\varphi \}=\mathcal{F}_{0}$, while for $T<T_{c}$ we have to assume that
coefficients $a,$ $A,$ etc. depend upon $T,P$ and concentration $c$. Such an
expansion is non analytic, however, since the derivatives of the free energy
with respect to $P$ and $T$ are discontinuous at $T=T_{c}$. This fact allows
us to define the order of transition using standard rules of thermodynamics.
\ The system is undergoing a \textsl{first order} transition if the first
order derivatives of $\mathcal{F}\{\varphi \}$ \ with respect \ to $T$ \
and/or $P$ are discontinuous at $\ T_{c}$, while it undergoes a \textsl{%
second order} transition\ if the second derivatives of $\mathcal{F}\{\varphi
\}$ \ with respect to $P$ and/or $T$ are discontinuous at $T_{c}$.

\ To actually use the Landau theory under such circumstances, $\mathcal{F}%
\{\varphi \}$ is minimized with respect to $\varphi $. \ Next, the first
derivatives of $\mathcal{F}\{\varphi \}$ are calculated with respect to $P$
and/or $T$ .\ This minimum value of the order parameter $\varphi $ is
substituted back into the derivatives in order to check their continuity at $%
T=T_{c}$. This requires making some additional assumptions on coefficients
in the Landau expansion. For instance, for the second order transition, one
should require $a=0$ \ in Eq.(1.1) while $A=A(P,T)$ is expected to change
sign at $T=T_{c}$ so that, by symmetry, it should become exactly zero at $%
T_{c}.$ This implies that for temperatures $T$ close to $T_{c}$ the
coefficient A is expected to behave as $A(P,T)=\alpha
(P)(T-T_{c})/T_{c}\equiv \alpha \tau .$ As for the coefficient $C$, it is
also expected that it can be a function of $P$ and $T$ . Its presence is not
mandatory, however. It depends upon the symmetry of the system. The details
were worked out by\ Lifshitz in 1941 [\textbf{4}].

\ 

\subsection{\textbf{\ Refinements by E.M. Lifshitz}}

\ 

Although the textbook by Landau and Lifshitz [\textbf{6}] contains only one
paragraph describing Lifshitz refinements, there are monographs, e.g. Refs. [%
\textbf{9-11}] and the references therein, elaborating on works by Lifshitz.
Such elaborations are quite sophisticated, so we refer our readers to these
monographs for details. Here we would like only to emphasize the key points
essential for our work. \ Lifshitz improvements of Landau's theory were
designed to explain phase transitions in \textit{superlattices}. The $\beta
^{\prime }$ phase of CuZn mixture is an example of a superlattice. By
definition, the \textit{superlattice}, (or superstructure)\textit{,} is a
lattice made of atoms of different kinds occupying existing lattice sites in
an orderly fashion. When this happens, the lattice is subdivided into
different sublattices, as depicted in Fig 2, so that different sublattices
are occupied by different atoms in the alloy. To date, an enormous number of
superlattices is known empirically [\textbf{7,8}]. In principle, Lifshitz
theory allows these predictions\textit{\ provided that the symmetry of the
high temperature phase is assigned. \ Then, the symmetry of the low
temperature phase} \textit{(superlattice) can be predicted}. Many practical
examples of such calculations can be found in Ref.[\textbf{9-11}]. It should
be kept in mind, however, that such a theory can explain diagrams like that
in Fig.1. only qualitatively. In its present form, it cannot make
predictions about the exact location of the phase boundaries, which is also
true of the method developed by Oganov. \ For this one needs a more detailed
microscopic theory. Attempts to develop such theory were made (to our
knowledge) only for specific systems, e.g. see Ref.[\textbf{10,12}]. \ It is
hoped that the results of this paper and that by Oganov \textit{et al} might
stimulate further research in this area.

The probabilistic interpretation of the order parameter $\varphi $ suggests
that it should be coordinate-independent. Yet, according to Lifshitz
original work [\textbf{4}], it is allowed to be coordinate-dependent.
Specifically, Lifshitz argues that the order parameter $\varphi $ is
proportional to the crystalline density $\rho $ which can be presented as $%
\rho =\rho _{0}+\delta \rho $, provided that $\rho _{0}$ serves as the basis
of representation of the \textit{prescribed} symmetry group \textbf{G}$_{0}$
of the high temperature phase, and $\delta \rho $ can be represented as 
\begin{equation}
\delta \rho =\sum\limits_{\substack{ n  \\ n\neq 1}}^{{}}\sum\limits_{i}%
\varphi _{i}^{(n)}\Psi _{i}^{(n)}.  \tag{1.2}
\end{equation}

The first summation takes place over all irreducible representations (except
trivial) while the second summation takes place over the bases of these
irreducible representations. Clearly, $\varphi _{i}^{(n)}$ is the analog of
the earlier introduced $\varphi $ while $\rho _{0}$ corresponds to the basis
of unit representation for \textbf{G}$_{0}.$ By design, $\delta \rho $ is
invariant under group $\mathbf{G}$ of \textit{lesser} symmetry expected for
the low temperature phase. For brevity, following Lifshitz, we shall
suppress the superscript $n$ in what follows. To make a connection with
Eq.(1.1) Lifshitz suggests writing $\varphi _{i\text{ }}$as $\varphi _{i%
\text{ }}=\varphi \gamma _{i\text{ }}$ where the functions $\gamma _{i}$ are
chosen in such a way that 
\begin{equation}
\varphi ^{2}=\sum\limits_{i}\varphi _{i\text{ }}^{2}\text{ }  \tag{1.3}
\end{equation}%
implying that $\sum\limits_{i}\gamma _{i}^{2}=1.$ Under such conditions, the
expansion, Eq.(1.1), acquires the following form 
\begin{equation}
\mathcal{F}\{\varphi \}=\mathcal{F}_{0}+A(P,T)\varphi ^{2}+\varphi
^{3}\sum\limits_{j}C_{j}(P,T)f_{j}^{(3)}(\{\gamma _{i}\})+\varphi
^{4}\sum\limits_{j}B_{j}(P,T)f_{j}^{(4)}(\{\gamma _{i}\})+...  \tag{1.4}
\end{equation}%
where $f_{j}^{(3)},$ $f_{j}^{(4)}...$ are the third, forth and higher order
invariants of the group \textbf{G} made of $\gamma _{i\text{ }}^{{}}s$. The
sums over $j$ count all such respective invariants. This refinement by
Lifshitz explains the group-theoretic nature of the coefficients in the
original Landau expansion, Eq.(1.1). Thus, some of these coefficients are
excluded from the expansion based on symmetry in addition to the
thermodynamic stability requirements, discussed by Landau\footnote{%
It should be noted that in Landau's work these symmetry arguments are
present already but without details.}. The results obtained thus far are
still not too restrictive to determine the group \textbf{G} in the low
temperature phase. Mathematically rigorous arguments leading to full
determination of \textbf{G} can be found in the monograph by Lyubarsky [%
\textbf{9}] and are too long to be used for these introductory remarks. To
get a feeling for these arguments we provide some less rigorous (physical)
arguments in spirit of the original paper by Lifshitz.

To this purpose, we assume that the order parameter is weakly
coordinate-dependent in the following sense. The representation of any of
230 space groups (Appendix A) acts in the vector space whose basis is made
of functions of the form 
\begin{equation}
\Psi _{\mathbf{k}\alpha }(\mathbf{r})=u_{\mathbf{k}\alpha }e^{i\mathbf{k}%
\cdot \mathbf{r}}\text{.}  \tag{1.5}
\end{equation}%
For a fixed $\mathbf{k,}$ functions $u_{\mathbf{k}\alpha }$ are invariant
under translations in the \textit{direct} lattice, $\mathbf{r}\rightarrow 
\mathbf{r}+\mathbf{a}$, while the combination $\mathbf{k}\cdot \mathbf{r}$
in the exponent changes to $\mathbf{k}\cdot \mathbf{r+k}\cdot \mathbf{a,}$
so that if \ $\mathbf{H}$ is some vector of the \textit{reciprocal} lattice,
then \ the vectors $\mathbf{k}$ and $\mathbf{k}+\mathbf{H}$ are \textit{%
equivalent.} \ This equivalence leads to the exponent $\exp (i\mathbf{a}%
\cdot \mathbf{H})$ being equal to 1. The subscript $\alpha $ in Eq.(1.5)
numbers all functions with fixed $\mathbf{k}$. Such a set of functions forms
a basis for representation for the point group. The above expression for $%
\Psi _{\mathbf{k}\alpha }(\mathbf{r})$ can be simplified for superlattices
according to the following arguments by Lifshitz. He noticed that lattices
(expected to become superlattices) at\ higher temperature (in the disordered
phase) should contain only one atom per unit cell $\Lambda \footnote{%
It is known from solid state physics that the elementary cells displayed in
Fig.s 2 and 3 should \textit{not }be confused with the unit cells. The
difference is well explained in the book by Ziman [\textbf{13}] for instance.%
}.$ At lower temperatures the points in such a lattice in different cells
become \textit{non-equivalent} thus signalling the formation of a
superlattice. Because of this, the number of functions $u_{\mathbf{k}\alpha
} $ is reduced to one (i.e. $\alpha =1).$ These arguments allow us to
replace as well all $u_{\mathbf{k}\alpha }$ s \ in Eq.(1.5) by unity. Next,
applying all rotations related to the discrete subgroup\ (of \textbf{G}) of
point symmetries at fixed $\mathbf{k}$\textbf{\ }creates an orbit for such a
subgroup, called a \textit{star}. Since the expansion, Eq.(1.2), is made for
a given star (the number of elements in a star thus forming a basis), it is
only natural to require the coefficient $A(P,T)$ to be k-depenent. Suppose
now that the low temperature phase corresponds to some \textbf{k}=\textbf{k}$%
_{0}$. Thermodynamically, for such a phase to be stable, one should look at $%
A(P,T)$ as a function of \textbf{k}. When considered as a function of 
\textbf{k,} the coefficient $A(P,T)$ should possess a minimum determining 
\textbf{k}$_{0}$.\ Lifshitz recognized that such a condition is not
constructive enough. In support of the necessity of such a condition the
following chain of arguments can be used. If \textbf{k}$_{0}$ would
determine the structure, the density $\delta \rho $ \ would be periodic with
periodicity induced by \textbf{k}$_{0}$. In view of Eq.s (1.2) and (1.5),\
such a requirement apparently looks completely satisfactory. Nevertheless,
it is incomplete, since by itself, it is not sufficient for the lattice to
be stable. For this to happen, one should take into account nearby
structures with \textbf{k}'s close to \textbf{k}$_{0}$. Since the
periodicity of $\delta \rho $ is built into the expansion, Eq.(1.2), through
its dependence on $\Psi _{\mathbf{k}\alpha }$, the only option allowing
structures with nearby \textbf{k}'s to be considered lies in making $\varphi
_{i}s$ \ in Eq. (1.2) weakly coordinate-dependent. In this case the
expression $A(P,T)\varphi ^{2}$ should be replaced by a more general
expression of the type 
\begin{equation}
\hat{A}=\int d\mathbf{r}\int d\mathbf{r}^{\prime }\delta \rho (\mathbf{r})h(%
\mathbf{r},\mathbf{r}^{\prime })\delta \rho (\mathbf{r}^{\prime })  \tag{1.6}
\end{equation}%
where the kernel is some function of $P,T$ symmetric with respect to
interchange of its spatial arguments. By construction, the $\mathbf{k}$%
\textbf{-}dependence of $A(P,T)$ is built into the transformation properties
of the kernel $h(\mathbf{r},\mathbf{r}^{\prime }).$ Specifically, \ for $h(%
\mathbf{r},\mathbf{r}^{\prime })$ consider an eigenvalue problem of the type 
\begin{equation}
\int d\mathbf{r}^{\prime }h(\mathbf{r},\mathbf{r}^{\prime })\phi (\mathbf{r}%
^{\prime })=\lambda \phi (\mathbf{r}).  \tag{1.7}
\end{equation}%
The choice of the kernel $h(\mathbf{r},\mathbf{r}^{\prime })$ is made in
such a way that the function, Eq.(1.5), is identified with the eigenfunction 
$\phi (\mathbf{r})$ in Eq.(1.7). Keeping in mind that for superlattices, $u_{%
\mathbf{k}\alpha }$ can be put equal to one and, taking into account the
translational invariance, the above eigenvalue problem can be rewritten
explicitly as 
\begin{equation}
\sum\limits_{l}\sum_{\mathbf{r}^{\prime }}h_{jl}(\mathbf{r}-\mathbf{r}%
^{\prime })e^{i\mathbf{k}_{l}\mathbf{\cdot r}^{\prime }}=\lambda _{j}e^{i%
\mathbf{k}_{j}\mathbf{\cdot r}}\text{ \ ,}  \tag{1.8}
\end{equation}%
where summation over $\mathbf{r}^{\prime }$ is made over vectors of the 
\textit{direct} lattice, while that over $l$ is made over the \textbf{k}
vectors forming a star. From here we obtain, 
\begin{equation}
\delta _{mj}\lambda _{j}=\sum\limits_{\mathbf{R}}h_{jm}(\mathbf{R})e^{-i%
\mathbf{k}_{m}\cdot \mathbf{R}}  \tag{1.9}
\end{equation}%
where $\mathbf{R}=\mathbf{r}-\mathbf{r}^{\prime }$. For $\mathbf{k}=\mathbf{k%
}_{0}$, this lattice is expected to be stable. To check that this is the
case, we use expansion, Eq.(1.2), for $\delta \rho $ in Eq.(1.6). To
actually do so, we \ need to be more specific about the expansion given by
Eq.(1.2). To this purpose we write 
\begin{equation}
\delta \rho (\mathbf{r})=\sum_{i}\varphi _{i}(\mathbf{r})e^{i\mathbf{k}_{i}%
\mathbf{\cdot r}}  \tag{1.10}
\end{equation}%
where $\mathbf{k}=\mathbf{k}_{0}+\mathbf{\kappa }$ \ with $\mathbf{\kappa }$
being an infinitesimally small vector. Our order parameter function $\varphi
_{i}(\mathbf{r})$ can also be Fourier expanded in such a way that 
\begin{equation}
\varphi _{i}(\mathbf{r})=\sum\limits_{\mathbf{\kappa }}c_{\mathbf{k}_{i0}}(%
\mathbf{\kappa })e^{i\mathbf{\kappa }\cdot \mathbf{r}}.  \tag{1.11}
\end{equation}%
Using expansions Eq.s (1.10) and (1.11) in Eq.(1.6) produces 
\begin{equation}
\hat{A}=\sum_{i}\sum\limits_{\mathbf{\kappa }}\sum_{j}\sum\limits_{\mathbf{%
\kappa }^{\prime }}c_{\mathbf{k}_{i0}}(\mathbf{\kappa })c_{\mathbf{k}%
_{i^{\prime }0}}(\mathbf{\kappa }^{\prime })\int d\mathbf{r}\int d\mathbf{r}%
^{\prime }e^{i(\mathbf{k}_{i}+\mathbf{\kappa )\cdot r}}h(\mathbf{r}-\mathbf{r%
}^{\prime })e^{i(\mathbf{k}_{j}+\mathbf{\kappa }^{\prime }\mathbf{)\cdot r}%
^{\prime }}.  \tag{1.12}
\end{equation}%
To simplify this expression we note that if $\mathbf{\kappa }^{\prime }s$ in
the double integral would be zero, the integral would reduce to the result
given by Eq.(1.9). Since the free energy at equilibrium is minimal, by
selecting the smallest eigenvalue $\lambda (\mathbf{k}_{0})$ from the set $%
\{\lambda _{j}\}$ we obtain the following result for $\hat{A},$%
\begin{equation}
\hat{A}=\lambda (\mathbf{k}_{0})\sum\limits_{i}\sum\limits_{\mathbf{\kappa }%
}c_{\mathbf{k}_{i0}}(\mathbf{\kappa })c_{\mathbf{k}_{i0}}^{\ast }(\mathbf{%
\kappa }),  \tag{1.13}
\end{equation}%
where $c_{\mathbf{k}_{i0}}^{\ast }=c_{-\mathbf{k}_{i0}}^{{}}$ . In arriving
at this result we kept only the diagonal (in $\mathbf{\kappa })$ term and
assumed (as Lifshitz did) that $c_{\mathbf{k}_{i0}}(\mathbf{\kappa }%
)^{\prime }s$ are very weakly dependent on $\mathbf{\kappa }$. Expanding the
exponents in the double integral in Eq.(1.12) in powers of $\mathbf{\kappa }%
, $ keeping only terms of the first order in $\mathbf{\kappa }$, and taking
into account the perturbative nature of such a calculation leads to the
following result: 
\begin{eqnarray}
&&\int d\mathbf{r}\int d\mathbf{r}^{\prime }e^{-i(\mathbf{k}_{i}+\mathbf{%
\kappa )\cdot r}}h(\mathbf{r}-\mathbf{r}^{\prime })e^{i(\mathbf{k}_{i}+%
\mathbf{\kappa )\cdot r}^{\prime }}  \notag \\
&=&\lambda _{i}+i\mathbf{\kappa }\cdot \int d\mathbf{r}\int d\mathbf{r}%
^{\prime }(\mathbf{r}^{\prime }-\mathbf{r})e^{-i\mathbf{k}_{i}\mathbf{\cdot r%
}}h(\mathbf{r}^{\prime }-\mathbf{r})e^{i\mathbf{k}_{i}\mathbf{\cdot r}%
^{\prime }}+O(\mathbf{\kappa }^{2})  \notag \\
&=&\lambda _{i}+i\mathbf{\kappa }\cdot \int d\mathbf{r}\int d\mathbf{r}%
^{\prime }\mathbf{r}^{\prime }h(\mathbf{r}^{\prime }-\mathbf{r})\{e^{-i%
\mathbf{k}_{i}\mathbf{\cdot r}}e^{i\mathbf{k}_{i}\mathbf{\cdot r}^{\prime
}}-e^{-i\mathbf{k}_{i}\mathbf{\cdot r}^{\prime }}e^{i\mathbf{k}_{i}\mathbf{%
\cdot r}}\}  \notag \\
&&+O(\mathbf{\kappa }^{2}).  \TCItag{1.14}
\end{eqnarray}%
While the terms of order $O(\mathbf{\kappa }^{2})$ will be considered in the
next subsection, here we would like to combine Eq.s (1.12), (1.14) in order
to present our results in a physically suggestive form. Taking into account
Eq.(1.9) we obtain$:$%
\begin{equation}
\hat{A}=\sum\limits_{i}\sum\limits_{\mathbf{\kappa }}\{\lambda (\mathbf{k}%
_{0})+i(\mathbf{\kappa }\cdot \frac{\partial }{\partial \mathbf{k}}\lambda (%
\mathbf{k})\mid _{\mathbf{k}=\mathbf{k}_{0}}+O(\mathbf{\kappa }^{2})\}c_{%
\mathbf{k}_{i0}}(\mathbf{\kappa })c_{\mathbf{k}_{i0}}^{\ast }(\mathbf{\kappa 
})  \tag{1.15}
\end{equation}%
where, clearly, 
\begin{equation}
\frac{\partial }{\partial \mathbf{k}}\lambda (\mathbf{k})\mid _{\mathbf{k}=%
\mathbf{k}_{0}}=\sum\limits_{\mathbf{r}}\sum\limits_{\mathbf{r}^{\prime }}%
\mathbf{r}^{\prime }h(\mathbf{r}^{\prime }-\mathbf{r})\{e^{-i\mathbf{k}_{i}%
\mathbf{\cdot r}}e^{i\mathbf{k}_{i}\mathbf{\cdot r}^{\prime }}-e^{-i\mathbf{k%
}_{i}\mathbf{\cdot r}^{\prime }}e^{i\mathbf{k}_{i}\mathbf{\cdot r}}\} 
\tag{1.16}
\end{equation}%
with integrals being replaced by lattice sums. Thus, in accord with Lifshitz
[\textbf{4,6}],\ who only suggests (\textit{without derivation}) that the
requirement $\frac{\partial }{\partial \mathbf{k}}\lambda (\mathbf{k})\mid _{%
\mathbf{k}=\mathbf{k}_{0}}=0$ is equivalent to the requirement of vanishing
of invariants of the type \{$e^{-i\mathbf{k}_{i}\mathbf{\cdot r}}e^{i\mathbf{%
k}_{i}\mathbf{\cdot r}^{\prime }}-e^{-i\mathbf{k}_{i}\mathbf{\cdot r}%
^{\prime }}e^{i\mathbf{k}_{i}\mathbf{\cdot r}}\}$, we have just demonstrated
that this is indeed the case. \ This equivalence is absolutely essential for
determination of the symmetry of the low temperature phase. Its existence
also implies the vanishing of the coefficient $C$ in the Landau expansion,
Eq.(1.1), [\textbf{10}]. These two conditions are the necessary and
sufficient conditions for the second order (continuous)\ phase transition to
take place. Violation of these conditions causes the order of phase
transition change from second to first. The details can be found in
previously cited literature. From the preceding discussion, it should be
clear that \textit{both} options depend crucially on the assigned symmetry
of the disordered phase which should be artificially introduced into the
theory.

\bigskip

\subsection{\textbf{Refinements by Ginzburg, Landau and Wilson. LGW free
energy\ functional}}

\ 

Ginzburg and Landau [\textbf{5}] applied the Landau theory for study of
superconductivity in restricted geometries. Subsequently (albeit in a
different context) the Ginzburg-Landau \ (G-L) functional was reobtained by
Kenneth Wilson, who not only recovered the G-L functional, but explained
what approximations should be made in order to recover this functional from
the microscopic (Ising) model. In view of his refinements, the G-L
functional is frequently referred to as the Landau-Ginzburg-Wilson (LGW)
functional [\textbf{14}]. It is written as

\begin{equation}
\mathcal{F}\{\varphi \}=\mathcal{F}_{0}+\frac{1}{2}\int d\mathbf{x}\{c\left(
\nabla \varphi (\mathbf{x})\right) ^{2}+a\varphi ^{2}(\mathbf{x})+\frac{b}{2}%
\varphi ^{4}(\mathbf{x})-2h(\mathbf{x})\varphi (\mathbf{x})\},  \tag{1.17}
\end{equation}%
where in the case of the Ising model the last term indirectly describes the
coupling of the\ Ising spins to the external magnetic field. In the absence
of such a field the above functional describes the second order phase
transition as discussed earlier\footnote{%
In this work we shall call \ the functional $\mathcal{F}\{\varphi \}-%
\mathcal{F}_{0}$ \ (with $h=0$) as Ginzburg-Landau (or G-L) and we shall
call it LGW if it is used in the exponent of the path integral, Eq.(1.27)
below. \ Such distinction is needed for mathematical reasons : for the G-L
functional the order parameter $\varphi $ is stictly nonnegative as we had
explained already in the main text. This allows us to relate the G-L and
Yamabe functionals (Section 3) to each other.}. Usually, it is rather
difficult to bring the coefficients of the \ Landau expansion in exact
correspondence with parameters of respective lattice models [\textbf{15,16}%
]. This is especially true for the constant $b$. It is expected, however,
that under any circumstance, $b$ is only very weakly temperature and
pressure dependent, and normally it is greater than zero. The case when both 
$a$ and $b$ are zero is a special case. It determines the so called
tricritical point in the $P$,$T$ parameter space. Physically, there can be
only one tricritical point ($P_{t}$, $T_{t})$ determined by the equations 
\begin{equation}
a(P_{t},T_{t})=0;b(P_{t},T_{t})=0.  \tag{1.18}
\end{equation}%
Below the tricritical temperature $T_{t}$ the parameter $b$ becomes negative
and the phase transition is of first order. In this case, it is necessary to
introduce an extra term $g\varphi ^{6}(\mathbf{x})$ into the LGW functional,
Eq.(1.17), with $g>0$. It is normally expected that the parameter $g$ is
practically constant below $T_{c}$ [\textbf{17}]$.$

Eq.(1.17) contains the magnetic field $h$ whose role we would like to
discuss now. To this purpose, let us consider the extremum of the LGW
functional near $T_{c}.$We obtain, 
\begin{equation}
\frac{\delta \mathcal{F}\{\varphi \}}{\delta \varphi (\mathbf{x})}%
=-c\bigtriangledown ^{2}\varphi +a\varphi +b\varphi ^{3}-h=0.  \tag{1.19}
\end{equation}%
Consider a special case of a constant (coordinate-independent) field $%
\varphi $, $\varphi (\mathbf{x})=\varphi _{0}$. In this case, Eq.(1.19) is
reduced to 
\begin{equation}
a\varphi _{0}+b\varphi _{0}^{3}=h.  \tag{1.20}
\end{equation}%
For $a>0$ and a weak magnetic field $h$, we obtain $\varphi _{0}=h/a$. For $%
a<0$ $\ $we obtain solutions even for $h=0$. These are $\varphi _{0}=\pm
(\left\vert a\right\vert /b)^{\frac{1}{2}}$ and $\varphi _{0}=0.$ The
solution $\varphi _{0}=0$ is not a minimum of the free energy and, hence,
should be discarded. Of the two other solutions, the system chooses one of
them (which is interpreted in the literature as \textit{spontaneous symmetry
breaking}). They both have the same free energy\footnote{%
Once the choice is made, the order parameter can be considered in all
subsequent calculations as positive.}. Let now $\varphi (\mathbf{x})=\varphi
_{0}+\delta \varphi \equiv \varphi _{0}+\eta (\mathbf{x}),$ so that we can
Taylor series expand $\mathcal{F}\{\varphi \},$ thus obtaining 
\begin{equation}
\mathcal{F}\{\varphi \}=\mathcal{F}\{\varphi _{0}\}+\frac{1}{2}\int dV\int
dV^{\prime }\left[ \frac{\delta ^{2}}{\delta \varphi (\mathbf{x})\delta
\varphi (\mathbf{x}^{\prime })}\mathcal{F}\{\varphi \}\mid _{\substack{ %
\varphi =\varphi _{0}  \\ h=0}}\right] \eta (\mathbf{x})\eta (\mathbf{x}%
^{\prime })+...,  \tag{1.21}
\end{equation}%
where 
\begin{equation}
\frac{\delta ^{2}}{\delta \varphi (\mathbf{x})\delta \varphi (\mathbf{x}%
^{\prime })}\mathcal{F}\{\varphi \}\mid _{\substack{ \varphi =\varphi _{0} 
\\ h=0}}=\left( -c\bigtriangledown ^{2}+a+3b\varphi _{0}^{2}\right) \delta (%
\mathbf{x}-\mathbf{x}^{\prime }).  \tag{1.22}
\end{equation}%
Clearly, we are interested only in the phase with lowest free energy. In
this case $a<0$ and, therefore, we obtain, $-\left\vert a\right\vert
+3b\varphi _{0}^{2}=2\left\vert a\right\vert $. This result implies that,
indeed, the selected minimum is stable since the spectrum of the operator
given above contains only nonnegative eigenvalues. This fact was used by
Levanyuk and then by Ginzburg in order to determine the limits of validity
of the Landau theory. Details of their arguments can be found, for example,
in Refs[\textbf{14,18}]. Following these references, the Ginzburg number 
\textbf{Gi }is\textbf{\ }defined\textbf{\ }in 3 dimensions\textbf{\ }as 
\begin{equation}
\mathbf{Gi=}b^{2}T_{c}^{2}/\alpha (P)c^{3},  \tag{1.23}
\end{equation}%
where $\alpha (P)$ was defined earlier, after Eq.(1.1). In order to explain
its physical meaning, we need to introduce two characteristic lengths. This
is accomplished by the use of \ Eq.s (1.21) and (1.22). Let $%
m^{2}=a+3b\varphi _{0}^{2},$ and consider the correlators defined by 
\begin{equation}
\left\langle \eta (\mathbf{x})\eta (\mathbf{x}^{\prime })\right\rangle =%
\frac{\int D[\eta (\mathbf{x})]\eta (\mathbf{x})\eta (\mathbf{x}^{\prime
})\exp (-\beta S[\eta (\mathbf{x})])}{\int D[\eta (\mathbf{x})]\exp (-\beta
S[\eta (\mathbf{x})])},  \tag{1.24}
\end{equation}%
where $S[\eta (\mathbf{x})]=\frac{1}{2}\int d^{d}\mathbf{r}\{c\left(
\bigtriangledown \eta (\mathbf{x})\right) ^{2}+m^{2}\eta ^{2}(\mathbf{x})\}$
and $\beta =1/T,$ with $T$ being temperature (in the system of units in
which Boltzmann's constant $k_{B}=1$). Calculation of such averages for
temperatures $T$ close to $T_{c}$ is described in any textbook on path
integrals [\textbf{18,19}] and is readily accomplished \ in any dimension $d$
due to the Gaussian nature of the respective path integrals in the numerator
and denominator. In particular, for $d=3,$ the result \ is given by 
\begin{equation}
\left\langle \eta (\mathbf{x})\eta (\mathbf{x}^{\prime })\right\rangle
\simeq \frac{T_{c}e^{-r/\xi }}{4\pi cr}\text{, \ }r=\left\vert \mathbf{x}-%
\mathbf{x}^{\prime }\right\vert ,\text{ }r/\xi >>1.  \tag{1.25}
\end{equation}%
The above expression defines the correlation length $\xi $ via $\xi =\left(
c/m^{2}\right) ^{1/2}.$ Consider the volume average of such a correlator 
\begin{equation}
\frac{1}{V^{2}}\int d^{3}x\int d^{3}x^{\prime }<\eta (\mathbf{x})\eta (%
\mathbf{x}^{\prime })>\sim \frac{\xi ^{2}}{cr_{0}^{3}}  \tag{1.26}
\end{equation}%
where the characteristic length $r_{0}$ is given by $r_{0}^{3}\sim V$ and
can be estimated by equating this average with the previously obtained
Landau result, $\varphi _{0}^{2}=\left\vert a\right\vert /b\sim m^{2}/b.$
This produces the following estimate for $r_{0}$: $r_{0}^{-3}\sim
m^{4}/T_{c}b$.\ Consider now the ratio $\left( r_{0}/\xi \right) ^{6}\sim
\tau ^{-1}\mathbf{Gi,}$ with $\tau $ defined after Eq.(1.1). Clearly, the
original Landau theory makes sense only if $m^{2}/b>>\xi ^{2}/cr_{0}^{3}$,
that is, when fluctuations are negligible. This happens when $\left\vert
\tau \right\vert >>\mathbf{Gi,}$ i.e. for the range of temperatures: $%
\left\vert T-T_{c}\right\vert >>T_{c}\mathbf{Gi.}$ When \ approaching $%
T_{c}, $ this inequality is inevitably violated. In this case, fluctuation
corrections should be taken into account as explained by Wilson, Fisher,
Kadanoff and many others [\textbf{14,18}]. Theoretically, however, one can
look at these results in different space dimensions. This allows one to
introduce the upper $d_{u}$ and the lower $d_{l}$ critical dimensions. For $%
d>d_{u}$, the Landau-Lifshitz theory can be used with confidence since
fluctuations become unimportant. For $d<d_{l}$ fluctuations are so
significant at any non zero temperatures that phase transitions do not
occur. Exactly at $d_{u}$ fluctuations are still important. Technically
speaking, $d_{u}$ is the dimensionality at which LGW model is
renormalizable. This means that the functional integral of the type 
\begin{equation}
I=\int D[\varphi (\mathbf{x})]\exp \{-\beta \hat{S}[\varphi (\mathbf{x})]\} 
\tag{1.27}
\end{equation}%
with $\hat{S}[\varphi (\mathbf{x})]=\mathcal{F}\{\varphi \}-\mathcal{F}%
\{\varphi _{0}\}$ can be consistently calculated perturbatively by expanding
the exponent in the above path integral in powers of \ the coupling constant 
$b$, which is dimensionless exactly at $d_{u}$\footnote{%
In the case of LGW functional $d_{u}=4$ while for the tricritical GLW
functional, $d_{u}=3$ [\textbf{17}].} .\textquotedblright
Consistently\textquotedblright\ here means in such a way that the
singularities occurring in perturbative calculations can be consistently
removed, thus making the LGW model renormalizable. Details can be found in
the literature [\textbf{15,19}]. The sophisticated machinery of the
renormalization group method, a by-product of such a renormalization
procedure, allows one to calculate the critical exponents associated with
various thermodynamic observables.

\ 

\subsection{\textbf{Description of the} \textbf{\ problems to be solved and
organization of the rest of the paper}}

\bigskip

The results just described allow us to formulate the following problems to
be considered in the rest of our paper. \ 

1. As the Fig. 1 and \ the Table 1 indicate, symmetry of the high
temperature (disordered) phase for CuZn alloy is fcc as depicted in Fig.2a.
The existing G-L theory needs this information \textbf{to be supplied in
advance as an input} in order to determine the cascade of symmetry-breaking
changes taking place either by lowering temperature at fixed concentration
or by changing concentration at fixed temperature. Can such information be
obtained by the proper re-interpretation of the G-L theory?

2. There is well a developed string-inspired formalism of conformal field
theories in two dimensions. It takes into account full conformal invariance
at criticality [\textbf{20}]. Can the G-L theory be modified in order to be
in accord with the existing two dimensional formalism ? \ That is to say, in
higher dimensions the LGW functional is normally used in the exponent of the
path integral, e.g. see Eq.(1.27). Can LGW theory be modified \ (to account
for full conformal invariance) to be used in the exponent of such a
string-inspired path integral? To what extent \ does such a modification
allow us to develop the exact higher dimensional analogs of \ two
dimensional manifestly full conformal invariant theories? By doing so, under
what conditions can we recover known higher dimensional perturbative results
for the LGW model?

In this work we provide affirmative answers to these problems. \ Based on
this, we are able to consider the whole spectrum of physically interesting
problems whose mathematical description is based entirely on solutions to
problems just formulated. Specifically,

a) In Section 2 we perform scaling analysis of the G-L model. We argue that
in dimensions 3 and higher such scaling analysis is insufficient to insure
the full conformal invariance of the G-L functional \textit{even at
criticality }(i.e. for $T=T_{c}$). Nevertheless, we argue that such
analysis, when properly (re) interpreted, is helpful in finding the correct
form of a conformally invariant G-L functional. It is found in Section 3,
where it is shown to coincide with the Yamabe functional known in
mathematics literature for some time[\textbf{21-25}].

b) In section 4 we argue that with respect to conformal variations, this
functional is equivalent to the Hilbert-Einstein (H-E) functional for pure
gravity in the presence of \ the cosmological term. The physical meaning of
this cosmological term in the context of the G-L theory is explained in
Sections 3 and 4. Its presence is of importance for both statistical and
high energy physics.

c) To avoid confusion, only statistical physics applications are treated in
this paper. In particular, to obtain better insight \ into the nature of the
Yamabe functional (which can be used only in dimensions 3 and higher), in
Section 5 we describe in some detail its two dimensional analog. This
Yamabe-like functional was considered in detail in the work by Osgood,
Phillips and Sarnak (OPS), Ref.[\textbf{26}]. Section 5 is not merely a
review of the OPS work. In it, we accomplish several tasks. First, we
connect the OPS results with those\textit{\ independently} obtained in
physics literature [\textbf{27},\textbf{28}], where the OPS-type functional
can be found in the exponents of string-theoretic path integrals. Such path
integrals are used for calculation of observables (e.g. averages of the
vertex operators) in CFT. They establish the most efficient direct link
between the string and the CFT formalisms. Next, we argue that the OPS
functional \textit{is the exact two dimensional analog of the Yamabe
functional. The Yamabe should be used }(\textit{instead of the OPS})\textit{%
\ in higher dimensions.} This observation allows us to extend (later, in
Section 6) the available two dimensional string-theoretic results to higher
dimensions.

d) In particular, as a precursor to these 3 dimensional calculations, \ we
reinterpret thermodynamically the inequalities obtained in OPS paper. Such a
reinterpretation is essential in determining the symmetry of the high
temperature phase thus allowing us \ to complete the work by Lifshitz. The
results by Chowla and Selberg [\textbf{29}] were used in order to
recalculate exactly the values of functional determinants for 2 dimensional
lattices of different symmetry. The logarithms of these determinants are
associated with the respective free energies. The obtained \textit{exact}
spectrum of free energies in two dimensions provides us with useful
reference for analogous 3 dimensional calculations done in the next section.

e) Even though the analytical expression for such a 3 dimensional spectrum
can also be obtained formally in closed form, as we demonstrate, in
practice, some numerical calculations were required in Section 6. The
results of these calculations are in complete qualitative accord with the
exact two dimensional results obtained earlier.\ Some details of our
calculations are\ presented\ in appendices A through D.

\bigskip

\section{\protect\bigskip Scaling analysis and conformal invariance of the
Ginzburg-Landau functional}

\bigskip

The conventional scaling analysis of the G-L functional routinely used in
physics literature can be found, for example, in the book by Amit [\textbf{15%
}]. This analysis differs somewhat from that for the $\phi ^{4}$ model as
described in the monograph by Itzykson and Zuber [\textbf{30}]. For the sake
of uninterrupted reading we would like to provide a sketch of arguments for
both cases now.

We begin with the $\phi ^{4}$ model. Let $\mathcal{L}(x)$ be the Lagrangian
of this scalar field model whose action functional in $d$ dimensions $S[\phi
]$ is given by $S[\phi ]=\int d^{d}x\mathcal{L}(x)$. Let furthermore $%
\lambda $ be some nonnegative parameter. Then the requirement that $S[\phi ]$
is independent of $\lambda $, i.e. $\int d^{d}x\mathcal{L}(x)=\int
d^{d}x\lambda ^{d}\mathcal{L}(\lambda x),$ leads to the constraint 
\begin{equation}
\int d^{d}x(x\cdot \frac{\partial }{\partial x}+d)\mathcal{L}(x)=0  \tag{2.1}
\end{equation}%
obtained by differentiation \ of \ $S[\phi ]$ with respect to $\lambda $
with $\lambda =1$ at the end of calculation. For $\mathcal{L}(x)$ given by 
\begin{equation}
\mathcal{L}(x)=\frac{1}{2}\left( \bigtriangledown \phi \right) ^{2}+\frac{%
m^{2}}{2}\phi ^{2}+\frac{\hat{G}}{4!}\phi ^{4}  \tag{2.2}
\end{equation}%
the change of $\mathcal{L}(x)$ under the infinitesimal scale transformation
is given by\footnote{%
In arriving at this result we took into account Eq.(13-40) of Ref. [\textbf{%
30}] along with condition $D=\frac{d}{2}-1$, with $D$ defined in Eq.(13-38)
of the same reference. Also, we had changed signs (as compared to the
original sourse) in $\mathcal{L}(x)$ to be in accord with accepted
conventions for the G-L functional.} 
\begin{equation}
\frac{\delta \mathcal{L}}{\delta \varepsilon }=(x\cdot \frac{\partial }{%
\partial x}+d)\mathcal{L}(x)\text{ }\mathcal{+}(d-4)\frac{\hat{G}}{4!}\phi
^{4}-m^{2}\phi ^{2}.  \tag{2.3}
\end{equation}%
Comparison between Eq.s(2.1) and (2.3) implies that the action $S[\phi ]$ is
scale invariant if $d=4$ and $m^{2}=0.$ The result just obtained raises the
immediate question: given that the massless G-L action is scale invariant
for $d=4$, will it be also conformally invariant under the same conditions?
We provide the answer to this question in several steps.

First, let $M$ be some Riemannian manifold whose metric is $g$. Then, any
metric $\tilde{g}$ conformal to $g$ can be written as $\tilde{g}=e^{f}g$
with $f$ being a smooth real valued function on $M$ [\textbf{23}]. Let $%
\Delta _{g}$ be the Laplacian associated with metric $g\footnote{%
That is $\Delta _{g}\Psi $ $=-(\det g)^{-\frac{1}{2}}\partial
_{i}(g^{ij}(\det g)^{\frac{1}{2}}\partial _{j}\Psi )$ for some scalar
function $\Psi (x).$}$ and, accordingly, let $\Delta _{\tilde{g}}$ be the
Laplacian associated with metric $\tilde{g}.$ Richardson [\textbf{31}]
demonstrated that 
\begin{equation}
\Delta _{\tilde{g}}=e^{-f}\Delta _{g}-\frac{1}{2}(\frac{d}{2}%
-1)fe^{-f}\Delta _{g}-\frac{1}{2}(\frac{d}{2}-1)e^{-f}(\Delta _{g}f)+\frac{1%
}{2}(\frac{d}{2}-1)e^{-f}\Delta _{g}\circ f  \tag{2.4}
\end{equation}%
(here $\left( \Delta _{g}\circ f\right) \Psi $ should be understood as $%
\Delta _{g}(f(x)\Psi (x)).$ \ Using basic facts from \ bosonic string string
theory, we notice at once that only for $d=2$ does one obtain the
conformally invariant (string-type) functional $S[\mathbf{X}]=\int_{M}d^{2}x%
\sqrt{g}(\bigtriangledown _{g}\mathbf{X})\cdot (\bigtriangledown _{g}\mathbf{%
X})$ (to be discussed later in Section 5). For $d>2$, in view of Eq.(2.4),
the conformal invariance of the $\phi ^{4}$ model is destroyed. In
particular, this means that Eq.(1.19) (for $h=0$) is not conformally
invariant \textit{even at criticality} ($a=0$)! Thus, using conventional
field-theoretic perturbational methods one encounters a problem already at
the zeroth order (in the coupling constant $\hat{G}$)\ level, unless $d=2$.
This problem formally does not occur if \ one requires our physical model to
be only scale invariant. This is undesirable however in view of the fact
that in 2 dimensions an arbitrary conformal transformation of the metric
tensor $g$ of the underlying two dimensional manifold $M$ is permissible at
criticality [\textbf{20}]. \textit{Abandoning the requirement of general
conformal invariance in 2 dimensions in favor of scale invariance in higher
dimensions would destroy all known string-theoretic methods of obtaining
exact results in two dimensions.} Our general understanding of critical
phenomena depends crucially on our ability to solve two dimensional models
exactly. All critical properties for the same type of models in higher
dimensions are expected to hold even in the absence of exact solvability.
Fortunately, the situation can be considerably improved by reanalyzing and
properly reinterpreting the scaling results for dimensions higher than 2.

This \ observation leads to the next step in our arguments. Using the book
by Amit [\textbf{15}], we consider first scaling of the non interacting
(free) G-L theory whose action functional is given by 
\begin{equation}
S[\phi ]=\int d^{d}x\{(\bigtriangledown \phi )^{2}+m^{2}\phi ^{2}\}. 
\tag{2.5}
\end{equation}%
Suppose now that, upon rescaling, the field $\phi $ transforms according to
the rule\footnote{%
Notice that this is already a special kind of conformal transformation}: 
\begin{equation}
\tilde{\phi}(Lx)=L^{\omega }\phi (x).  \tag{2.6}
\end{equation}%
If we require 
\begin{equation}
\int d^{d}x\{(\bigtriangledown \phi )^{2}+m^{2}\phi ^{2}\}=\int
d^{d}xL^{d}\{(\tilde{\bigtriangledown}\tilde{\phi})^{2}+\tilde{m}^{2}\tilde{%
\phi}^{2}\}  \tag{2.7}
\end{equation}%
and use Eq.(2.6) \ we obtain, 
\begin{equation}
S[\phi ]=\int d^{d}xL^{d}\{(\bigtriangledown \phi )^{2}L^{2\omega -2}+\tilde{%
m}^{2}\phi ^{2}L^{2\omega }\}.  \tag{2.8}
\end{equation}%
In order for the functional $S[\phi ]$ to be scale invariant the mass $m^{2}$
should scale as: $\tilde{m}^{2}=m^{2}L^{-2}$. With this requirement the
exponent $\omega $ is found to be $\omega =1-\frac{d}{2}$. This scaling of
the mass is in accord with Amit [\textbf{15}], page 26, Eq.(2.72). It comes
directly from the rescaling of the G-L functional, Eq.(1.17), causing the
coefficient of the gradient term to become one. In the standard
field-theoretic $\phi ^{4}$-type model, the coefficient of the gradient term
is one already by the existing convention and, hence, the free field model $(%
\tilde{g}=0$) is scale-invariant only if the mass term is zero in accord
with Eq.(2.3). Hence, although the field-theoretic methods are applicable to
both $S[\phi ]$ and the G-L \ functionals, the G-L functional is \textsl{not}
entirely equivalent to the standard $\phi ^{4}$ model functional with
respect to scaling properties.

\ The above scaling can be done a bit differently. This is also discussed in
the book by Amit. We would like to use such different scaling to our
advantage. Following Amit [\textbf{15}], we notice that although the action $%
S[\phi ]$ is scale invariant, there is some freedom of choice for the
dimensionality of the field $\phi $. For instance, instead of $S[\phi ]$ we
consider 
\begin{equation}
S[\phi ]=\frac{1}{a^{d}}\int d^{d}x\{(\bigtriangledown \phi )^{2}+m^{2}\phi
^{2}\}  \tag{2.9}
\end{equation}%
with $a^{d}$ being some volume, say, $a^{d}=\int d^{d}x$. Then, by repeating
arguments associated with Eq.(2.8) we obtain $\omega =1$ (instead of
previously obtained $\omega =1-\frac{d}{2}),$ e.g. see Eq.(2-66) in the book
by Amit [\textbf{15}]$.$ Although, from the point of view of scaling
analysis both results are actually equivalent, they become quite different
if we want to extend such scaling analysis by considering general conformal
transformations. Although, in view of Eq.(2.4), such a task seems impossible
to accomplish, fortunately, this is not true, as we demonstrate below.

The next step can be made by noticing that the mass term scales as scalar
curvature $R$ \ for some Riemannian manifold $M$, i.e. the scaling $\tilde{m}%
^{2}=m^{2}L^{-2}$ is \textit{exactly the same} as the scaling of $R$ given
by 
\begin{equation}
\tilde{R}=L^{-2}R.  \tag{2.10}
\end{equation}%
This result can be found, for example, in the book by Wald, Ref. [\textbf{32}%
], Eq.(D.9), page 446. In general, the scalar curvature $R(g)$ changes under
the conformal transformation $\tilde{g}=e^{2f}g$ according to the rule [%
\textbf{23}] 
\begin{equation}
\tilde{R}(\tilde{g})=e^{-2f}\{R(g)-2(d-1)\Delta _{g}f-(d-1)(d-2)\left\vert
\bigtriangledown _{g}f\right\vert ^{2}\}  \tag{2.11}
\end{equation}%
where $\Delta _{g}f$ is the Laplacian of $f$ and $\bigtriangledown _{g}f$ is
the covariant derivative defined with respect to metric $g$. \ From here we
see that, indeed, for constant $f$ 's the scaling takes place in accord with
Eq.(2.10). Now, however, we can do more.

Following Lee and Parker [\textbf{23}], we would like to simplify the above
expression for $R$. To this purpose we introduce a substitution: $%
e^{2f}=\varphi ^{p-2}$, where $p=\frac{2d}{d-2},$ so that $\tilde{g}=\varphi
^{p-2}g.$ With such a substitution, Eq.(2.11) acquires the following form: 
\begin{equation}
\tilde{R}(\hat{g})=\varphi ^{1-p}(\alpha \Delta _{g}\varphi +R(g)\varphi ), 
\tag{2.12}
\end{equation}%
with $\alpha =4\frac{d-1}{d-2}.$ Clearly, such an expression makes sense
only for $d\geq 3$ and breaks down for $d=2$. But we know already the action 
$S[\mathbf{X}]$ which is both scale and conformally invariant in $d=2$. It
is given after Eq.(2.4) and will be discussed further in Section 5. The
results of this section will allow us to obtain similar actions, which are
both scale and conformally invariant in dimensions 3 and higher. This is
described in the next section.

\bigskip

\section{Ginzburg-Landau functional and the Yamabe problem}

\bigskip

We begin with the following observation. Let $\tilde{R}(\tilde{g})$ in
Eq.(2.11) be some constant (that this is indeed the case will be demonstrate
shortly below). Then Eq.(2.12) acquires the following form 
\begin{equation}
\alpha \Delta _{g}\varphi +R(g)\varphi =\tilde{R}(\tilde{g})\varphi ^{p-1}. 
\tag{3.1}
\end{equation}%
By noticing that $p-1=\frac{d+2}{d-2}$ we obtain at once: $p-1=3$ (for $d=4$%
) and $p-1=5$ (for $d=3$). These are familiar Ginzburg-Landau values for the
exponents of \ interaction terms for critical, Eq.(1.19), and tricritical
G-L theories\footnote{%
E.g. read the discussion after Eq.(1.18) and take into account that in the
tricritical case $d_{u}=3.$}. Once we recognize these facts, the action
functional producing the G-L-type Eq.(3.1) can be readily constructed. For
this purpose it is sufficient to rewrite Eq.(2.9) in a manifestly covariant
form. We obtain, 
\begin{equation}
S[\varphi ]=\frac{1}{\left( \int_{M}d^{d}x\sqrt{g}\varphi ^{p}\right) ^{%
\frac{2}{p}}}\int_{M}d^{d}x\sqrt{g}\{\alpha (\bigtriangledown _{g}\varphi
)^{2}+R(g)\varphi ^{2}\}\equiv \frac{E[\varphi ]}{\left\Vert \varphi
\right\Vert _{p}^{2}}.  \tag{3.2}
\end{equation}%
Minimization of this functional produces the following Euler-Lagrange
equation 
\begin{equation}
\alpha \Delta _{g}\bar{\varphi}+R(g)\bar{\varphi}-\lambda \bar{\varphi}%
^{p-1}=0  \tag{3.3}
\end{equation}%
with constant $\lambda $ denoting the extremum value for the ratio: 
\begin{equation}
\lambda =\frac{E[\bar{\varphi}]}{\left\Vert \bar{\varphi}\right\Vert _{p}^{p}%
}=\inf \{S[\varphi ]:\tilde{g}\text{ conformal to }g\text{\}.}  \tag{3.4}
\end{equation}%
In accord with Landau theory [\textbf{3}], it is expected that the conformal
factor $\varphi $ is a smooth nonnegative function on $M$ achieving its
extremum value $\bar{\varphi}$ . Comparison between Eq.s(3.1) and (3.3)
implies that $\lambda =\tilde{R}(\tilde{g})$ as required. These results
belong to Yamabe, who obtained the\ Euler-Lagrange G-L-type Eq.(3.3) upon
minimization of the functional $S[\varphi ]$ without knowlege of Landau
theory. The constant $\lambda $ is known in literature as the \textit{Yamabe
invariant }[\textbf{23,33}]. Its value is an invariant of the conformal
class $(M,g)$. The \textit{Yamabe problem} lies in finding a compact
Riemannian manifold (\textit{M,g}) of dimension $n\geq 3$\ whose metric is
conformal to metric $\tilde{g}$\ producing constant scalar curvature.

\ Subsequent developments, e.g. that given in Ref.[\textbf{34,35}] extended
this problem to manifolds with boundaries and to non compact manifolds. It
is not too difficult to prove that the (Yamabe-Ginzburg-Landau-like)
functional is manifestly conformally invariant. To this purpose, we need to
rewrite Eq.(2.12) in the following equivalent form 
\begin{equation}
\varphi ^{p}\tilde{R}(\tilde{g})=(\alpha \varphi \Delta _{g}\varphi
+R(g)\varphi ^{2}).  \tag{3.5}
\end{equation}%
It can be used in order to rewrite $E[\varphi ]$ as follows: $E[\varphi
]=\int d^{d}x\sqrt{\tilde{g}}\tilde{R}(\tilde{g})$. Next, by noting that $%
\int d^{d}x\sqrt{\tilde{g}}=\int d^{d}x\sqrt{g}\varphi ^{p},$ we can rewrite
the Yamabe functional in the Hilbert-Einstein form 
\begin{equation}
S[\varphi ]=\frac{\int d^{d}x\sqrt{\tilde{g}}\tilde{R}(\hat{g})}{\left( \int
d^{d}x\sqrt{\tilde{g}}\right) ^{\frac{2}{p}}}  \tag{3.6}
\end{equation}%
where both the numerator and the denominator are invariant with respect to
changes in the metric. This becomes especially clear if we recall that $%
\tilde{R}(\tilde{g})$ is a constant.

\ In order to use these results in statistical mechanics, we require that
the extremum of the Yamabe functional $S[\varphi ]$ is realized for
manifolds $M$ whose scalar curvature $R(g)$ in Eq.(3.3) is also constant. In
view of the relation $\int d^{d}x\sqrt{\tilde{g}}=\int d^{d}x\sqrt{g}\varphi
^{p},$ it is clear that for the fixed background metric $g$ Eq.(3.3) can be
obtained alternatively using the following variational functional 
\begin{equation}
\tilde{S}[\varphi ]=\int d^{d}x\sqrt{g}\{\alpha (\bigtriangledown
_{g}\varphi )^{2}+R(g)\varphi ^{2}\}-\widetilde{\lambda }\int d^{d}x\sqrt{g}%
\varphi ^{p}  \tag{3.7}
\end{equation}%
where the Lagrange multiplier $\widetilde{\lambda }$\footnote{%
Actually, $\widetilde{\lambda }=$ $\lambda \frac{2}{p}\equiv $ $\lambda 
\frac{d-2}{d}$} is responsible for the volume constraint. This form of the
functional $\tilde{S}[\varphi ]$ brings this higher dimensional result in
accord with that appropriate for two dimensions (e.g. see below Section 5,
Eq.(5.24)).

\ 

Apart from the normalizing denominator, Eq.(3.6) represents the
Hilbert-Einstein action for pure gravity in $d$ dimensions. In the
denominator, the volume $V$ taken to power $\frac{2}{p},$ serves to make $%
S[\varphi ]$ conformally invariant [\textbf{22}], page 150.

\section{\protect\bigskip Ginzburg-Landau from Hilbert-Einstein}

In this section we analyze significance of the cosmological constant in the
Hilbert-Einstein action for gravity from the point of view of the G-L model.
To this purpose, following Dirac [\textbf{36}], consider the extended
Hilbert-Einstein action functional defined for some pseudo Riemannian
manifold $M$ of total space-time dimension $d$, without boundary: 
\begin{equation}
S^{c}(g)=\int_{M}R\sqrt{g}d^{d}x+C\int_{M}d^{d}x\sqrt{g}\text{.}  \tag{4.1.}
\end{equation}%
The (cosmological) constant $C$ is determined by the following arguments.
Let $R_{ij}$\ be the Ricci curvature tensor, so that the $Einstein$\ $space$%
\ is defined as a solution of the following vacuum Einstein equation 
\begin{equation}
R_{ij}=\lambda g_{ij}  \tag{4.2}
\end{equation}%
with $\lambda $\ being constant\textit{. }From this definition it follows
that 
\begin{equation}
R=d\lambda .  \tag{4.3}
\end{equation}%
Following Dirac [\textbf{36}], variation of the action $S^{c}(g)$ produces 
\begin{equation}
R_{ij}-\frac{1}{2}g_{ij}R+\frac{1}{2}Cg_{ij}=0.  \tag{4.4}
\end{equation}%
Combined use of Eq.s(4.3) and (4.4) produces as well, 
\begin{equation}
C=\lambda (d-2).  \tag{4.5}
\end{equation}%
Again, using Eq.s(4.3) and (4.5) we can rewrite Eq.(4.4) as follows 
\begin{equation}
R_{ij}-\frac{1}{2}g_{ij}R+\frac{1}{2d}(d-2)Rg_{ij}=0.  \tag{4.6}
\end{equation}

These results remain unchanged if we use the Riemanninan manifold $M$
instead of pseudo Riemannian. This observation allows us to use many facts
from the Yamabe theory [\textbf{22-24}], originally developed for Riemannian
manifolds.

\ 

In view of this, we argue that Eq.(4.6) can be obtained as well by varying
the Yamabe functional, Eq.(3.6). Indeed, following Aubin [\textbf{22}] and
Schoen [\textbf{24}], let $t$ be some small parameter labeling the family of
metrics $g_{ij}(t)=g_{ij}+th_{ij}$. Then, these authors demonstrate that 
\begin{equation}
\left( \frac{dR_{t}}{dt}\right) _{t=0}=\bigtriangledown ^{i}\bigtriangledown
^{j}h_{ij}-\bigtriangledown ^{j}\bigtriangledown _{j}h_{i}^{i}-R^{ij}h_{ij} 
\tag{4.7}
\end{equation}%
and 
\begin{equation}
\left( \frac{d}{dt}\sqrt{\left\vert g_{t}\right\vert }\right) _{t=0}=\frac{1%
}{2}\sqrt{\left\vert g\right\vert }g^{ij}h_{ij}  \tag{4.8}
\end{equation}%
where, as usual, $\left\vert g\right\vert $ $=$ $\left\vert \text{det }%
g_{ij}\right\vert $.

Consider now the Yamabe functional, Eq.(3.6), but this time, written for the
family of metrics. We have 
\begin{equation}
\mathcal{R}(g(t))=\left( V(t)\right) ^{\frac{-2}{p}}\int_{M}R(g(t))DV(t) 
\tag{4.9}
\end{equation}%
where the volume is given by $V(t)=\int_{M}d^{d}x\sqrt{g(t)}$ and,
accordingly, $DV(t)$=$d^{d}x\sqrt{g(t)}.$ Using Eq.s(4.7) and (4.8) in
Eq.(4.9) and taking into account that the combination $\bigtriangledown
^{i}\bigtriangledown ^{j}h_{ij}-\bigtriangledown ^{j}\bigtriangledown
_{j}h_{i}^{i}$ is the total divergence, produces the following result: 
\begin{eqnarray}
\left( \frac{d}{dt}\mathcal{R}(g(t))\right) _{t=0} &=&V(0)^{-2\frac{n-1}{n}%
}[\int_{M}(Rg^{ij}/2-R^{ij})h_{ij}DV(0)\text{ }\int_{M}DV(0)  \notag \\
&&-(\frac{1}{2}-\frac{1}{d})\int_{M}DV(0)\int_{M}h_{ij}g^{ij}DV(0)]. 
\TCItag{4.10}
\end{eqnarray}%
If the metric $g$ is the critical point of $\mathcal{R}(g(t)),$ then 
\begin{equation}
\left[ R_{ij}-\frac{R}{2}g_{ij}\right] \int_{M}DV(0)+(\frac{1}{2}-\frac{1}{d}%
)(\int_{M}RDV(0))g_{ij}=0.  \tag{4.11}
\end{equation}%
From here, multiplication of both sides by $g^{ij}$ and subsequent summation
produces at once 
\begin{equation}
R-\frac{R}{2}d+(\frac{1}{2}-\frac{1}{d})<R>d=0\text{,}  \tag{4.12}
\end{equation}%
where $<R>=\frac{1}{V(0)}\int RDV(0)$ is the average scalar curvature.
Eq.(4.12) can be rewritten as $R=<R>.$ But this condition is exactly
equivalent to the Einstein condition, Eq(4.2), in view of Eq.(4.3)! Hence,
under such circumstances, Eq.s (4.11) and (4.12) are equivalent. For the set
of metrics of fixed conformal class (i.e. related to each other by $\tilde{g}%
=e^{2f}g)$ the variational problem for the G-L functional, Eq.(3.7), is
exactly equivalent to the variational problem for the Hilbert-Einstein
functional, Eq.(4.1), for gravity field in the vacuum in the presence of the
cosmological constant. Moreover, the result $R=<R>$ jsut obtained coincides
with the proviously obtained Eq. 92.12) (in view of Eq. (3.4)). This
equivalence is of major importance for application discussed in Sections 5
and 6.

\ 

The results displayed above become trivial for $d=2$. Physically, however,
the case $d=2$ is important since it is relevant to all known statistical
mechanics exactly solvable models treatable by methods of conformal field
theories. Hence, we would like to discuss some\ modifications of the above
results required for treatment of two dimensional analogs of the G-L theory.

\section{Ginzburg-Landau-like theory in two dimensions}

\subsection{\protect\bigskip General remarks}

\ \ From field-theoretic treatments of the G-L model [\textbf{19}] we know
that straightforward analysis based on asymptotic $\varepsilon -$expansions
from the critical dimension (4) to the target dimension (2) is impractical.
At the same time, known results for CFT and exactly solvable models are
useful thus far only in $d=2$. The question arises: is there an analog of
the G-L Eq.(3.1) in two dimensions? And, if such an analog does exist, what
use can be made of it? In the rest of this section we provide affirmative
answers to these questions. We demonstrate that: a) indeed, a two
dimensional analog of the G-L equation \textit{does exist} and is given by
the Liouville Eq.(5.19) below, b) the fuctional, Eq.(5.14), whose
minimization produces such an equation is the exact two dimensional analog
of the G-L-Yamabe functional, Eq.(3.2), c) these results can be (re)obtained
from the existing string-theoretic formulations of CFT\ developed entirely
independently, d) in view of the noted correspondence, these
string-theoretic CFT results can be extended to account for the
Lifshitz-type problems discussed in the Introduction. This allows us to
obtain a positive answer to the 1st Problem formulated in the Introduction.

\subsection{Designing the two dimensional G-L-Yamabe functional}

\QTP{Body Math}
To discuss topics related to items a) and b) just mentioned, we begin with
the observation that in two dimensions, Eq.(2.4) acquires a very simple form 
\begin{equation}
\Delta _{\tilde{g}}=e^{-2f}\Delta _{g}\text{ ,}  \tag{5.1}
\end{equation}%
where we use a factor of 2 to be in accord with Eq.(2.11) for scalar
curvature. According to Eq.(2.11), the scalar curvature in two dimensions
transforms like 
\begin{equation}
\tilde{R}(\tilde{g})=e^{-2f}\{R(g)-\Delta _{g}2f\}  \tag{5.2}
\end{equation}%
while the area $dA=d^{2}x\sqrt{g}$ transforms like 
\begin{equation}
d\hat{A}=e^{-2f}dA.  \tag{5.3}
\end{equation}%
These facts immediately suggest that the previously introduced action
functional 
\begin{equation}
S[\mathbf{X}]=\int_{M}d^{2}x\sqrt{g}(\bigtriangledown _{g}\mathbf{X})\cdot
(\bigtriangledown _{g}\mathbf{X})  \tag{5.4}
\end{equation}%
is conformally invariant. Using results by Polyakov [\textbf{37}] and noting
that $(\bigtriangledown _{g}\mathbf{X})\cdot (\bigtriangledown _{g}\mathbf{X}%
)=g^{\alpha \beta }\partial _{\alpha }X^{\mu }\partial _{\beta }X_{\mu }$,
we need to consider the following path integral\footnote{%
Without loss of generality, we would like to consider the case of a one
component field $\phi $ only} 
\begin{equation}
\exp (-\mathcal{F(}\mathit{g}))\equiv \int D[\phi ]\exp (-\frac{1}{2}%
\int_{M}d^{2}x\sqrt{g}g^{\alpha \beta }\partial _{\alpha }\phi \partial
_{\beta }\phi )  \tag{5.5}
\end{equation}%
\textit{where} $\mathcal{F}$($g$) is the \textquotedblright free
energy\textquotedblright \footnote{%
Usually, instead of $\mathcal{F(}\mathit{g)}$ one writes $\mathcal{F(}%
\mathit{g)/k}_{B}\mathit{T,}$where $T$ is the temperature and $k_{B}$ is the
Bolzmann's constant. In the present case the problems we are studying do not
require specific values for these constants. For this reason they will be
suppressed. Also, one should keep in mind that the free energy is always
defined with respect to some reference state. This will be the case in our
calculations as well.}. Fortunately, this integral was calculated by
Polyakov [\textbf{37]} for two dimensional manifolds $M$ \textit{without}
boundaries and by Alvarez [\textbf{38}] for manifolds \textit{with}
boundaries. In this work we shall be mainly concerned with manifolds \textit{%
without} boundaries. Although in the original work by Polyakov, one can find
the final result of calculation of the above path integral, the details of
this calculation can be found only elsewhere. In particular, we shall follow
pedagogically written papers by Weisberger [\textbf{39,40]} and Osgood,
Phillips and Sarnak [\textbf{26,41, 42}] (OPS).

\ To begin, let $\tilde{g}_{\alpha \beta }$ be some reference metric, and
let $g_{\alpha \beta }$ be a metric conformally related to it, i.e. $%
g_{\alpha \beta }=\exp (-2\varphi )\tilde{g}_{\alpha \beta }$. Should the
above path integral be for the flat (i.e.$g_{\alpha \beta }$=$\delta
_{\alpha \beta }$) two dimensional manifold, one would have at once the
result: $\mathcal{F}$=$\frac{1}{2}\ln \det^{\prime }\Delta _{0}$,\ where the
prime indicates that the zero mode is omitted. Because to assume such \
flatness in general is too restrictive, it is appropriate to pose a problem:
how is the path integral for the metric $g$ related to that for the metric $%
\tilde{g}$? The paper [\textbf{26}] by OPS provides an answer, e.g. see
Eq.(1.13) of this reference. To connect this equation with the free energy,
we replace it by the equivalent expression 
\begin{equation}
\ln \left( \frac{\det^{\prime }\Delta _{\hat{g}}}{A_{\hat{g}}}\right) -\ln
\left( \frac{\det^{\prime }\Delta _{g}}{A_{g}}\right) =-\frac{1}{6\pi }[%
\frac{1}{2}\int\limits_{M}dA_{g}\{\left\vert \bigtriangledown _{g}\varphi
\right\vert ^{2}+R(g)\varphi \}]  \tag{5.6}
\end{equation}%
useful in applications to strings and CFT, e.g see Ref.[\textbf{43}], page
637.

It is worthwhile to provide a few computational details leading to Eq.(5.6).
Therefore, let $\hat{\psi}_{i}$ be eigenfunctions of the Laplacian $\Delta
_{g}$ with eigenvalues $\lambda _{i}$ arranged in such a way that $0=\hat{%
\lambda}_{0}<\hat{\lambda}_{1}\leq \hat{\lambda}_{2}\leq \cdot \cdot \cdot ,$
i.e. 
\begin{equation}
-\Delta _{g}\hat{\psi}_{i}+\hat{\lambda}\hat{\psi}_{i}=0.  \tag{5.7}
\end{equation}%
Then, we construct the zeta function 
\begin{equation}
\zeta _{g}(s)=\sum\limits_{i=1}^{\infty }\hat{\lambda}_{i}^{-s}  \tag{5.8}
\end{equation}%
in such a way that 
\begin{equation}
\det \text{ }^{\prime }\Delta _{g}=\exp (-\zeta _{g}^{^{\prime }}(0)) 
\tag{5.9}
\end{equation}%
with $\zeta _{g}^{^{\prime }}(0)=\left( \frac{d}{ds}\zeta _{g}(s)\right)
_{s=0}.$ Using Eq.(5.1), we obtain as well 
\begin{equation}
-e^{-2\varphi }\Delta _{g}\psi _{i}+\lambda \psi _{i}=0.  \tag{5.10}
\end{equation}%
In particular, for a constant $\varphi =\bar{\varphi}$ we obtain 
\begin{equation}
\zeta _{\hat{g}}(s)=\sum\limits_{i=1}^{\infty }\left( e^{-2\bar{\varphi}}%
\hat{\lambda}_{i}\right) ^{-s}=e^{2s\bar{\varphi}}\zeta _{g}(s).  \tag{5.11}
\end{equation}%
Use of Eq.(5.9) in Eq.(5.11) produces: 
\begin{equation}
\zeta _{\hat{g}}^{^{\prime }}(0)=\zeta _{g}^{^{\prime }}(0)+2\bar{\varphi}(%
\frac{\chi (M)}{6}-1).  \tag{5.12}
\end{equation}%
This result was obtained with help of the known fact, Ref.[\textbf{26}],
Eq.(1.9), that 
\begin{equation}
\zeta _{g}(0)=\frac{\chi (M)}{6}-1  \tag{5.13}
\end{equation}%
with $\chi (M)$ being the Euler characteristic of two dimensional manifold $%
M $ without boundaries. In view of the definition, Eq.(5.9), and the
Gauus-Bonnet theorem,\ we observe that Eq.(5.6) is reduced to Eq.(5.12) for
the case of constant conformal factor $\varphi =\bar{\varphi}$, as required.
\ 

\ 

\textit{It should be mentioned that }Eq.(5.6) was used earlier in finite
size scaling calculations [\textbf{44-46}]. Unfortunately, the authors of [%
\textbf{45,46}] did not take into account that the above formula is valid,
strictly speaking, for manifolds $\mathit{M}$ without boundaries while the
topology considered in these works was that for the punctured disc and/or
annulus. Rigorous results for such topologies were obtained by Weisberger [%
\textbf{40}] and more general case of surfaces of higher genus with
boundaries and/or punctures was discussed in detail by OPS [\textbf{42}]. To
focus on the main goals of this paper, we shall not discuss the finite size
scaling analysis further.

\ 

\QTP{Body Math}
Using Eq.(5.6) and, following OPS [\textbf{26}], we would like consider the
related functional $\mathcal{F}$($\varphi )$ defined by\footnote{%
When comparing with the OPS paper, it should be noted that OPS use the
Gaussian curvature $K$ while we use the scalar curvature $R=2K$} 
\begin{equation}
\mathcal{F}(\varphi )=\frac{1}{2}\int\limits_{M}dA_{g}\{\left\vert
\bigtriangledown _{g}\varphi \right\vert ^{2}+R(g)\varphi \}-\pi \chi (M)\ln
\int\limits_{M}dA_{g}\text{ }e^{2\varphi }.  \tag{5.14}
\end{equation}%
This functional is the exact analog of the Yamabe functional, Eq.(3.2), in
dimensions 3 and higher as we shall demonstrate shortly below. In the
meantime, in view of Eq.(5.6), it can be rewritten in the following
equivalent form\footnote{%
Here one should understand the word "equivalence" in the sense that both
functionals produce the same critical metrics upon minimization.} 
\begin{equation}
\mathcal{F}(\varphi )=-6\pi \ln \det \text{ }^{\prime }\Delta _{g}+\pi
(6-\chi (M))\ln A.  \tag{5.15}
\end{equation}%
Let $a$ be some constant, then 
\begin{equation}
\mathcal{F}(\varphi +a)=\mathcal{F}(\varphi ).  \tag{5.16}
\end{equation}%
This signifies that the above action is scale invariant. This property is in
complete accord with Eq.s (2.7) and (2.8) discussed earlier. If we impose
the constraint (fix the gauge): $A=1$, then we end up with the
Liouville-like action used in CFT. We would like to explain this in some
detail now.

Following OPS, it is convenient to replace the constraint $A=1$ by the
alternative constraint on the field $\varphi $: 
\begin{equation}
\int\limits_{M}\varphi dA_{g}=0.  \tag{5.17}
\end{equation}%
To demonstrate that such an imposed constraint is equivalent to the
requirement $A=1$, we note that, provided that the field $\psi $ minimizes $%
\mathcal{F}(\psi )$ subject to the constraint Eq.(5.17), the field 
\begin{equation}
\varphi =\psi -\frac{1}{2}\ln \int\limits_{M}\exp (2\psi )dA_{g}  \tag{5.18}
\end{equation}%
minimizes $\mathcal{F}(\varphi )$ subject to the constraint $A=1$. Using
these facts, minimization of $\mathcal{F}(\psi )$ produces the following
Liouville equation 
\begin{equation}
-\Delta _{g}\psi +\frac{1}{2}R(g)-\frac{2\pi \chi (M)\exp (2\psi )}{%
\int\limits_{M}\exp (2\psi )dA_{g}}=0.  \tag{5.19}
\end{equation}%
Comparing this result with Eq.(5.2) and taking into account that 
\begin{equation}
\frac{2\pi \chi (M)}{\int\limits_{M}\exp (2\psi )dA_{g}}=\tilde{R}(\tilde{g}%
)=const,  \tag{5.20}
\end{equation}%
we conclude that, provided that the background metric $g$ is given so that
the scalar curvature $R(g)$ (not necessarily constant) can be calculated,
the Liouville Eq.(5.19) is exactly analogous to the previously obtained
Eq.(3.3), or equivalently, Eq.(2.12). In view of this analogy, use of Eq.s
(3.4) and (3.6) as well as Eq.s(5.14),(5.15) causes the functional $\mathcal{%
F}(\varphi )$ to attain its extremum for metric $\tilde{g}$ of constant
scalar curvature $\tilde{R}(\tilde{g})$. To decide if the extremum is
minimum or maximum, we have to consider separately cases $\chi (M)>0$ and $%
\chi (M)\leq 0.$ In the case of $\chi (M)>0$ we have only to consider
manifolds homeomorphic to $S^{2}$ so that $\chi (M)=2$. Fortunately, this
case was considered in detail by Onofri [\textbf{47}]\footnote{%
Generalization of these two dimensional results to higher dimensional
manifolds of even dimensions can be found in the paper by Beckner [\textbf{48%
}].}. Using his work, the following inequality 
\begin{equation}
\ln \int\limits_{S^{2}}dA_{\tilde{g}}\exp (\psi )\leq \int\limits_{S^{2}}dA_{%
\tilde{g}}\psi +\frac{1}{4}\int\limits_{S^{2}}dA_{\tilde{g}}\left\vert
\bigtriangledown \psi \right\vert ^{2}  \tag{5.21}
\end{equation}%
attributed to Aubin [\textbf{22}] and inspired by previous results by
Trudinger and Moser [\textbf{48}], is helpful for deciding whether the
obtained extremum is minimum or maximum. Here $\tilde{g}$ is the metric of
the unit sphere $S^{2}$ with constant Gaussian curvature $1$. The metric $g$
conformal to $\tilde{g}$ is given by $g=\exp (2\psi )\tilde{g}$ , with $\psi 
$ obeying the Liouville equation (just like Eq.(5.19)), where both $R(g)$
and $\tilde{R}(\tilde{g})$ are constant by virtue of the initial choice of $%
\tilde{g}$. By combining Eq.s(5.6),(5.14) and (5.15) with the inequality
(5.21) \ and taking into account that \ by design ln$A_{\tilde{g}}=0$ we
obtain, 
\begin{equation}
-3\pi \ln \frac{\det^{\prime }\Delta _{g}}{\det^{\prime }\Delta _{\tilde{g}}}%
=\frac{1}{4}\int\limits_{S^{2}}dA_{\tilde{g}}\{\left\vert \bigtriangledown _{%
\hat{g}}\psi \right\vert ^{2}+2\psi \}-\ln \int\limits_{S^{2}}dA_{\tilde{g}%
}\exp (2\psi )\geq 0,  \tag{5.22}
\end{equation}%
with equality occurring only at the extremum $\psi =\psi ^{\ast },$ with the
function $\psi ^{\ast }$ being a solution of the Liouville Eq.(5.19). It can
be shown [\textbf{47}] that: a) such a solution involves only M\"{o}bius
transformations of the sphere $S^{2}$ and that, b) the functional $\mathcal{F%
}(\varphi )$ is invariant with respect to such transformations. \ The case
of $\chi (M)\leq 0$ \ is treated in Section 2.2 of the OPS paper, Ref.[%
\textbf{26}] and leads to the same conclusions about extremality of the
functional $\mathcal{F}(\varphi )$.

\ 

The two dimensional results just obtained\ are in accord \textsl{by design}
with results obtained in higher dimensions (discussed in Sections 3 and 6).
In particular,\textsl{\ the functional} $\mathcal{F}(\varphi )$ \textsl{is
the exact analog of the Yamabe functional} $S$[$\varphi ],$ Eq.(3.2). Since
in both cases the functionals are \textquotedblright
translationally\textquotedblright\ (actually, scale) invariant, e.g. compare
Eq.(3.6) with Eq.(5.16), in both cases, the extremum is realized for metric
conformal to the metric of constant scalar curvature, e.g. compare Eq.(3.3)
with the Liouville Eq.(5.19).

\bigskip

\subsection{\ \ Connections with string and CFT}

\bigskip\ 

The results just obtained allow us now to discuss topic c) listed at the
begining of this section. Using known facts from string and conformal field
theories, [\textbf{27,28}], it is of interest to consider averages of the
vertex operators 
\begin{equation}
\left\langle \prod\limits_{i=1}^{n}\exp (\beta _{i}\phi
(z_{i}))\right\rangle \equiv \int D\left[ \phi \right] \exp \{-S_{L}(\phi
)\}\prod\limits_{i=1}^{n}\exp (\beta _{i}\phi (z_{i})),  \tag{5.23}
\end{equation}%
where the Liouville action $S_{L}(\phi )$ is given (in notation adopted from
these references) by 
\begin{equation}
S_{L}(\phi )=\frac{1}{8\pi }\int_{M}dA_{\tilde{g}}[\left\vert
\bigtriangledown \phi \right\vert ^{2}-QR(\hat{g})\phi +8\pi \bar{\mu}\exp
(\alpha _{+}\phi )].  \tag{5.24}
\end{equation}%
The actual values and the meaning of constants $Q$, $\bar{\mu}$ and $\alpha
_{+}$ are explained in these references and are of no immediate use for us.
Clearly, upon proper rescaling, we can bring $S_{L}(\phi )$ to the form
which agrees with $\mathcal{F}(\varphi ),$ defined by Eq.(5.14), especially
in the trivial case when both $\chi (M)$ and $\bar{\mu}$ are zero. When they
are not zero, the situation in the present case becomes totally analogous to
that discussed earlier for the Yamabe functonal. In particular, in Section 3
we noticed that the G-L Euler-Lagrange Eq.(3.3) can be obtained either by
minimization of the Yamabe functional, Eq.(3.2) (or (4.9)), or by
minimization of the G-L functional, Eq.(3.7), where the coupling constant $%
\widetilde{\lambda }$ plays the role of the Lagrange multiplier enforcing
the volume constraint. In the present case, variation of the Liouville
action $S_{L}(\phi )$ will produce the Liouville equation, e.g. see Eq.s
(5.19)-(5.20), which is the two dimensional analog of the G-L Eq.(3.3). \
This variation is premature, however, since we can reobtain $\mathcal{F}%
(\varphi )$ exactly using the path integral, Eq.(5.23). This procedure then
will lead us directly to the Liouville Eq.(5.19).

To this purpose, we need to consider the path integral, Eq.(5.23), in the
absence of sources, i.e. when all $\beta _{i}=0.$ Following ideas of Ref.s [%
\textbf{27,28}], we take into account that: a) 
\begin{equation}
\frac{1}{4\pi }\int\limits_{M}dA_{\tilde{g}}R(\hat{g})=\chi (M)=2-2g 
\tag{5.26}
\end{equation}%
with $g$ being genus of $M$ and, b) the field $\phi $ can be decomposed into 
$\phi =\phi _{0}+\varphi $ in such a way that $\phi _{0\text{ }}$is
coordinate-independent and $\varphi $ is subject to the constraint given by
Eq.(5.17). Then, use of the identity 
\begin{equation}
\int\limits_{-\infty }^{\infty }dx\exp (ax)\exp (-b\exp (\gamma x))=\frac{1}{%
\gamma }b^{-\frac{a}{\gamma }}\Gamma (\frac{a}{\gamma })  \tag{5.27}
\end{equation}%
(with $\Gamma (x)$ being Euler's gamma function) in the path integral,
Eq.(5.23), requires us to evaluate the following integral 
\begin{eqnarray}
I &=&\int\limits_{-\infty }^{\infty }d\phi _{0}\exp (\phi _{0}\frac{Q}{2}%
\chi (M))\exp ((\bar{\mu}\int\limits_{M}dA_{\tilde{g}}\exp (\alpha
_{+}\varphi ))\exp (\alpha _{+}\phi _{0})  \notag \\
&=&\frac{\Gamma (-s)}{\alpha _{+}}(\bar{\mu}\int\limits_{M}dA_{\tilde{g}%
}\exp (\alpha _{+}\varphi ))^{s}  \TCItag{5.28}
\end{eqnarray}%
with $s$ given by 
\begin{equation}
s=-\frac{Q}{2\alpha _{+}}\chi (M).  \tag{5.29}
\end{equation}%
Using this result in Eq.(5.23), we obtain the following path integral (up to
a constant) 
\begin{equation}
Z[\varphi ]=\int D[\varphi ]\exp (-\mathcal{\hat{F}}(\varphi ))  \tag{5.30}
\end{equation}%
with functional $\mathcal{\hat{F}}(\varphi )$ given by 
\begin{equation}
\mathcal{\hat{F}}(\varphi )=S_{L}(\varphi ;\bar{\mu}=0)-\frac{Q}{2\alpha _{+}%
}\chi (M)\ln [\bar{\mu}\int\limits_{M}dA_{\tilde{g}}\exp (\alpha _{+}\varphi
)].  \tag{5.31}
\end{equation}%
This functional (up to rescaling of the field $\varphi $) is just the same
as the Yamabe-like functional $\mathcal{F}(\varphi )$ given by Eq.(5.14).

Define now the free energy $\mathcal{F}$ in the usual way via $\mathcal{F}%
=-\ln Z[\varphi ]$ (as was done after Eq.(5.5)) and consider the saddle
point approximation to the functional integral $Z[\varphi ].$ Then,\ for a
spherical topology, in view of Eq.(5.22), we (re)obtain $\mathcal{F}\geq 0$
with equality obtained when $\varphi =\psi ^{\ast }$. \ Sources (or the
vertex operators) can be taken into account also, especially in view of the
results of [\textbf{42}].\footnote{%
Since the results of this reference are useful as well as well for extension
of the finite size scaling results given in works by Affleck [\textbf{46}]
and Bl\"{o}te et al [\textbf{45}], the detailed analysis of this case is
left for further study.}

To better understand the physical significance of the obtained results, it
is useful to reobtain them using a somewhat different method. The results
obtained with this alternative method are also helpful when we shall discuss
their higher dimensional analogs in the next section. To this purpose we
write (up to a normalization constant) 
\begin{equation}
\int D\left[ \phi \right] \exp \{-S_{L}(\phi )\}=\int\limits_{0}^{\infty
}dAe^{-\bar{\mu}A}Z_{L}(A)  \tag{5.32}
\end{equation}%
where 
\begin{equation}
Z_{L}(A)=\int D[\phi ]\delta (\int_{M}dA_{\hat{g}}\exp (\alpha _{+}\phi
)-A)\exp (S_{L}(\phi ;\bar{\mu}=0)).  \tag{5.33}
\end{equation}%
If, as before, we assume that $\phi =\phi _{0}+\varphi ,$ then an elementary
integration over $\phi _{0}$ produces the following explicit result for $%
Z_{L}(A)$:%
\begin{equation}
Z_{L}(A)=\frac{-1}{\alpha _{+}}A^{\omega }\int D[\varphi ]\left[ \int_{M}dA_{%
\hat{g}}\exp (\alpha _{+}\varphi )\right] ^{-(\omega +1)}\exp
(-S_{L}(\varphi ;\bar{\mu}=0))\text{,}  \tag{5.34}
\end{equation}%
where the exponent $\omega $ is given by 
\begin{equation}
\omega =\frac{\chi (M)Q}{2\alpha _{+}}-1.  \tag{5.35}
\end{equation}%
Finally, using Eq.(5.34) in Eq.(5.32) produces back Eq.s (5.30) and (5.31)
(again, up to a constant factor). An overall \textquotedblright
-\textquotedblright\ sign can be removed by proper normalization of the path
integral. These results can be used for computation of the correlation
functions of conformal field theories (CFT). Details can be found in [%
\textbf{28}].

\subsection{Completion of the work by Lifshitz in 2 dimensions}

Since in this work our main interest is investigation of higher dimensional
analogs of the results just obtained, no further computational details
related to two dimensional CFT will be presented in this work. Instead, we
would like to complete our investigation related to item d) mentioned at the
begining of this section. It will enable us to develop similar treatments
for higher dimensions to be considered in the next section.

We beging with combining inequalities given by Eq.s(5.21) and (5.22). This
produces, 
\begin{equation}
-\ln \det \text{ }^{\prime }\Delta _{g}-(-\ln \det \text{ }^{\prime }\Delta
_{\tilde{g}})\geq 0,  \tag{5.36}
\end{equation}%
with equality taking place only when the metric $g$ is equal to that for the
unit round sphere $S^{2},$ i.e. to $\tilde{g}$. This means that 
\begin{equation}
\det \text{ }^{\prime }\Delta _{g}\leq \det \text{ }^{\prime }\Delta _{%
\tilde{g}}  \tag{5.37}
\end{equation}%
implying that the determinant of the Laplacian for the round sphere provides
the \textit{upper} bound for determinants of Laplacians whose metric is
conformally equivalent to that for the round sphere $S^{2}.$ Calculation of
the determinant of the Laplacian for the round sphere can be done with help
of Eq.s(5.8) and (5.9). Specifically, for this case we need to calculate $Z_{%
\hat{g}}^{\prime }(0)$ where 
\begin{equation}
Z_{\tilde{g}}(s)=\sum\limits_{n=1}^{\infty }\frac{2n+1}{(n(n+1))^{s}}. 
\tag{5.38}
\end{equation}%
This calculation is rather difficult and can be found, for example, in [%
\textbf{49}] along with its multidimensional generalization. For $S^{2}$ the
final result is given by 
\begin{equation}
Z_{\tilde{g}}^{\prime }(0)=4\zeta ^{\prime }(-1)-\frac{1}{2}.  \tag{5.39}
\end{equation}%
From here we obtain, $\det $ $^{\prime }\Delta _{g}\leq \exp \{\frac{1}{2}%
-4\zeta ^{\prime }(-1)\}.$ We deliberately avoid long discussion leading to
this result since we are more interested\ in similar calculations for torus
topology. \ In this last case the reference metric $\hat{g}$ is that for the
flat torus $T^{2}=\mathbf{C}/\Lambda ,$ where $\mathbf{C}$ is the complex
plane and $\Lambda $ is some lattice. Evidently, $\chi (T^{2})=0$, so that $%
R(\tilde{g})=0.$ These facts produce at once 
\begin{equation}
-\ln \det \text{ }^{\prime }\Delta _{g}-(-\ln \det \text{ }^{\prime }\Delta
_{\tilde{g}})=\frac{1}{12\pi }\int\limits_{M}dA_{\tilde{g}}\left\vert
\bigtriangledown _{\tilde{g}}\varphi \right\vert ^{2}\geq 0  \tag{5.40}
\end{equation}%
with equality when $\varphi =const.$

From the point of view of applications to statistical mechanics (or better,
to CFT), the above inequality is useful when calculating the path integral,
Eq.(5.30), using the saddle point method. In this case, since $\mathcal{\hat{%
F}}(\varphi )$ plays a role of the LGW free energy, in accord with general
requirements of thermodynamics, the free energy attains its minimum at
equilibrium, i.e. for $\varphi =\varphi ^{\ast }.$ Here $\varphi ^{\ast }$
is a solution of the Liouville Eq.(5.19) in accord with previously obtained
result, Eq.(5.22), for the sphere. In the present case of toral topology, it
is known that $\chi (M)=0,$ so that $\varphi ^{\ast }=const$ is an
acceptable solution. Such a two dimensional result is formally in complete
accord with the higher dimensional G-L result, Eq.(1.20), ( $h=0$) of
Section 1. In the present case, the choice of the $const$ is dictated by
Eq.(5.18)\footnote{%
From the arguments related to Eq.(5.18) it should be clear that the
restriction $A=1$ is not essential but convenient}. For such a choice, $%
\mathcal{\hat{F}}(\varphi ^{\ast })=0.$ This is clearly the lowest possible
value for the free energy, which in G-L theory corresponds to the free
energy at criticality (i.e. at $T=T_{c}$). Expanding $\mathcal{\hat{F}}%
(\varphi )$ around the equilibrium value of the field $\varphi $ in the path
integral in Eq.(5.30) produces quadratic and higher order terms as usual [%
\textbf{19]}. If one ignores terms higher than quadratic, one obtains the
standard Gaussian-type path integral discussed in the Introduction. It plays
a major role in many two dimensional CFT\ models [\textbf{20}].
Justification of such a truncation is discussed in the next section. In the
meantime, we would like to discuss calculation of this type of path integral
in some detail. Since the path integral is Gaussian, its calculation is
essentially the same as calculation of $\det $ $^{\prime }\Delta _{\tilde{g}%
} $ . In view of Eq.(5.9), such a calculation involves use of the zeta
function of the following type 
\begin{equation}
Z_{\tilde{g}}(s)=\sum\nolimits_{\mathit{l}\mathsf{\in \Lambda }^{\ast
}}^{\prime }\frac{1}{\left( 4\pi ^{2}\left\vert \mathbf{l}\right\vert
^{2}\right) ^{s}}.  \tag{5.41}
\end{equation}%
This result can be easily understood using some basic information from solid
state physics [\textbf{13}]. Indeed, the eigenfunctions $f(\mathbf{l})$ of
the Laplace operator for $T^{2}$ are given by $f(\mathbf{l})=\exp (i2\pi
(l_{1}n_{1}+l_{2}n_{2}))$ with vector $\mathbf{l}=\{\mathit{l}%
_{1},l_{2}\}\in \Lambda ^{\ast }$ being some vector of the dual (reciprocal)
lattice $\Lambda ^{\ast }$, while the numbers $\{n_{1},n_{2}\}$ are related
to some vector of the direct lattice $\Lambda $. Accordingly, the
corresponding eigenvalues are $4\pi ^{2}\left\vert \mathbf{l}\right\vert ^{2}%
\footnote{%
Since this result can be extended immediately to 3 and higher dimensions it
will be used in the next section as well.}.$

The method of calculation of $Z_{\tilde{g}}(s)$ presented in Ch 10 of Ref.[%
\textbf{20}] while being straightforward is not too illuminating, especially
if one is contemplating its extension to dimensions higher than 2. We follow
therefore the approach taken by OPS, Ref.[\textbf{26}], where the first
Kronecker limit formula is used for evaluation of $Z_{\tilde{g}}^{\prime
}(0).$ To avoid duplications, our arguments (leading to the same results)
are somewhat different than those used in the OPS paper. These arguments
allow us to make additional useful connections with some facts from number
theory.\footnote{%
Some of these facts have been presented already in our recent work, Ref.[%
\textbf{50}].}

It is well known [\textbf{51}] that for each torus $T^{2},$ the modular
lattice $\Lambda $ is given by 
\begin{equation}
\Lambda =\mathbf{Z}\omega _{1}+\mathbf{Z}\omega _{2}\text{ or, symbolically, 
}\Lambda =[\omega _{1},\omega _{2}],  \tag{5.42}
\end{equation}%
where the periods $\omega _{1}$ and $\omega _{2}$ are such that $\tau =\func{%
Im}\frac{\omega _{2}}{\omega _{1}}>0.$ That is, at least one of the periods
should be complex. Different tori are related to each other by a modular
transformation of the type 
\begin{equation}
\tau ^{\prime }=\frac{a\tau +b}{c\tau +d}  \tag{5.43}
\end{equation}%
provided that $ad-bc=1$ with $a,b,c$ and $d$ being some integers. The
requirement $ad-bc=1$ guarantees that the inverse transformation (i.e. $\tau
^{\prime }\rightarrow \tau )$ also looks like Eq.(5.43), with respective
constants being also integers. Invariance of the observables of CFT with
respect to modular transformations is widely emphasized [\textbf{20}].
However, such invariance is too broad and leads to some inconsistencies in
CFT and string theories discussed in Ref.[\textbf{50}]. These
inconsistencies can be removed if we restrict values of the modular
parameter $\tau $ to those originating from complex multiplication (CM). The
concept of CM can be easily understood based on the following example.
Consider two lattices $\Lambda $ and $\Lambda ^{\prime }$ such that there is
a matrix $\mathbf{A}$%
\begin{equation}
\mathbf{A}=\left( 
\begin{array}{cc}
a & b \\ 
c & d%
\end{array}%
\right)  \tag{5.44}
\end{equation}%
so that 
\begin{eqnarray}
\omega _{2}^{^{\prime }} &=&a\omega _{2}+b\omega _{1},  \TCItag{5.45} \\
\omega _{1}^{^{\prime }} &=&c\omega _{2}+d\omega _{1}.  \notag
\end{eqnarray}%
Clearly, if we require $ad-bc=1$ and then form a ratio of the above two
equations, we would obtain Eq.(5.43). This time, however, we would like to
keep the requirement $ad-bc=1$ but avoid forming the ratio. This move is
motivated by the fact that by making this ratio, we would loose the option: $%
\omega _{2}^{^{\prime }}=\alpha \omega _{2}^{^{\prime }}$ and $\omega
_{1}^{^{\prime }}=\alpha \omega _{1}$ for some, yet unknown, $\alpha .$
Substitution of these relations into Eq.(5.45) leads to the following
eigenvalue problem 
\begin{equation}
\alpha ^{2}-\alpha tr\mathbf{A+}\det \mathbf{A=}0.  \tag{5.46}
\end{equation}%
But, we know already that $\det \mathbf{A=}1.$ Hence, Eq.(5.46) can be
rewritten as 
\begin{equation}
\alpha ^{2}-\alpha (a+d)\text{ }\mathbf{+}1\mathbf{=}0.  \tag{5.47}
\end{equation}%
This is the equation for an integer in the quadratic number field. It is
easy to demonstrate that such an integer must belong to the $imaginary$
quadratic number field. To prove this fact, we write for the roots 
\begin{equation}
\alpha _{1,2}=\frac{a+d}{2}\pm \frac{1}{2}\sqrt{(a+d)^{2}-4}.  \tag{5.48}
\end{equation}%
In order for $\alpha $ to be an integer belonging to the imaginary quadratic
field, the following set of options should be explored first. These are:

a) $a=d=0$, thus producing $\alpha _{1,2}=\pm i;$

b) $a=\pm 1,d=0$ (or $a=0,d=\pm 1),$ thus producing $\alpha _{1,2}=\frac{1}{2%
}(\pm 1\pm \sqrt{-3});$

c) $a=d=\pm 1,$ thus producing $\alpha _{1,2}=\pm 1.$ \ \ \ \ \ \ \ \ \ \ \
\ 

\ 

Before analyzing these results it is helpful to recall the following
definition from CM. \textit{\ }A torus\textit{\ T}$^{2}$\textit{\ }admits CM
if it admits an automorphism\textit{\ }$\Lambda \rightarrow c\Lambda $%
\textit{\ }so that\textit{\ }$c\Lambda \subseteq \Lambda $\textit{\ }for some%
\textit{\ }$c\neq Z.$\textit{\ }The case when\textit{\ }$c\Lambda \subset
\Lambda $\textit{\ }requires us to\textit{\ }replace Eq.(5.48) by 
\begin{equation}
\alpha _{1,2}=\frac{a+d}{2}\pm \frac{1}{2}\sqrt{(a+d)^{2}-4n}  \tag{5.49}
\end{equation}%
where the nonnegative integer $n$\ should be strictly greater than one%
\textit{.}

\ \ 

To understand the physical implications of these results, recall that if $%
a(\Lambda )$ denotes the area of the period parallelogram associated with $%
\Lambda ,$ then it can be shown [\textbf{51}] that 
\begin{equation}
a(\Lambda )=\frac{1}{2}\left\vert \omega _{1}\bar{\omega}_{2}-\omega _{2}%
\bar{\omega}_{1}\right\vert .  \tag{5.50}
\end{equation}%
If we now rescale $\omega ^{\prime }s$: $\omega _{2}^{^{\prime }}=\alpha
\omega _{2}^{^{\prime }},$ $\omega _{1}^{^{\prime }}=\alpha \omega _{1},$
this produces: $a(\alpha \Lambda )=\left\vert \alpha \right\vert
^{2}a(\Lambda ).$ If we require that the area upon rescaling remains
unchanged, we are left with $\left\vert \alpha \right\vert ^{2}=1.$ This is
the only acceptable option in view of the constraint $A=1$ on the area,
which was imposed earlier. Above we obtained two types of integers (also
units) of the imaginary quadratic field : a) the Gaussian units: $\pm 1,\pm
i $ of the Gaussian ring $\mathbf{Z}[i]$ which are the 4-th roots of unity,
i.e. $1,$ $i,$ $i^{2}=-1$ and $i^{3}=-i$, and, b) the units of the ring $%
\mathbf{Z}[j]$ which are the 6-th roots of unity, i.e. $1,-1,$ $j,$ $-j,$ $%
j^{2}$ and $-j^{2}$ , where $j=\frac{1}{2}(-1+\sqrt{-3}).$ Since these are
units in the respective rings, automatically we get $\left\vert \alpha
\right\vert ^{2}=1.$ In view of the constraint $ad-bc=1,$ no units from
other imaginary quadratic fields can be used. Indeed, in the most general
case of CM, we have to use Eq.(5.49) instead of Eq.(5.48). This is
equivalent to replacing the constraint $ad-bc=1$ by $ad-bc=n$ and leads to $%
a(\alpha \Lambda )=na(\Lambda ).$ Clearly, since by definition $n\neq 1,$
for physical reasons we are left only with the units of the two rings just
discussed.

\ 

CM along with the area constraint is sufficient for determination of all
regular lattices in two dimensions. These are: the square (associated with $%
i $), the hexagonal (associated with $j$) and the triangular (dual of the
hexagonal). These are described in Refs[\textbf{52, 53}]. No other
translationally invariant lattices exist in two dimensions [\textbf{53}].

\ 

Number-theoretically, these results can be reformulated as follows (e.g. see
Appendix B). For the Gaussian lattice $\Lambda _{i}$ we can take as basis $%
\Lambda _{i}=[1,i]$ while for the cube root of unity (hexagonal) lattice $%
\Lambda _{j}=[1,j].$ It should be clear that the usual complex numbers $%
z=x+iy$ belong to $\Lambda _{i}$[\textbf{53}] so that the Gaussian integers
are made of $x$ and $y$ being integral. Analogously, the \textquotedblright
hexagonal\textquotedblright\ numbers are made of $z=x+jy$ with
\textquotedblright hexagonal integers" made accordingly from the integral $x$%
'$s$ and $y$'$s$. The norm $N(z)$ can be defined now as is done in complex
analysis, i.e. 
\begin{equation}
N(z)=z\bar{z}.  \tag{5.51}
\end{equation}%
The norm has a very useful property which can be formulated as follows. If $%
z^{\prime \prime }=z^{\prime }z$, then $N(z^{\prime \prime })=N(z^{\prime
})N(z)$. In particular, if $z$ is the unit of the complex imaginary
quadratic field, then there is some $z^{\prime }$ in the same field such
that $\ zz^{\prime }=1$. This produces $N(z^{\prime })N(z)=1$, so that $%
N(z)=1$. In the case of a Gaussian field this leads to the equation $%
n^{2}+m^{2}=1$, producing 4 Gaussian units: $\pm 1,\pm i.$ At this point it
is useful to keep in mind that for the field of real numbers there are only
2 units:$\pm 1.$ In view of the results just presented we would like to
rewrite the zeta function in Eq.(5.41) in the number-theoretic form. We
obtain, 
\begin{equation}
Z_{\hat{g}}(s)=\frac{1}{\left( 4\pi ^{2}\right) ^{s}}\sum\nolimits_{\mathit{z%
}\mathsf{\in \Lambda }^{\ast }}^{\prime }\frac{1}{N(z)^{s}}  \tag{5.52}
\end{equation}%
Apart from the factor $\left( 4\pi ^{2}\right) ^{-s}$, the obtained result
corresponds to the Dedekind zeta function. In the case of other (higher than
quadratic order) number fields, the situation is more complicated as
explained in the book by Terras [\textbf{54}], so that other methods should
be used. These will be discussed in the next section. In the case of two
dimensions, connections between the exactly solvable statistical mechanical
models and number theory have been known for some time\footnote{%
These can be found \ for example at the Google database, e.g. see
\textquotedblright Number theory, Physics and Geometry", Les Houches, March
2003.} \ All these connections are based on the known fact that the Dedekind
zeta function is related to the Dirichlet L-function via 
\begin{equation}
\sum\nolimits_{\mathit{z}\mathsf{\in \Lambda }_{i}^{\ast }}^{\prime }\frac{1%
}{N(z)^{s}}=\mu \zeta (s)\sum\limits_{n=1}^{\infty }\frac{\chi (x)}{n^{s}} 
\tag{5.53}
\end{equation}%
where $\mu $ is the number of units in the number field and $\chi (n)$ is
the Dirichlet character. In particular, for the Gaussian number field we
obtain 
\begin{equation}
\sum\nolimits_{\mathit{z}\mathsf{\in \Lambda }_{j}^{\ast }}^{\prime }\frac{1%
}{N(z)^{s}}=4\zeta (s)\sum\limits_{n=1}^{\infty }\frac{\chi _{4}(n)}{n^{s}} 
\tag{5.53}
\end{equation}%
with 
\begin{equation}
\chi _{4}(n)=\left\{ 
\begin{array}{c}
0\text{ \ \ \ \ \ \ \ \ \ \ \ \ \ if }n\text{ is even} \\ 
\left( -1\right) ^{\left( n-1\right) /2}\text{ if }n\text{ is odd}%
\end{array}%
\right.  \tag{5.54}
\end{equation}%
while for the hexagonal number field we get 
\begin{equation}
\sum\nolimits_{\mathit{z}\mathsf{\in \Lambda }^{\ast }}^{\prime }\frac{1}{%
N(z)^{s}}=6\zeta (s)\sum\limits_{n=1}^{\infty }\frac{\chi _{3}(n)}{n^{s}} 
\tag{5.55}
\end{equation}%
with 
\begin{equation}
\chi _{3}(n)=\left\{ 
\begin{array}{c}
0\text{ if }n\equiv 0\text{ }\func{mod}3 \\ 
1\text{ if }n\equiv 1\text{ }\func{mod}3 \\ 
-1\text{ if }n=-1\func{mod}3%
\end{array}%
\right. .  \tag{5.56}
\end{equation}%
To obtain $\frac{d}{ds}Z_{\hat{g}}(s)\mid _{s=0}$ using Eq.s (5.53)-(5.55)
is possible, but is not very illuminating. It is much better to use the 1st
Kronecker limit formula valid for any quadratic number field [\textbf{55}].
In order to reveal the meaning of this formula and its relevance to our
needs a few steps are required. First, we note that the area form given by
Eq.(5.50) can be equivalently rewritten as 
\begin{equation}
a(\Lambda )=\frac{1}{2}\left\vert \omega _{1}\bar{\omega}_{2}-\omega _{2}%
\bar{\omega}_{1}\right\vert =\left\vert \omega _{1}\right\vert ^{2}\func{Im}%
\tau  \tag{5.57}
\end{equation}%
where $\tau =\frac{\omega _{2}}{\omega _{1}}.$ Second, if \textbf{l} is the
vector of the reciprocal lattice $\Lambda ^{\ast }$, then it can be
decomposed as $\mathbf{l}=l_{1}\omega _{1}+l_{2}\omega _{2}$ with $l_{1}$
and $l_{2}$ being some integers. Accordingly, 
\begin{eqnarray}
\left\vert \mathbf{l}\right\vert ^{2} &=&(l_{1}\omega _{1}+l_{2}\omega
_{2})(l_{1}\bar{\omega}_{1}+l_{2}\bar{\omega}_{2})  \notag \\
&=&\left\vert \omega _{1}\right\vert ^{2}(l_{1}+\tau l_{2})(l_{1}+\bar{\tau}%
l_{2}).  \TCItag{5.58}
\end{eqnarray}%
Next, since we have fixed the area $a(\Lambda )$ so that in our case, it is
equal to one, we obtain 
\begin{equation}
\func{Im}\tau =\frac{1}{\left\vert \omega _{1}\right\vert ^{2}}.  \tag{5.59}
\end{equation}%
By denoting $\func{Im}\tau =y$ we can rewrite Eq.(5.41) as follows 
\begin{equation}
Z_{\tilde{g}}(s)=\sum\nolimits_{\mathit{l}\mathsf{\in \Lambda }^{\ast
}}^{\prime }\frac{1}{\left( 4\pi ^{2}\left\vert \mathbf{l}\right\vert
^{2}\right) ^{s}}\equiv \frac{1}{\left( 2\pi \right) ^{2s}}%
\sum\nolimits_{l_{1},l_{2}}^{\prime }\frac{y^{s}}{\left\vert l_{1}+\tau
l_{2}\right\vert ^{2s}}.  \tag{5.60}
\end{equation}%
In mathematics literature [\textbf{55}] the derivation of the Kronecker 1st
limit formula is given for the function 
\begin{equation}
E(\tau ,s)=\sum\nolimits_{l_{1},l_{2}}^{\prime }\frac{y^{s}}{\left\vert
l_{1}+\tau l_{2}\right\vert ^{2s}}  \tag{5.61}
\end{equation}%
where the prime indicates that the summation takes place over all integers $%
(l_{1},l_{2})\neq (0,0).$ For many applications, it is more convenient to
consider the combination 
\begin{equation}
E^{\ast }(\tau ,s)=\pi ^{-s}\Gamma (s)\frac{1}{2}E(\tau ,s)  \tag{5.62}
\end{equation}%
which possess the nice analytical property [\textbf{56}] 
\begin{equation}
E^{\ast }(\tau ,s)=E^{\ast }(\tau ,1-s).  \tag{5.63}
\end{equation}%
Kronecker found the Laurent expansion of $E^{\ast }(\tau ,s)$ near $s=0.$ It
is given by [\textbf{57}] by 
\begin{equation}
E^{\ast }(\tau ,s)=\frac{-1}{2s}+\frac{\gamma }{2}-\ln 2\sqrt{\pi y}-\ln
\left\vert \eta (\tau )\right\vert ^{2}+O(s)  \tag{5.64}
\end{equation}%
where $\gamma $ is Euler's constant and 
\begin{equation}
\eta (\tau )=\exp (i\pi \tau /12)\prod\limits_{n=1}^{\infty }(1-\exp \{i2\pi
n\tau \})  \tag{5.65}
\end{equation}%
is the Dedekind eta function. By combining Eq.s (5.60)-(5.64) and taking
into account that for $\varepsilon \rightarrow 0^{+}$ 
\begin{equation*}
\Gamma (\varepsilon )=\frac{1}{\varepsilon }(1-\gamma \varepsilon
+O(\varepsilon ^{2}))
\end{equation*}%
and 
\begin{equation*}
\pi ^{\varepsilon }=1+\varepsilon \ln \pi +O(\varepsilon ^{2})
\end{equation*}%
after some straightforward algebra we obtain (for $s\rightarrow 0^{+}),$ 
\begin{equation}
Z_{\tilde{g}}(s)=-1-s\ln y\left[ \bar{\eta}(\tau )\eta (\tau )\right] ^{2}, 
\tag{5.66}
\end{equation}%
in agreement with Weil [\textbf{58}], page 75. From here we get 
\begin{equation}
\frac{d}{ds}Z_{\tilde{g}}(s)\mid _{s=0}=-\ln y\left[ \bar{\eta}(\tau )\eta
(\tau )\right] ^{2}  \tag{5.67}
\end{equation}%
in agreement with OPS [\textbf{26}]. Eq.(5.9), when combined with Eq.(5.67),
produces 
\begin{equation}
\det \text{ }^{\prime }\Delta _{\tilde{g}}=y\left\vert \eta (\tau
)\right\vert ^{4}.  \tag{5.68}
\end{equation}%
Comparison of this result with that given by Eq.s (10.29), (10.30) of Ref.[%
\textbf{20}] indicates that the free energy $\mathcal{F}$ for the Gaussian
model on the torus is given by 
\begin{equation}
\mathcal{F=}\frac{1}{2}\ln y\left\vert \eta (\tau )\right\vert ^{2}. 
\tag{5.69}
\end{equation}%
For the hexagonal lattice $\Lambda $ OSP found numerically that $\det $ $%
^{\prime }\Delta _{\tilde{g}}=\frac{\sqrt{3}}{2}\left\vert \eta
(j)\right\vert ^{4}$ =0.35575 \footnote{%
We shall recalculate this result below using different methods.}. In
Appendix C it is shown that the hexagonal lattice provides the highest
packing fraction: $q_{2}^{\triangle }=0.9069.$ \ We anticipate that there is
a correlation between the packing fraction and the numerical value of the
determinant (or, alternatively, the value of the free energy). For this to
happen, according to Eq.(5.40) we must have 
\begin{equation}
\det \text{ }^{\prime }\Delta _{g}\leq \det \text{ }^{\prime }\Delta _{%
\tilde{g}}=0.35575  \tag{5.70}
\end{equation}%
for any lattice \textit{other} than the hexagonal. In terms of the free
energy defined by Eq.(5.69) this means that the free energy of the hexagonal
lattice is the \textit{highest} possible\footnote{%
After Eq.(5.3) we have noticed that the free energy is defined with respect
to some reference state. To be in accord with the G-L theory we can choose
as the reference state the state at criticality. In this case the value of
the free energy for the hexagonal lattice should be taken as zero so that
the square lattice will have lower free energy typical for the lower
temperature phase in accord with the G-L theory. Inequality (5.70) indicates
that transformation from the higher symmetry hexagonal lattice to the lower
symmetry square lattice is conformal in accord with OSP [\textbf{26}].}.This
makes sense thermodynamically. Indeed, according to the G-L theory [\textbf{6%
}], at criticality, the symmetry of the system is higher than that in the
low temperature phase. The free energy should be \textit{lower} in the low
temperature phase to make such a transition to lower symmetry
thermodynamically favorable.\footnote{%
It is tempting to conjecture that the values of respective determinants are
proportional to the packing fractions. Unfortunately, this is not the case
as we shall demonstrate below.} In the next section we shall obtain
analogous 3d results for the hexagonal close packed (hcp) and the face
centered cubic lattices (fcc) whose symmetry is the highest. Such symmetry
is typical for the high temperature phase (e.g. see Fig.1) and in accord
with group-theoretic classification of successive second order phase
transitions as predicted (in part) by Indenbom [\textbf{59}], e.g. see Fig.4
in his work. Further details are provided in Section 6 below.

The above inequality means that we have to prove that it holds for the
square, hexagonal and triangular lattices. To accomplish this task, we
follow the classical paper by Chowla and Selberg [\textbf{29}] (C-S)
discussed also in our earlier work [\textbf{50}] in connection with the
Veneziano amplitudes. To facilitate the reader's understanding we note that
from the point of view of algebraic geometry every torus $T^{2}$ can be
associated with the projective version of the elliptic curve whose standard
form is given by [\textbf{51,55,60}] 
\begin{equation}
y^{2}=4(x-e_{1})(x-e_{2})(x-e_{3}).  \tag{5.71}
\end{equation}%
Its periods $\omega _{1}$ and $\omega _{2}$ are given by

$\qquad \qquad \qquad \qquad \qquad \qquad \qquad \qquad \qquad $%
\begin{equation}
\omega _{1}=\int\limits_{e_{1}}^{\infty
}dx[(x-e_{1})(x-e_{2})(x+e_{1}+e_{2})]^{-\frac{1}{2}}=\frac{2K(k)}{\sqrt{%
e_{1}-e_{3}}},  \tag{5.72a}
\end{equation}

\begin{equation}
\frac{\omega _{2}}{\sqrt{-1}}=\int%
\limits_{e_{2}}^{e_{3}}dx[(e_{1}-x)(x-e_{2})(x+e_{1}+e_{2})]^{-\frac{1}{2}}=%
\frac{2\sqrt{-1}K^{\prime }(k)}{\sqrt{e_{1}-e_{3}}},  \tag{5.72b}
\end{equation}%
where $k^{2}=\frac{e_{2}-e_{3}}{e_{1}-e_{3}}$ . For the sake of comparison
with the results of\ C-S it is useful to keep in mind that the values of
integrals $K(k)$ and $K^{\prime }(k)$ can be rewritten in the alternative
(Legendre) form, e.g. 
\begin{equation}
K(k)=\int\limits_{0}^{\pi /2}\frac{d\varphi }{1-k^{2}\sin ^{2}\varphi } 
\tag{5.73a}
\end{equation}%
and 
\begin{equation}
K^{\prime }(k)=K(k^{\prime })=\int\limits_{0}^{\pi /2}\frac{d\varphi }{%
1-k^{^{\prime }2}\sin ^{2}\varphi },  \tag{5.73b}
\end{equation}%
where $0<k<1$ and $k^{2}+k^{\prime 2}=1.$ Nevertheless, in the actual
calculations shown below it is sufficient to use Eq.s(5.72a), (5.72b) as
defining equations for $K(k)$ and $K^{\prime }(k).$ Using these definitions
we obtain as well, 
\begin{equation}
\tau =\sqrt{-1}\frac{K^{\prime }}{K}=\frac{\omega _{2}^{{}}}{\omega _{1}}. 
\tag{5.74}
\end{equation}%
Naturally, we are interested only in $\tau $ belonging to the imaginary
quadratic field. In our case it is either $i$ or $j$. \ Hence, if $\tau $ is
assigned, it is sufficient to know only one period in order to determine
another. By means of straightforward manipulations with elliptic functions
C-S demonstrate that for $\Delta (\tau )=\left[ \eta (\tau )\right] ^{24}$
the following equality holds 
\begin{equation}
\Delta (\tau )=\left( \frac{2K}{\pi }\right) ^{12}2^{-8}\left( kk^{\prime
}\right) ^{4}.  \tag{5.75a}
\end{equation}%
From here, in view of Eq.(5.70), we obtain 
\begin{equation}
\left( \Delta (\tau )\right) ^{\frac{1}{6}}=\left( \frac{2K}{\pi }\right)
^{2}2^{-\frac{4}{3}}\left( kk^{\prime }\right) ^{\frac{2}{3}}.  \tag{5.75b}
\end{equation}%
In the case of the square lattice the associated elliptic curve is $%
y^{2}=x^{3}-x$ [\textbf{60}]$.$ This allows us to obtain the period $\omega
_{1}=\int\limits_{1}^{\infty }\frac{dx}{\sqrt{x^{3}-x}}.$ Use of
substitutions $x=1/y$ and $1-y^{2}=z,$ produces $\omega _{1}=\frac{1}{2}%
\int\limits_{0}^{1}$ $dzz^{-\frac{1}{2}}(1-z)^{-\frac{3}{4}}$\ =$\frac{1}{2}%
\frac{\Gamma (1/4)\Gamma (1/2)}{\Gamma (3/4)}.$ The remaining calculations
are straightforward and are based on definitions made in Eq.s(5.72) and next
to it.\ This allows us to obtain $k^{2}=k^{\prime 2}=1/2$ and, accordingly, $%
K(1/\sqrt{2})=\frac{1}{\sqrt{2}}\omega _{1}=\frac{1}{4}\frac{1}{\sqrt{\pi }}%
\left[ \Gamma (1/4)\right] ^{2}$. Collecting terms and using Eq.(5.75b) we
obtain, 
\begin{equation}
\left( \Delta (\tau )\right) ^{\frac{1}{6}}=\frac{1}{16}\frac{1}{\pi ^{3}}%
\left[ \Gamma (1/4)\right] ^{4}  \tag{5.76}
\end{equation}%
and, since for the square lattice $y=1,$ in view of Eq.s(5.70) and (5.75),
we obtain, $\det $ $^{\prime }\Delta _{g}=0.3464$ . Hence, for the square
lattice, inequality Eq.(5.70) indeed holds. The OPS result, Eq.(5.70), is
given for the hexagonal lattice whose dual lattice $\Lambda _{j}^{\ast \text{
}}$ is triangular. Hence, the result, Eq.(5.70), represents the determinant
for the hexagonal lattice, while the actual calculations were made using the
reciprocal (dual) lattice, which is triangular. For the triangular lattice
the associated elliptic curve is $y^{2}=x^{3}-1$ [\textbf{60}]. Therefore,
one finds $k^{2}=\frac{1}{2}(1-\sqrt{-3})$ and, because $k^{\prime
2}=1-k^{2},$ one finds $kk^{\prime }=1$, which is used in Eq.(5.75). Also,
for the period $\omega _{1}$ one obtains $\omega
_{1}=\int\limits_{1}^{\infty }\frac{dx}{\sqrt{x^{3}-1}}.$ Substitutions $%
x=1/y$ and $1-y^{3}=z$ produce $\omega _{1}=\frac{1}{3}\int%
\limits_{0}^{1}dzz^{\frac{-1}{2}}(1-z)^{\frac{5}{6}}.$ Accordingly, 
\begin{equation}
K(k)=\frac{1}{6}\sqrt{\frac{\sqrt{3}}{2}(\sqrt{3}-\sqrt{-1})}\frac{\Gamma
(1/6)\Gamma (1/2)}{\Gamma (2/3)}.  \tag{5.77}
\end{equation}%
Using Eq.(5.75b) and, taking into account that for the triangular lattice $y=%
\frac{\sqrt{3}}{2},$ we obtain 
\begin{equation}
\det \text{ }^{\prime }\Delta _{\tilde{g}}=\frac{\sqrt{3}}{2}\frac{1}{36}%
\frac{4}{\pi ^{2}}\frac{\sqrt{3}}{2^{4/3}}\left[ \frac{\Gamma (1/6)\Gamma
(1/2)}{\Gamma (2/3)}\right] ^{2}=0.357567,  \tag{5.78}
\end{equation}%
to be compared with Eq.(5.70)\footnote{%
Apparently, there is a minor typographical error in the bound, Eq.(5.70),
taken from the OPS paper [\textbf{26}].}

The packing fraction $q_{2}^{\Box }$ for the square lattice (e.g., see
Appendix c) can be obtained using simple arguments presented in Ref.[\textbf{%
61}] and is given by: $q_{2}^{\Box }=0.7853.$ Accordingly, the ratio $%
q_{2}^{\triangle }/q_{2}^{\Box }=1.154845$ while the ratio of determinants
is $\frac{\det \text{ }^{\prime }\Delta _{\tilde{g}}}{\det \text{ }^{\prime
}\Delta _{g}}=1.032237.$ From here we conclude that, although the
inequalities between the packing fractions and the inequalities between the
respective determinants are in accord with each other, one \textit{cannot}
claim that the values of determinants are proportional to their respective
packing fractions. Also, there is no need to recalculate determinants for
the hexagonal lattice because the $\tau $ parameter is the same as for the
triangular lattice [\textbf{53}] and so is its imaginary part. We can,
loosely speaking, call such lattices self-dual\footnote{%
According to the accepted terminology [\textbf{52}], the lattice is
self-dual if its disciminant is equal to 1.The cubic lattice in \textbf{Z}$%
^{n}$ is trivially self-dual (type I lattice). But there are other less
trivial examples in higher dimensions (type II lattices in terminology of
Ref.[\textbf{52}] ).}. We shall encounter the same kind of self-duality in 3
dimensions as well. This will be discussed in the next section.

\bigskip

\section{Ginzburg-Landau theories in dimensions 3 and higher}

\bigskip

\subsection{\textit{\ }Designing higher dimensional CFT(s)}

\subsubsection{ General Remarks}

\textit{\ }In the previous section we\textit{\ }explained a delicate
iterrelationship between the path integrals Eq.(5.5) and (5.23). From the
literature on CFT cited earlier it is known that, actually, \textit{both}
are used for desingning of different CFT\ models in 2 dimensions. For
instance, if one entirely ignores the effects of curvature in Eq.(5.5), then
one ends up with the Gaussian-type path integral whose calculation for the
flat torus is discussed in detail in Ref.[\textbf{20}], pages 340-343, and,
by different methods, in our Section 5. By making appropriate changes to the
boundary conditions (or, equivalently, by considering appropriately chosen
linear combinations of modular invariants) it is possible to build partition
functions for all exiting CFT models. For the same purpose one can use the
path integral given by Eq.(5.23), but the calculation proceeds differently
as we explained in Section 5. Since in 2 dimensions conformal invariance is
crucial in obtaining exact results, use of the Gaussian-type path integrals
is, strictly speaking, not permissible. Fortunately, saddle point-type
calculations made for the path integral, Eq.(5.23), produce the same results
since the extremal metrics happens to be flat. If one does not neglect
curvature effects in Eq.(5.5), one ends up with the integrand of the path
integral, Eq.(5.23). This result is a consequence of Eq.(5.6) known as a
conformal anomaly. If one would like to proceed \textit{in analogous fashion}
in dimensions higher than two one finds that there is a profound difference
between calculations done in odd and even dimensions. We would like to
explain this circumstance in some detail now. By doing so we shall provide a
positive solution to the 2nd Problem formulated in the Introduction.

In 2 dimensions the conformal invariance of the action, Eq.(5.4), has been
assured by the transformational properties of the 2 dimensional Laplacian
given by Eq.(5.1). In higher dimensions, the Laplacian is transformed
according to Eq.(2.4), so that even the simplest Gaussian model is not
conformally invariant! This observation makes use of traditional
string-theoretic methods in higher dimensions problematic. In two dimensions
these are based on a two stage process. First one calculates the path
integral, Eq.(5.5), exactly and, second, one uses the result of that
calculation (the conformal anomaly) as an input in another path integral,
e.g. Eq.(5.23), which is obtained by integrating this input over all members
of the conformal class. Since in odd dimensions there is no conformal
anomaly as we shall demonstrate momentarily, such a two stage process cannot
be used. Moreover, absence of a conformal anomaly in odd dimensions also
affects the results of finite scaling analysis\footnote{%
As mentioned in Section 5.2, even in 2 dimensions the results of
calcultations of conformal anomalies were used incorrectly in the finite
size scaling analysis.}. The situation can be improved considerably if we 
\textit{do not} rely on the two stage process just described. We would like
to explain this fact in some detail now.

Even though the transformational properties (with respect to conformal
transformations) of the Laplacian, Eq.(2.4), in dimensions higher than 2 are
rather unpleasant, they can be considerably improved if, instead of the
usual Laplacian, one uses the conformal (Yamabe) Laplacian $\square _{g}$
defined by 
\begin{equation}
\square _{g}=\Delta _{g}+\hat{\alpha}R(g)  \tag{6.1}
\end{equation}%
where $\hat{\alpha}=\alpha ^{-1}=\frac{1}{4}\frac{d-2}{d-1}.$ By
construction, in 2 dimensions it becomes the usual Laplacian. In higher
dimensions its transformational properties are much simpler than those for
the usual Laplacian (e.g see Eq.(2.4)). Indeed, it can be shown [\textbf{62}%
] that 
\begin{equation}
\square _{e^{2f}g}=e^{-(\frac{d}{2}+1)f}\square _{g}e^{(\frac{d}{2}-1)f}. 
\tag{6.2a}
\end{equation}%
This result can be easily understood if we use results of Section 2. Indeed,
since $e^{2f}=\varphi ^{p-2}$ and $p=\frac{2d}{d-2}$, we obtain at once $e^{(%
\frac{d}{2}-1)f}=\varphi $, while the factor $e^{-(\frac{d}{2}+1)}$ is
transformed into $\varphi ^{1-p}.$ From here, it is clear that Eq.(2.12) for
scalar curvature can be equivalently rewritten as 
\begin{equation}
\tilde{R}(\tilde{g})=\alpha \varphi ^{1-p}(\Delta _{g}\varphi +\alpha
^{-1}R(g)\varphi ),  \tag{6.3}
\end{equation}%
where $\tilde{g}=e^{2f}g$ , so that 
\begin{equation}
\tilde{R}(\tilde{g})=\alpha \square _{e^{2f}g}\text{ .}  \tag{6.4}
\end{equation}%
In practical applications it could be more useful to consider two successive
conformal transformations made with conformal factors $e^{2f}$ and $e^{2h}.$
If $e^{2f}=\varphi ^{p-2}$ and $e^{2h}=\psi ^{p-2}$ then, we obtain, 
\begin{equation}
\Delta _{\tilde{g}}\psi +\alpha ^{-1}R(\tilde{g})\psi =\varphi ^{1-p}[\Delta
_{g}\left( \varphi \psi \right) +\alpha ^{-1}R(g)\left( \varphi \psi \right)
].  \tag{6.2b}
\end{equation}%
This result is, of course, equivalent to Eq.(6.2a).

Consider now the following path integral 
\begin{equation}
\exp \left( -\mathcal{F}(g)\right) =\int D\left[ \varphi \right] \exp
\{-S_{\Box _{g}}(\varphi )\}  \tag{6.5}
\end{equation}%
where 
\begin{eqnarray}
S_{\Box _{g}}(\varphi ) &=&\int_{M}d^{d}x\sqrt{g}\{(\bigtriangledown
_{g}\varphi )^{2}+\alpha ^{-1}R(g)\varphi ^{2}\}  \notag \\
&=&\int_{M}d^{d}x\sqrt{g}[\varphi \square _{g}\varphi ]=\int_{M}d^{d}x\sqrt{%
\tilde{g}}\tilde{R}(\tilde{g}).  \TCItag{6.6}
\end{eqnarray}%
The conformal factor $\varphi ^{-p}$ in Eq.(6.3) is eliminated by the
corresponding factor coming from the volume factor of $\sqrt{\tilde{g}}=%
\sqrt{\exp (2f)g}=\varphi ^{p}\sqrt{g}$, as explained after Eq.(3.5). Thus,
Eq.(6.5)\textit{\ is the exact higher dimensional analog of the two
dimensional path integral}, Eq.(5.5). Thus, problems related to higher
dimensional CFT are those of Riemannian (quantum) gravity \textbf{[25]. }In
this paper we are not considering the pseudo Riemannian case associated with
Einsteinian gravity.

The question arises: If the path integral, Eq.(6.5), is such an analog, is
there a higher dimensional analog of Eq.(5.6)? The answer is
\textquotedblright yes\textquotedblright , if the dimension of space is even
and \textquotedblright no\textquotedblright\ if the dimension of space is
odd [\textbf{62}].

Because Eq.(5.6) is used heavily in finite size calculations [\textbf{44}],
the absence of similar results in 3 dimensions should be taken into account.
Earlier attempts to generalize these two dimensional results to higher
dimensions [\textbf{44}] were made without such consideration.

\ 

\subsubsection{\textbf{\ Lack of conformal anomaly in odd dimensions}}

\textbf{\ }

In view of its importance, we would like to provide a sketch of the
arguments leading to the\ answer "no" \ in dimension 3, important for
developments in this paper. Clearly, the same kind of arguments \ will be of
use in other odd dimensions. In doing so, although we follow arguments of
Refs. [\textbf{62,63}], some of our derivations are original. We begin by
assuming that there is a one parameter family of metrics: $\tilde{g}(x)$\
=exp$(2xf)g$. Next, we define the operator $\delta _{f}$\ via 
\begin{equation}
\delta _{f}\square _{g}=\frac{d}{dx}\mid _{x=0}\square _{\text{exp}(2xf)g}%
\text{ .}  \tag{6.7}
\end{equation}%
Taking into account Eq.(6.2) we obtain explicitly 
\begin{equation}
\delta _{f}\square _{g}=-2f\square _{g}  \tag{6.8}
\end{equation}%
and, accordingly, 
\begin{equation}
\delta _{f}e^{-t\square _{g}}=-t(\delta _{f}\square _{g})e^{-t\square _{g}}.
\tag{6.9}
\end{equation}%
These results allow us to write for the zeta function (Appendix B) 
\begin{eqnarray}
\delta _{f}\zeta _{\square _{g}}(s) &=&\frac{1}{\Gamma (s)}%
\int\limits_{0}^{\infty }dtt^{s-1}\delta _{f}Tr(e^{-t\square _{g}})  \notag
\\
&=&\frac{1}{\Gamma (s)}\int\limits_{0}^{\infty }dtt^{s}Tr(-2f\square
_{g}e^{-t\square _{g}})  \notag \\
&=&\frac{-2s}{\Gamma (s)}\int\limits_{0}^{\infty }dtt^{s-1}Tr(fe^{-t\square
_{g}}).  \TCItag{6.10}
\end{eqnarray}%
The last line was obtained by performing integration by parts. Since for
small $t^{\prime }$s it is known that, provided that $\dim \ker \square
_{g}=0\footnote{%
For path integrals this is always assumed since zero modes of the
corresponding operators are associated with some kind of translational,
rotaional, etc. symmetry. To eliminate the undesirable dilatational
symmetry, one actually should use the Yamabe functional, Eq.(3.6), as
explained in Section 3. Although this is silently assumed thus far,
arguments additional to those in Section 3 will be introduced further below.}%
,$\ 
\begin{equation}
Tr(fe^{-t\square _{g}})\simeq \sum\limits_{k=0}^{\infty }\left(
\int\limits_{M}f(x)u_{k}(x)dvol\right) t^{k-d/2}.  \tag{6.11}
\end{equation}%
Using this result in Eq.(6.10) produces 
\begin{eqnarray}
\delta _{f}\zeta _{\square _{g}}(0) &=&-\frac{2s}{\Gamma (s)}%
(\int\limits_{0}^{1}dtt^{s-1}Tr(fe^{-t\square _{g}})+\int\limits_{1}^{\infty
}dtt^{s-1}Tr(fe^{-t\square _{g}}))  \notag \\
&=&-\frac{2s}{\Gamma (s)}(\sum\limits_{k=0}^{\infty }\frac{%
\int\nolimits_{M}fu_{k}dvol}{s+k-d/2}+\int\limits_{1}^{\infty
}dtt^{s-1}Tr(fe^{-t\square _{g}}))\mid _{s=0}.  \TCItag{6.12}
\end{eqnarray}%
Since for $s\rightarrow 0^{+}\ $we have$\ \ \left( 1/\Gamma (s)\right) \sim
s,$\ the second (regular) term in brackets will become zero when multiplied
by the combination $\frac{2s}{\Gamma (s)}$, while the first term will become
zero even if it might acquire a pole (when $d=2k$). Hence, for all
dimensions $d\geq 3$\ we obtain $\delta _{f}\zeta _{\square _{g}}(0)=0$\ or 
\begin{equation}
\zeta _{\square _{g}}(0)=\zeta _{\square _{\tilde{g}}}(0).  \tag{6.13}
\end{equation}%
This result can be used further now. Indeed, if we write 
\begin{equation}
\delta _{f}\left[ \Gamma (s)\zeta _{\square _{g}}(s)\right] =\Gamma
(s)[\delta _{f}\zeta _{\square _{g}}(0)+s\delta _{f}\zeta _{\square
_{g}}^{\prime }(0)+O(s^{2})]  \tag{6.14}
\end{equation}%
and take into account Eq.(6.13) and the fact that $s\Gamma (s)=1$\ we
obtain, 
\begin{eqnarray}
\delta _{f}\zeta _{\square _{g}}^{\prime }(0) &=&\delta
_{f}\int\limits_{0}^{\infty }dtt^{s-1}Tr(e^{-t\square _{g}})\mid _{s=0} 
\notag \\
&=&-2s(\sum\limits_{k=0}^{\infty }\frac{\int\nolimits_{M}fu_{k}dvol}{s+k-d/2}%
+\int\limits_{1}^{\infty }dtt^{s-1}Tr(fe^{-t\square _{g}}))\mid _{s=0}. 
\TCItag{6.15}
\end{eqnarray}%
Applying the same arguments to Eq.(6.15) as those which were used for
Eq.(6.12) we conclude that, provided that $\dim \ker \square _{g}=0,$\ in
odd dimensions, $\delta _{f}\zeta _{\square _{g}}^{\prime }(0)=0$. That is%
\begin{equation}
\zeta _{\square _{g}}^{\prime }(0)=\zeta _{\square _{\tilde{g}}}^{\prime
}(0).  \tag{6.16}
\end{equation}%
In view of Eq.(5.9) this leads also to 
\begin{equation}
\det \square _{g}=\det \square _{\tilde{g}},  \tag{6.17}
\end{equation}%
QED.

\subsubsection{\protect\bigskip 3d CFT\ path integrals}

The previously obtained results can be refined further if we recall
Eq.s(5.9)-(5.11). In particular, let $e^{2\bar{\varphi}}$ in Eq.(5.11) be
rewritten as some nonnegative constant $l$. \ Then we obtain 
\begin{equation}
\zeta _{\tilde{g}}(s)=l^{s}\zeta _{g}(s).  \tag{6.18}
\end{equation}%
This result is consistent with Eq.(6.13) for $s=0.$ Differentiation with
respect to $s$ produces 
\begin{equation}
\zeta _{\tilde{g}}^{\prime }(0)=\zeta _{g}(0)\ln l+\zeta _{g}^{\prime }(0). 
\tag{6.19}
\end{equation}

In view of Eq.(5.9), this result is equivalent to lndet $\square _{\tilde{g}%
}=\ln $det $\square _{g}-\zeta _{g}(0)\ln l$. This result apparently
contradicts Eq.(6.17), but the contradiction is only apparent in view of the
earlier footnote. The situation is easily correctable if in the path
integral, Eq.(6.5), we replace the action functional $S_{\Box }(\varphi )$
by that of Yamabe given by Eq.s(3.2) (or (3.6)). This, by the way, allows us
to fix the value of $\zeta _{g}(0)$ in Eq.(6.19): provided that we identify
the constant $l$ with the volume $V$, the value of $\zeta _{g}(0)=\frac{2}{p}%
.$ After this, the situation\ in the present case becomes similar to that
encountered in the previous section, e.g. see Eq.s (5.12 and (5.13). Now,
instead of the functional $\mathcal{F}(\varphi )$ given by Eq.(5.15), we
consider the related functional given by 
\begin{eqnarray}
\mathcal{F}(\varphi ) &=&\ln \text{det}\square _{g}-\zeta _{g}(0)\ln V 
\notag \\
&\equiv &\ln \text{det}\square _{g}-\frac{2}{p}\ln V.  \TCItag{6.20}
\end{eqnarray}%
For the path integral calculations, a functional defined in such a way is
not yet sufficient. To repair this deficiency we have to impose a volume
constraint. That is we need to consider the path integral of the type 
\begin{equation}
Z_{Y}(V)=\int D[\varphi ]\delta (\int_{M}d^{d}x\sqrt{g}\varphi ^{p}-V)\exp
(-S[\varphi ])  \tag{6.21}
\end{equation}%
with $S[\varphi ]$ given by Eq.(3.2) (or (3.6)).

Clearly, the path integral $Z_{Y}(V)$ ($Y$ in honor of Yamabe) is the exact
higher dimensional analog of the \textquotedblright
stringy\textquotedblright\ path integral $S_{L}(A)$ given by Eq.(5.33). In
view of Eq.(3.7), it also can be viewed as the path integral for pure
gravity in the presence of the cosmological constant. Because of this, the
standard path integral for the self interacting scalar $\varphi ^{4}$ (or
LGW) field theory is obtainable now in complete analogy with Eq.(5.32), i.e.:

\ 

\begin{equation}
\int D\left[ \phi \right] \exp \{-S_{LGW}(\phi )\}=\int\limits_{0}^{\infty
}dVe^{-bV}Z_{Y}(V)  \tag{6.22}
\end{equation}%
But, since the variation of the Yamabe functional produces the same Eq.(3.3)
as can be obtained with LGW functional, $S_{LGW}(\phi ),$ one can develop
things differently, but surely equivalently. \ To this purpose, instead of
the functional $S[\varphi ]$ given in Eq.(3.2) we use 
\begin{equation}
S_{V}[\varphi ]=\frac{1}{V^{\frac{2}{p}}}\int_{M}d^{d}x\sqrt{g}%
\{(\bigtriangledown _{g}\varphi )^{2}+R(g)\varphi ^{2}\}  \tag{6.23}
\end{equation}%
and replace $S[\varphi ]$ in the exponent of the path integral in Eq.(6.21)
by $S_{V}[\varphi ]$ from Eq.(6.23). Then, instead of Eq.(6.22), we obtain, 
\begin{equation}
\int D\left[ \phi \right] \exp \{-S_{LGW}(\phi )\}\ddot{=}%
\int\limits_{0}^{\infty }dVZ_{Y}(V)\text{,}  \tag{6.24}
\end{equation}%
where the sign \"{=} means \textquotedblright supposedly\textquotedblright .
This is so, because, at the level of saddle point calculations the left hand
side and the right hand side produce the same G-L equation. Beyond the
saddle point, calculations are not necessarily the same. \ Although we plan
to discuss this issue in detail in subsequent publications, some special
cases are further discussed below in this section.

It should be clear, that at the level of saddle point calculations,
replacement of the functional $S[\varphi ]$ in Eq.(6.21) by $\mathcal{F}%
(\varphi )$ from Eq.(6.20) is completely adequate, so that the sequence of
steps in analysis performed for the two dimensional case in Section 5 are
transferable to higher dimensions without change. This can be summarized as
follows.

\ Although in 3 dimensions we have the result given by Eq.(6.17), which
forbids use of identities like that in Eq.(5.6), still, based on arguments
just presented, the functional $\mathcal{F}(\varphi )$ defined by Eq.(6.20)
should be used in the exponent of the corresponding path integral replacing
that given in Eq.(5.15) in 2 dimensions. Since by doing so one will be
confronted with the same type of minimization problems as discussed earlier
in Section 5\footnote{%
This conclusion had been reached without any reliance on path integrals and
on physical applications in Ref.[\textbf{64}]. In this and related Ref.[%
\textbf{31}] the extremal properties of determinants of $\square _{g}$ and $%
\Delta _{g}$ with respect to variations of the background metric were
studied.}, thus defined functional integral is an exact 3 dimensional analog
of the path integral, Eq.(5.30).

\ 

\subsection{\ \ \ Completion of the work by Lifshitz in 3 dimensions}

\subsubsection{\protect\bigskip\ \ \ \ General remarks}

The results just obtained allow us to proceed\ with the rest of our
developments in complete accord with results of Section 5. \ This means that
our task from now on will be to provide an affirmative answer to the 1st
Problem formulated in Section 1.4 following logical steps developed
previously. To this purpose, it is helpful to use some results from the
classical paper by Hawking on zeta function regularization of path integrals
in curved spacetime [\textbf{65}] in view of the fact that, typically, the
calculation of path integrals is done by the saddle point method. In the
present case, the saddle point \ level of approximation for the path
integral, Eq.(6.21), is equivalent to minimization of the functional $%
\mathcal{F}(\varphi )$ given by Eq.(6.20). In turn, this is equivalent to
minimization of the Yamabe functional, Eq.(3.2). The following theorem
(attributed to Aubin, Trudinger and Yamabe) is very helpful for this task.

\textit{For any d-dimensional compact manifold} $M$%
\begin{equation}
\lambda (M)\leq \lambda (S^{d})  \tag{6.25}
\end{equation}%
\textit{where }$\lambda (S^{d})$\textit{\ is the Yamabe invariant for
d-dimensional sphere}.

The proof of this result can be found in Refs.\textbf{[22,23]}.

\textbf{\ }

The previously obtained inequality, Eq.(5.37), indicates that, at least in
two dimensions, the same type of inequality exists for determinants. These
two dimensional results for determinants cannot be readily extended to
higher dimensions however since, in view of Eq.(6.17), there is no conformal
anomaly in 3 dimensions (but there is in\ dimension 4 [\textbf{66,67}]!).
Therefore, in Refs. [\textbf{66,67}] extremal properties of these
determinants with respect to changes in the background metric have been
studied. These studies produced\ inequalities analogous to that given in
Eq.(6.25), which take place only \textit{locally}, i.e. in the vicinity of
some point on $p\in M$.

The Yamabe constant $\lambda (S^{d})$ has been calculated. In particular,
for $S^{3}$ it is found to be $\lambda (S^{3})=6(2\pi ^{2})^{\frac{2}{3}},$
Ref.[\textbf{33}]\footnote{%
In $d$-dimensions this result is replaced by $\lambda (M)\leq d(d-1)\left[
Vol(S^{d})\right] ^{\frac{2}{d}}.$}. Calculation of \ $\lambda (M)$ for
manifolds other than those diffeomorphic to $S^{3}$ has been found to be
surprisingly difficult. Only few explicit examples are known to date [%
\textbf{33}]. Fortunately for us, they are just sufficient for our current
purposes. In particular, Gromov and Lawson [\textbf{68}] have demonstrated
that $\lambda (T^{d})=0$ for $d\geq 3$ and, moreover, the same result holds
for the connected sum ($T^{d}$ $\#$ $T^{d}$, etc.) of $d-$dimensional tori.
Aubin Ref.[\textbf{22}], pages 150-152, proved the following theorem

\textit{Let M be d-dimensional }$C^{\infty }$\textit{\ compact Riemannian
manifold, then there is a conformal metric whose scalar curvature is either
a nonpositive constant or is everywhere positive.}

From the results of Section 1 and 2 it should be clear that in the case of
the G-L functional, based on physical arguments (e.g. read the discussion
related to Eq.(1.20)), one should look at cases of \textit{nonpositive
constant scalar curvatures}. That is, one should look at the mass term $%
m^{2} $ representing constant negative scalar curvature for temperatures 
\textit{below} criticality (and zero scalar curvature at criticality) in
accord with Eq.(2.10).\footnote{%
In our recent work, Ref.[\textbf{69}], we have studied the connections
between the hyperbolicity and conformal invariance. Development of such
connections is linked with the study of properties of hyperbolic 3-manifolds
and orbifolds. Examples of such connections can be found in the same
reference in the context of AdS-CFT correspondence.}.

\ Using the above Theorem by Aubin and taking into account details of its
proof we conclude that for manifolds admitting constant curvature (spherical(%
$s$), hyperbolic($h$) or flat ($0$)) one should expect

\begin{equation}
\lambda _{h}\leq \lambda _{0}\leq \lambda _{s}.  \tag{6.26}
\end{equation}

This result is consistent with that known in two dimensions. Indeed,
formally using Eq.(3.6) in two dimensions and employing the Gauss-Bonnet
theorem we obtain 
\begin{equation}
\lambda (M)=4\pi \chi (M).  \tag{6.27}
\end{equation}

As discussed in Section 1 in this work, we are interested mainly in
determining symmetries of the high temperature phase. In this phase the G-L
order parameter is zero and so is the free energy at the saddle point level
of approximation (e.g. read the comments after Eq.(5.31)). This implies that
the Yamabe constant is zero, i.e. that $\lambda (T^{d})=0$. Next, using
Eq.(6.16), valid only if the volume constraint is imposed as we explained
above, for a given background metric $g$ we can choose the metric $\tilde{g}$
as flat. With such a choice of metric we can use the results of Richardson,
Ref.[\textbf{31}], which we would like to summarize briefly now.

Let $\tilde{g}(u)=e^{\phi (u)}g$ be 1-parameter family of metrics of fixed
volume and such that $\tilde{g}(0)=g$. This implies that $\phi (0)=0$ and $%
\int\limits_{M}e^{\phi (u)}dV_{0}=V.$ In two dimensions, using results of
OPS [\textbf{26}], especially their Eq.(1.12), it is straightforward to
obtain the following result 
\begin{equation}
\frac{d}{du}(-\ln \det^{{}}\text{ }^{\prime }\Delta _{g(u)})\mid _{u=0}=\dot{%
\zeta}_{g}^{\prime }(0)=\frac{1}{12\pi }\int\limits_{M}\dot{\phi}K(g)dV_{g}%
\text{,}  \tag{6.28}
\end{equation}%
where $K(g)$ is the Gaussian curvature for the metric $g$. $\dot{\zeta}%
_{g}(0)$ and $\dot{\phi}$ represent $\frac{d}{du}\zeta _{g(u)}(0)\mid _{u=0}$%
and $\frac{d}{du}\phi (u)\mid _{u=0}$respectively. Volume conservation
implies$\ $%
\begin{equation}
\frac{d}{du}\int\limits_{M}e^{\phi (u)}dV_{g}\mid _{u=0}=\int\limits_{M}\dot{%
\phi}dV_{g}=0  \tag{6.29}
\end{equation}%
in accord with the earlier result, Eq.(5.17). If the Gaussian curvature $%
K(g) $ is constant, then Eq.(6.28) and Eq.(6.29) produce the same result.
This implies that $\frac{d}{du}\zeta _{g(u)}(0)\mid _{u=0}=0$. That is $g,$
is the \textquotedblright critical\textquotedblright\ (extremal) metric. In
view of Eq.(6.28), this also means that for such a metric the free energy\
attains its extremum. We encountered such extremal situations in the
previous section when we demonstrated that the triangular and hexagonal
lattices have the highest possible free energies as compared to other
lattices. We shall demonstrate below that the same holds in 3 dimensions
where the hcp and fcc lattices possess the highest possible free energies as
compared to other lattices. To this purpose, we need to generalize the above
two dimensional results to higher dimensions. Formally, this is not an easy
task mainly because of the differences in properties of the Laplacian under
conformal transformations in two and higher dimensions, e.g. see Eq.(2.4),
causing absence of the Moser-Trudinger type of inequalities in dimensions
higher than two. Nevertheless, some of the difficulties can be easily
overcome with help of the results we have already. For instance, by
combining Eq.s (6.10), (6.14) and (6.15) we obtain at once 
\begin{eqnarray}
\delta _{f}\zeta _{\square _{g}}^{\prime }(0) &=&\delta
_{f}\int\limits_{0}^{\infty }dtt^{s-1}Tr(e^{-t\square _{g}})\mid _{s=0} 
\notag \\
&=&\int\limits_{0}^{\infty }dtt^{s}Tr(-2f\square _{g}e^{-t\square _{g}})\mid
_{s=0}.  \TCItag{6.30}
\end{eqnarray}%
From here we obtain essentially the same result as the main theorem by
Richardson [\textbf{31}], i.e. 
\begin{equation}
\dot{\zeta}_{\square _{\hat{g}}}^{\prime }(0)\mid
_{u=0}=0=\int\limits_{M}dV_{g}\dot{\phi}(x)\square _{g}\zeta (1,x,x). 
\tag{6.31}
\end{equation}%
That is, provided that we replace $\square _{g}$ by $\Delta _{g}$ and
require that \textit{locally} $\zeta (1,x,x)=const$, with the heat kernel $%
\zeta (s,x,x)$ given (as usual) by 
\begin{equation}
\zeta (s,x,x)=\sum\limits_{k=1}^{\infty }\frac{\psi _{k}^{2}(x)}{\lambda
_{k}^{s}},  \tag{6.32}
\end{equation}%
we obtain the main result by Richardson, Ref.[\textbf{31}], e.g. see his
Theorem 1 and Corollary 1.1. Here $\psi _{k}(x)$ are eigenfunctions of the
Laplacian (or Yamabe Laplacian respectively) corresponding to eigenvalues $%
\lambda _{k}$ with $0=\lambda _{0}<\lambda _{1}\leq \lambda _{2}\leq \cdot
\cdot \cdot \leq \lambda _{k}\leq \cdot \cdot \cdot .$ In two dimensional
case the condition for criticality is given by $\dot{\zeta}_{g}^{\prime
}(0)=0$ is \textit{local} meaning that, provided the the volume is
constrained, the constancy of the Gaussian curvature $K(g)$ at given point
of $M$ is caused by the metric $g$ for which $\dot{\zeta}_{g}^{\prime }(0)$
is extremal. In 3 and higher dimensions, the constancy of curvature at the
point of $M$ is replaced by constancy of $\zeta (1,x,x)$ under the same
conditions of volume conservation. This condition is necessary but is not
sufficient now, since the analog of the Moser-Trudinger inequality (used to
prove sufficiency in 2 dimensions) does not exists. Instead, one should
study locally the second variation of $\zeta _{\tilde{g}}^{\prime }(0)$ with
respect to the underlying background metric in order to decide if such
(local) extremum is maximum or minimum. Fortunately, this task was
accomplished in Ref.s [\textbf{31,64}]. In particular, Richardson [\textbf{31%
}] obtained the following theorem of major importance for our work:

\ 

The Euclidean metric on a cubic 3-torus is a local \textsl{maximum} of
determinant of the Laplacian with respect to fixed-volume conformal
variations of the metric.

\ 

This Theorem is proven only for the cubic 3-torus. The word
\textquotedblright local\textquotedblright\ means that there could be (or,
there are, as we shall demonstrate) other 3-tori also providing local maxima
for determinants. In fact, according to the result by Chiu [\textbf{70}],
all determinants of flat 3-tori possess local maxima so that the determinant
for the face centered cubic (fcc) lattice has the largest determinant.
Unfortunately, his results are nonconstructive and hence, cannot be used in
physical applications. Therefore, below we provide an entirely different way
to reach the same conclusions.

\ The second variation of the Yamabe functional was calculated by Muto [%
\textbf{71}] (see also Ref.[\textbf{25}]) with the result: 
\begin{equation}
\left( \frac{d^{2}}{dt^{2}}\mathcal{R}(g(t))\right) _{t=0}=\frac{d-2}{2}%
[\int\limits_{M}dV_{g}(\sigma (\bigtriangledown _{g}\varphi
)^{2}-R(g)\varphi ^{2})],  \tag{6.33}
\end{equation}%
where the constant $\sigma =d-1.$ As in the case of quadratic actions in the
flat space [\textbf{15,19}] the second variation (with volume constrained to
be equal to one) looks very much the same as the original quadratic Yamabe
functional, except for the \textquotedblright wrong\textquotedblright\ sign
in front of scalar curvature\footnote{%
This \textquotedblright wrong\textquotedblright\ sign has significance,
however. It is in accord with existing calculations of the fluctuation
corrections to G-L theory in the low temperature phase, e.g. see Eqs.(1.20),
(1.22). Moreover, given that the low temperature \textquotedblright
mass\textquotedblright\ term in Eq.(1.22) is $2\left\vert a\right\vert $ \
and comparing the expansion, Eq.(1.21), with Eq.(6.33) we obtain (for $d=4$) 
\emph{exactly the same} fluctuation kernel using Eq.(6.33). This is so since
in $d=4$ the combination $\frac{4-2}{2}R(g)=2R(g)$ and $R(g)$ is \textsl{%
negative} in the low temperature phase. The constant $c$ in front of
Laplacian is 6 in the present case while it was left unspecified in
Eq.(1.22).}. Following Muto [\textbf{71}] we conclude that: a) if $R(g)$ is
positive, the second variation can be made positive for appropriately chosen 
$\varphi \footnote{%
This can be easily understood if we expand $\varphi $ into Fourier series
made of eigenfunctions of the Laplacian and take into account that for 
\textbf{any} closed Riemannian manifold the spectrum of the Laplacian is
nonnegative and nondecreasing [\textbf{22}].},b)$ if $R(g)$ is negative, it
is positive for the same reasons.

The positivity of second variation implies that the extremal \textit{%
constant curvature} metric $g$ provides a \textit{locally stable} \textit{%
minimum }for $\mathcal{R}(g(t))$ (that is, using results of Section 4, the
Einstein metric obtained as solution \ to Eq.(4.2) is stable among nearby
metrics).

It is interesting to notice that calculation of higher order fluctuation
corrections to the Yamabe path integral, Eq.(6.21), involves calculations on
the moduli space of Einsteinian metrics, Ref.[\textbf{72}]. This observation
provides a strong link between higher dimensional LGW theory and two
dimensional string inspired CFT discussed in the previous section.
Naturally, Eq.(6.24) can be used to investigate to what extent the final
results of conventional field-theoretic calculations, Ref.[\textbf{19}], may
differ from more sophisticated string-theoretic calculations in the style of
Ref.s [\textbf{27,28,72}]. This task is left for further study.

For the same reasons as in the two dimensional case considered in the
previous section, we are not interested in a positive constant curvature
metric typical for $S^{n}$even though such metric do provide a local maximum
for determinants of both Laplace and Yamabe operators [\textbf{64}]. Thus,
we are left \ only with zero and negative curvature metrics characteristic
for physical systems at and below criticality (i.e. below $T_{c}).$
Thermodynamically, the system at criticality should possess \textit{higher}
free energy than the system \textit{below} criticality. In our case, at
criticality the free energy $\mathcal{F}$=$\frac{1}{2}\ln \det \Delta _{%
\tilde{g}}$ (e.g. see Appendix D). Hence, the above cited theorem by
Richardson guarantees that for cubic 3-tori the free energy is larger as
compared to that below criticality (which corresponds to negative
curvatures) and Chiu's results [\textbf{70}] imply (by analogy with two
dimensions) that among 3-torus lattices the fcc lattice possess the highest
free energy. Below we shall provide more direct demonstration of this fact
and, in addition, we shall demonstrate that the spectra of both fcc and hcp
lattices possess the same value for determinants, thus implying the
possibility of phase transitions between these lattices. Such transitions
have been indeed observed in nature [\textbf{73}], but their description
falls outside of the scope of this paper.

Thus, in the rest of this section we shall concentrate on explicit
calculations of the toral determinants for different 3-dimensional lattices.
Such calculations involve the use of the 3-dimensional version of the zeta
function, Eq.(5.41). Unlike the two dimensional case considered earlier, we
cannot apply directly the 1st Kronecker limit formula in order to obtain the
corresponding values for determinants. We are also unable to extend the
ideas of complex multiplication in order to determine the types of allowable
lattices. For even dimensional spaces, where one can use the concept of the
Abelian variety such a task can be accomplished. Some of these varieties
possess complex multiplication and can be mapped into even dimensional tori [%
\textbf{74}]. We are not aware of similar results for odd dimensional
spaces. Thus, we have calculated determinants numerically using a procedure
to to be discussed below. \ The results obtained are in complete qualitative
agreement with those obtained in two dimensions.

In order to introduce our readers to issues involved in such calculation, we
would like to discuss the 3 dimensional analog of the 1st Kronecker limit
formula now. It was considered by Bump and Goldfield [\textbf{57}] and later
summarized in the paper by Efrat [\textbf{75}].

We begin with the following observation. As it is shown by Sarnak [\textbf{76%
}], the function $E(\tau ,s)$ in Eq.(5.61) is an eigenfunction of the
hyperbolic two dimensional Laplacian $\Delta =-y^{2}(\dfrac{\partial ^{2}}{%
\partial x^{2}}+\dfrac{\partial ^{2}}{\partial y^{2}}),$ i.e. 
\begin{equation}
\Delta E(\tau ,s)=s(1-s)E(\tau ,s).  \tag{6.34}
\end{equation}%
We discussed this type of eigenvalue equation extensively in our earlier
work, Ref.[\textbf{69}]. In the same work we discussed the $d-$dimensional
extension of such an eigenvalue problem. Unfortunately, these results cannot
be used in the present case. As it is demonstrated by Bump [\textbf{77}],
instead of looking for eigenfunctions of the 3 dimensional hyperbolic
Laplacian, one should consider a more complicated eigenvalue problem. It is
important to realize at this point that \textit{this eigenvalue problem
should be of the same relevance to all exactly solvable 3 dimensional
statistical mechanics models as two dimensional eigenvalue problem,
Eq.(6.34), to two dimensional exactly solvable models} discussed in Ref.[%
\textbf{20}]. Since we are not aware of exact solutions of non trivial 3
dimensional models, we believe that discussing the issues involved in such
calculations might shed some new light on the whole problem of exact
solvability in dimension 3. In addition, we would like to present these
results in order to compare them against our calculations of determinants
presented below.

To begin, we need to write the 3 dimensional analog of $E(\tau ,s)$ (e.g.
compare with Eq.(5.61)). It is given by [\textbf{78}] 
\begin{equation}
E(\tau ,s)=\sum\limits_{(m,n,k)=1}\frac{\left( y_{1}^{2}y_{2}\right) ^{s}}{%
[y_{1}^{2}\left\vert kz_{2}+m\right\vert ^{2}+(kx_{3}+mx_{1}+n)^{2}]^{\frac{%
3s}{2}}}.  \tag{6.35}
\end{equation}%
In Appendix B we introduce the Epstein zeta function, Eq.(B.5). It can be
demonstrated [\textbf{57,75}] that 
\begin{equation}
Z(\mathbf{A},s)=\left\vert \mathbf{A}\right\vert ^{-\frac{s}{2}}\zeta
(3s)E(\tau ,s)  \tag{6.36}
\end{equation}%
where $\left\vert \mathbf{A}\right\vert =\det \mathbf{A}$ and $%
z_{2}=x_{2}+iy_{2}$, $y_{1},y_{2}\in \mathbf{R}_{>0},x_{1},x_{2},x_{3}\in 
\mathbf{R.}$ By analogy with the two dimensional case, one can establish a
very important functional equation for the combination $E^{\ast }(\tau
,s)=\pi ^{\frac{-3s}{2}}\Gamma (\frac{3s}{2})\zeta (3s)E(\tau ,s).$ Namely, $%
E^{\ast }(\tau ,s)=E^{\ast }(\tau ,1-s)$ [\textbf{57}]. This equation allows
us to extract the $s\rightarrow 0^{+}$ limit for $E(\tau ,s)$ using the
associated result for $s\rightarrow 1^{+}$ obtained in Efrat's paper [%
\textbf{75}]. For the combination $\zeta (3s)E(\tau ,s),$ which he also
denotes as $E^{\ast }(\tau ,s),$ he obtains 
\begin{equation}
E^{\ast }(\tau ,s)=\frac{2/3}{s-1}+(C-2/3\ln \left( y_{1}y_{2}^{2}\right)
-4\ln g(\tau ))+O(s-1),  \tag{6.37}
\end{equation}%
where 
\begin{equation}
g(\tau )=\exp (-\frac{y_{1}^{1/2}y_{2}}{8\pi }E^{\ast
}(z_{1},s))\prod\limits_{(m,n)\neq 0}\left\vert 1-\exp (-2\pi
y_{2}\left\vert nz_{1}-m\right\vert +2\pi i(mx_{1}+nx_{4}))\right\vert 
\tag{6.38}
\end{equation}%
with $z_{1}=x_{1}+iy_{1}$ and $x_{4}=x_{3}-x_{1}x_{2}$ and $C$ being a known
constant. The limiting expression \ (for $s\rightarrow 0^{+})$ can be found
in the work by Chiu [\textbf{70}]. Since neither Chiu nor Efrat have
provided any explicit examples of actual calculations involving formulas
just presented, we have chosen another approach to the whole calculation of
these limits. Before discussing our calculations we would like to mention
that Bump [\textbf{77}] and, following him, Efrat [\textbf{75}] have
demonstrated that $E(\tau ,s)$ defined in Eq.(6.35) is an eigenfunction of
two operators $\Delta _{1}$ and $\Delta _{2}$ \ (whose explicit form is
rather complicated [\textbf{77}]) so that 
\begin{eqnarray}
\Delta _{1}E(\tau ,s) &=&3s(s-1)E(\tau ,s)  \TCItag{6.39a} \\
\Delta _{2}E(\tau ,s) &=&-s(s-1)(2s-1)E(\tau ,s)  \TCItag{6.39b}
\end{eqnarray}%
to be compared with Eq.(6.34).

\ 

\subsubsection{\textit{\ }Calculation of determinants: general discussion of
the numerical algorithm}

\ 

So far in this paper calculation of determinants has been done with the help
of zeta functions, e.g. see Eq.(5.9). This method required use of the first
Kronecker limit formula in two dimensional calculations. Generalization of
this result to higher dimensions was formally accomplished by Epstein at the
turn of 20th century [\textbf{79}]. However, his results were too general to
allow any practical calculations. This fact caused many attempts to
improve/simplify his calculations. The result, Eq.(6.37), is one of many [%
\textbf{70,80}], etc. It can be derived directly from the 3 dimensional
Epstein zeta function [\textbf{75,77}]. However, subsequent results by Chiu [%
\textbf{70}] stop short of actual use of Eq.(6.37) in order to produce
numerical values for lattice determinants, as it is done in two dimensions.
In view of this, we are going to develop another method of calculation of
lattice determinants, which does not involve the use of zeta function. The
reliability of this alternative method is tested against exactly known
results obtained,\ again, with the help of some ideas from physics. In
Appendix D we provide the simplest example of this type of calculation, so
that readers are encouraged to read this appendix before proceding with the
rest of this subsection.

Assuming this, we need to calculate the 3 dimensional version of Eq.(5.41),
i.e. 
\begin{equation}
Z_{\tilde{g}}(s)=\sum\nolimits_{\mathit{l}\mathsf{\in \Lambda }^{\ast
}}^{\prime }\frac{1}{\left( 4\pi ^{2}\left\vert \mathbf{l}\right\vert
^{2}\right) ^{s}}.  \tag{6.40}
\end{equation}%
As in Appendix D, we disregard the constant $4\pi ^{2}$ in the denominator
and consider the following regularized sum instead 
\begin{equation}
\mathcal{S}=\sum\limits_{\mathbf{l}^{\ast }}\frac{1}{\left\vert \mathbf{l}%
^{\ast }\right\vert ^{2}+\kappa ^{2}},  \tag{6.41}
\end{equation}%
where now the summation takes place over all lattice cites of the reciprocal
lattice $\Lambda ^{\ast }$ due to the presence of the small parameter $%
\kappa ^{2},$ which will be put to zero at the end of calculations.
Evidently, as in Appendix D, we have 
\begin{equation}
\beta \mathcal{F}=\lim_{\kappa ^{2}\rightarrow 0}\frac{1}{2}\ln \frac{\det
\Delta _{\tilde{g}}(\kappa ^{2})}{\det \Delta _{\tilde{g}}(0)}%
=\int\limits_{0}^{\kappa ^{2}}d\tilde{\kappa}^{2}\sum\limits_{\mathbf{l}%
^{\ast }}\frac{1}{\left\vert \mathbf{l}^{\ast }\right\vert ^{2}+\tilde{\kappa%
}^{2}}.  \tag{6.42}
\end{equation}%
Again, as in the Appendix D, to proceed, we need to evaluate somehow the sum
at the r.h.s. of Eq.(6.42). To this purpose we would like to take advantage
of the Poisson summation formula [\textbf{81}]. In the present case we
obtain, 
\begin{equation}
\sum\limits_{\mathbf{l}^{\ast }}\frac{1}{\left\vert \mathbf{l}^{\ast
}\right\vert ^{2}+\kappa ^{2}}=\sqrt{\frac{vol(\Lambda )}{vol(\Lambda ^{\ast
})}}\sum\limits_{\mathbf{l}}\frac{e^{-\kappa \left\vert \mathbf{l}%
\right\vert }}{\left\vert \mathbf{l}\right\vert },  \tag{6.43}
\end{equation}%
where we take into account that in 3 dimensions the summand on the l.h.s.
represents the Fourier transform of the screened Coulomb (also known as the
Debye-H\"{u}ckel (D-H)) potential displayed as the summand on the r.h.s. The
parameter $\kappa $ is known in the literature on electrolyte solutions [%
\textbf{6,82}] as the D-H inverse screening length. Treated from such
perspective, our calculations are reminiscent of those for the Madelung
constant in solid state physics [\textbf{13,83}].Using Eq.(6.43) in
Eq.(6.42) and taking into account results of the Appendix A we obtain upon
integration over $\kappa $ the following result: 
\begin{equation}
\beta \mathcal{F=}\lim_{\kappa \rightarrow 0}(-1)\kappa \sqrt{disc(\Lambda )}%
\sum\limits_{\mathbf{l}}\frac{\exp (-\kappa \left\vert \mathbf{l}\right\vert
)}{\left\vert \mathbf{l}\right\vert ^{2}}.  \tag{6.44}
\end{equation}%
This result is still very inconvenient to use. To simplify matters further
we make use of the expansion Eq.(B.9) for the theta function and take also
into account Eq.s(B.5)-(B.8). This produces the following result to be used
in numerical calculations: 
\begin{equation}
\beta \mathcal{F=}\lim_{\kappa \rightarrow 0}(-1)\kappa \sqrt{disc(\Lambda )}%
\sum\limits_{n=1}^{\infty }\frac{a(n)}{n}\exp (-\kappa \sqrt{n}).  \tag{6.45}
\end{equation}%
As in Eq.(B.7), we have subtracted one so that the series starts with $n=1.$
The coefficients $a(n)$ have been tabulated for various lattices [\textbf{%
52, 84}]. The difficulty in evaluating such sums lies in the fact that there
is only a finite number of $a(n)^{\prime }s$ which are available in
literature. The difficulty with calculations containing finite number of
terms can be seen already in evaluation of much simpler sums such as, for
example, 
\begin{equation}
S(\kappa )=\kappa \sum\limits_{n=0}^{\infty }\exp (-\kappa n).  \tag{6.46}
\end{equation}%
Clearly, evaluating the geometric progression and taking the limit$:\kappa
\rightarrow 0^{+}$ produces 1 as expected but, should we keep a finite
number of terms and take the same limit, we would obtain an entirely wrong
result. The results can be considerably improved if we correlate the number
of terms in the sum $N$ with the optimal value of $\kappa $ for such $N$.
This can be achieved by minimizing the above sum with respect to $\kappa .$
This leads to the following minimization equation: 
\begin{equation}
\sum\limits_{n=0}^{N}\exp (-\kappa ^{\ast }n)=\kappa ^{\ast
}\sum\limits_{n=0}^{N}n\exp (-\kappa ^{\ast }n)  \tag{6.47a}
\end{equation}%
or, equivalently, 
\begin{equation}
\frac{1}{\kappa ^{\ast }}=\left\langle n\right\rangle \text{.}  \tag{6.47b}
\end{equation}%
The best results are obtained with still additional refinement. Since we are
interested in the limit $\kappa \rightarrow 0^{+}$, it is convenient to look
at \ the obtained values of $\kappa ^{\ast }$ as function of $N$, i.e. we
look for $\kappa ^{\ast }(N).$ We expect that for $N_{2}>N_{1\text{ }}$the
optimal $\kappa ^{\prime }s$ should behave as $\kappa ^{\ast }(N_{2})<\kappa
^{\ast }(N_{1}),$ etc. This indeed happens. Moreover, we expect that $%
S(\kappa ^{\ast }(N))$ behaves in a similar way, i.e. if we have successive
sums including $N_{1}<N_{2}<N_{3\text{ \ }}<\cdot \cdot \cdot $, we expect
that $\left\vert S(\kappa ^{\ast }(N_{1}))-S(\kappa ^{\ast
}(N_{2}))\right\vert >\left\vert S(\kappa ^{\ast }(N_{2}))-S(\kappa ^{\ast
}(N_{3}))\right\vert >\cdot \cdot \cdot $ \ so that the series is converging
in the desired direction. For the sum like that given in Eq.(6.46) this is
easy to check, but for the sum, Eq.(6.45), this is less obvious (although
numerically we obtained exactly the same type of convergence). So,
additional arguments need to be invoked. For instance, the sum in Eq.(6.44)
can be analyzed as is typically done in solid state physics. Here, one
routinely replaces such sums by integrals [\textbf{13,83}]. In our case we
obtain,

\begin{equation}
\kappa \sum\limits_{\mathbf{l}}\frac{\exp (-\kappa \left\vert \mathbf{l}%
\right\vert )}{\left\vert \mathbf{l}\right\vert ^{2}}\sim \kappa
\int\limits_{0}^{\infty }dxx^{2}\frac{\exp (-\kappa x)}{x^{2}}\sim 1. 
\tag{6.48}
\end{equation}%
As plausible as it is, one can obtain a much better estimate based on the
work by Jones and Ingham [\textbf{85}]. It is more accurate when used for
the sum, Eq.(6.45). According to these authors, if one thinks about a
selected atom at the lattice site chosen as the origin, then at large
distances $R$ from the origin, it is permissible to assume that the atoms
are more or less homogeneously distributed on the sphere of radius $R.$ In
view of this, the factor $a(n)$ is proportional to the area of a sphere of
radius $R$ while the factor of $\ n\sim R^{2}.$ Hence, in our case the sum 
\begin{equation}
\tilde{S}(\kappa )=\kappa \sum\limits_{n=1}^{\infty }\exp (-\kappa \sqrt{n})
\tag{6.49}
\end{equation}%
provides an \textsl{upper} bound for the sum in Eq.(6.45).The \textsl{lower}
bound can be estimated if we replace all $a(n)^{\prime }s$ in Eq.(6.45) by 1
thus obtaining the sum 
\begin{equation}
\check{S}(\kappa )=\kappa \sum\limits_{n=1}^{\infty }\frac{1}{n}\exp
(-\kappa \sqrt{n}).  \tag{6.50}
\end{equation}

\subsubsection{Use of designed algorithm for calculation of various lattice
determinants}

\ 

The convergence procedure for the sums defined above can be easily tested
for any $N$ and, indeed, it satisfies the convergence criteria just
described. A numerical check provided assurance that the sum, Eq.(6.45), is
indeed bounded by the above two sums for lattices we are interested in. The
numerical values for coefficients $a(n)^{\prime }s$ are taken from Ref.[%
\textbf{84}]. Specifically, Table 7 of this reference provides 80 entries
for the cubic lattice, Table 11 provides 80 entries for the fcc lattice,
Table 16 provides 80 entries for the hcp lattice and Table 24 provides 80
entries for the bcc lattice. The book by Conway and Sloane (C-S), Ref. [%
\textbf{52}], provides the values of discriminants for lattices of unit edge
length. In particular, for the bcc lattice we find: $\sqrt{disc(\Lambda )}%
=4, $ and accordingly 1 for the cubic, 2 for the fcc and $\sqrt{2\text{ }}$%
for the hcp lattices. In actual computations the following information was
taken into account. First, both the cubic and hcp lattices are self-dual [%
\textbf{52}]. Second, the bcc lattice is dual to fcc [\textbf{13,52}]. This
means, for instance, that one should be careful when using the relation $%
vol(\Lambda )vol(\Lambda ^{\ast })=1$ (or, as in solid state \ physics, Ref.[%
\textbf{13}], $vol(\Lambda )vol(\Lambda ^{\ast })=\left( 2\pi \right) ^{3})$%
\ since, if one chooses $vol\Lambda _{bcc}=4$ then, one formally gets $%
vol\Lambda _{fcc}^{\ast }=1/4$ (or $\left( 2\pi \right) ^{3}/4)$ as compared
with 2 for fcc just presented. To resolve this \textquotedblright
paradox\textquotedblright\ we use results from the solid state physics [%
\textbf{13}]. For instance, the basis vectors of the fcc lattice are known
to be: $\mathbf{a}_{1}=\frac{a}{2}(-\mathbf{a}_{x}+\mathbf{a}_{z}),$ $%
\mathbf{a}_{2}=\frac{a}{2}(\mathbf{a}_{y}+\mathbf{a}_{z}),\mathbf{a}_{3}=%
\frac{a}{2}(-\mathbf{a}_{x}+\mathbf{a}_{y})$. From here the volume is
obtained as $V_{fcc}$=$\mathbf{a}_{1}\cdot (\mathbf{a}_{2}\times \mathbf{a}%
_{3})=\frac{a^{3}}{4}.$ Using this result for $a=1,$ we obtain: $%
V_{fcc}=1/V_{bcc}$ where $V_{bcc}=4$ in accord with C-S. At the same time,
C-S provide the value 2 for $V_{fcc}.$ As a final example, let us consider
the bcc lattice. In this case we have the basis vectors $\mathbf{a}_{1}=%
\frac{a}{2}(-\mathbf{a}_{x}-\mathbf{a}_{y}+\mathbf{a}_{z}),$ $\mathbf{a}_{2}=%
\frac{a}{2}(\mathbf{a}_{y}++\mathbf{a}_{y}+\mathbf{a}_{z}),$ $\mathbf{a}_{3}=%
\frac{a}{2}(-\mathbf{a}_{x}+\mathbf{a}_{y}-\mathbf{a}_{z})$ so that the
volume is obtained as $V_{bcc}$ =$\frac{a^{3}}{2}.$ From here, we obtain $%
V_{bcc}=1/V_{fcc}$ where $V_{fcc}=2$ in accord with C-S. Since the hcp
lattice is selfdual [\textbf{52}], we obtain $V_{hcp}=1/V_{hcp}=1/\sqrt{2}.$
These examples illustrate the relationships between the (C-S) and the
accepted solid state (SS) conventions. They are summarized in the Table 2
below.\ \ \ \ \ \ \ \ \ \ \ \ \ \ \ \ \ \ \ \ \ \ \ \ \ \ \ \ \ \ \ \ \ \ \
\ \ \ \ \ \ \ \ \ \ \ 

\ \ \ \ \ \ \ \ \ \ \ \ \ \ \ \ \ \ \ \ \ \ \ \ \ \ \ \ \ \ \ \ \ \ \ \ \ \
\ \ \ 

\ \ \ \ \ \ \ \ \ \ \ \ \ \ \ \ \ \ \ \ \ \ \ \ \ \ \ \ \ \ \ \ \ \ \ \ \ \
\ \ \ \ \ Table 2\ 

\bigskip

\ \ \ \ \ \ \ \ \ \ \ \ \ \ \ \ \ \ \ \ \ \ 
\begin{tabular}{|c|c|c|c|}
\hline
Lattice & Vol & Latice & Vol \\ \hline
$bcc(C-S)$ & $4$ & $fcc^{\ast }(SS)$ & $1/4$ \\ \hline
$fcc(C-S)$ & $2$ & $bcc^{\ast }(SS)$ & $1/2$ \\ \hline
$\mathbf{Z}^{3}(C-S)$ & $1$ & \textbf{Z}$^{3}(SS)$ & $1$ \\ \hline
$hcp(C-S)$ & $\sqrt{2}$ & $hcp^{\ast }(SS)$ & $1/\sqrt{2}$ \\ \hline
\end{tabular}

\bigskip

It should be clear from this table and the examples just presented that the
conventional solid state \textit{direct} lattice results are \textit{dual}
to that given in C-S monograph [\textbf{52}]. To make sense out of our
numerical results, we use the C-S results in our calculations compatible
with C-S data for $a(n)^{\prime }s$. Using the numerical procedure outlined
above the following results for the free energies are summarized in Table 3
below.

\ \ \ \ \ \ \ \ \ \ \ \ \ \ \ \ \ \ \ \ \ \ \ \ \ \ \ \ \ \ \ \ \ \ \ \ \ \ 

\ \ \ \ \ \ \ \ \ \ \ \ \ \ \ \ \ \ \ \ \ \ \ \ \ \ \ \ \ \ \ \ \ \ \ \ \ \
\ \ \ \ \ \ Table 3

\ \ \ \ \ \ \ \ \ \ \ \ \ \ \ \ \ \ \ 

\ \ \ \ \ \ \ \ \ \ \ \ \ \ \ \ \ \ \ \ 
\begin{tabular}{|l|l|l|}
\hline
$Lattice\Lambda ^{\ast }$ & $Free$ $energy$ & $Packing$ $fraction$ \\ \hline
$hcp$ & $-6.669431$ & $0.74$ \\ \hline
$fcc$ & $-6.65616$ & $0.74$ \\ \hline
$bcc$ & $-8.68137$ & $0.68$ \\ \hline
$\mathbf{Z}^{3}$ & $-9.27008$ & $0.52$ \\ \hline
\end{tabular}

\bigskip

These results are in qualitative agreement with those obtained earlier in
two dimensions where we observed that the respective free energies are
arranged in accordance with their respective packing fractions. The accuracy
of our calculations is supported by the fact that we have obtained
numerically practically the same free energies for the hcp and fcc lattices
whose packing fractions are the same (e.g. see Appendix C). As in two
dimensions, the earlier made assumption that the free energies are
proportional to their respective packing fractions happens to be wrong. The
obtained results provide an affirmative answer to the 1st Problem formulated
in Section 1.4.

\ The equality of free energies between the hcp and fcc lattices has been
established in recent computer simulation methods [\textbf{86}] and
apparently is of great practical interest. Such an equality between the free
energies might lead to the phase transition between these lattices and,
indeed, they were observed and theoretically discussed [\textbf{73}].

\ The equality between the hcp and fcc free energies has been achieved with
account of the following observation. On page 114 of C-S book, Ref.[\textbf{%
52}] it is said that: \textquotedblright The hcp is not itself a lattice,
but may be defined as a union of a lattice L....and the
translate...\textquotedblright . Such an opinion is not shared by solid
state physicists for whom hcp is a legitimate lattice [\textbf{8,13,83}]. If
one compares between C-S and SS definitions, then one should multiply $\sqrt{%
2}$ in Table 2 by a factor of 2. After such multiplication the free energies
of both hcp and fcc lattices become the same. The data of Table 3 reflect
this observation and are in accord with experimental results depicted in
Fig.1, which we would like to discuss now.

\subsubsection{\textbf{6.3. A brief walk across the CuZn phase diagram}.%
\textit{\ } \ }

As is mentioned in the Introduction, the CuZn\ phase diagram is a typical
equilibrium phase diagram for a binary alloy. Many such alloys make
superlattices under appropriate concentration/temperature conditions[\textbf{%
8,87}]. Apparently, this fact was a motivation for Lifshitz\ original work,
Ref.[\textbf{4}]. Taking this into account, we would like to complete this
section by connecting the phase diagram depicted in Fig.1 with the results
summarized in Table 3.

To do so, we would like to remind our readers about how such phase diagrams
are obtained in real life and what they are actually supposed to convey.
Initially, a typical phase diagram, like that depicted in Fig.1, is
determined experimentally by preparing an array\ of samples of varying
composition at room temperature and then heating each to temperatures above
that required for complete homogeneous melting. Upon cooling, detection of a
first order transition is possible by simple calorimetric methods. Much more
sophisticated methods (including calorimetric, X-ray diffraction, etc.) must
be used for detection of the second order transitions. Once the phase
diagram is complete, the fraction of each component in a given phase can be
extracted directly from the diagram at a specified temperature $T$\ and
composition $C_{0}.$

For the sake of illustration, we begin with the simplest typical case of a
two component (A and B) system existing in two phases, say $\alpha $ and $%
\beta $, as depicted in Fig.4.\ If such a system is kept under constant
pressure, depending on temperature, we expect it to be either in the $\alpha 
$ or the $\beta $ phases or in both.\ It is of interest to know the fraction
of a given phase in the overall system, but it is not always possible to
measure this directly. Normally, only the overall composition $C_{0}$ is
known. Although for a single-phase region on the phase diagram the
composition is $C_{0}$,\ for a two-phase region, the composition of each
phase is obtained by first placing a horizontal line $l$\ through the point (%
$C_{0}$,$T$) on the diagram. At each end, this horizontal line intersects
the boundary separating the two phase region from a single phase region. The
two-phase region is always bounded by two single-phase regions. Any larger
diagram, like that depicted in Fig.1, is assembled of alternating fragments
of this kind. The composition of a given phase is then obtained as an
abscissa of the above mentioned intersection point. Thus, the compositions
of the $\alpha $ and $\beta $\ phases are $C_{a}$ and $C_{\beta }$,
respectively as depicted in Fig.4.

%\FRAME{ftbpFU}{3.1816in}{2.7319in}{0pt}{\Qcb{The lever rule construction}}{}{%
%figure4recent45.jpg}{\special{language "Scientific Word";type
%"GRAPHIC";maintain-aspect-ratio TRUE;display "USEDEF";valid_file "F";width
%3.1816in;height 2.7319in;depth 0pt;original-width 3.5829in;original-height
%3.0727in;cropleft "0";croptop "1";cropright "1";cropbottom "0";filename
%'figure4recent45.JPG';file-properties "NPEU";}}

\begin{figure}[tbp]
\begin{center}
\includegraphics[width=3.85 in]{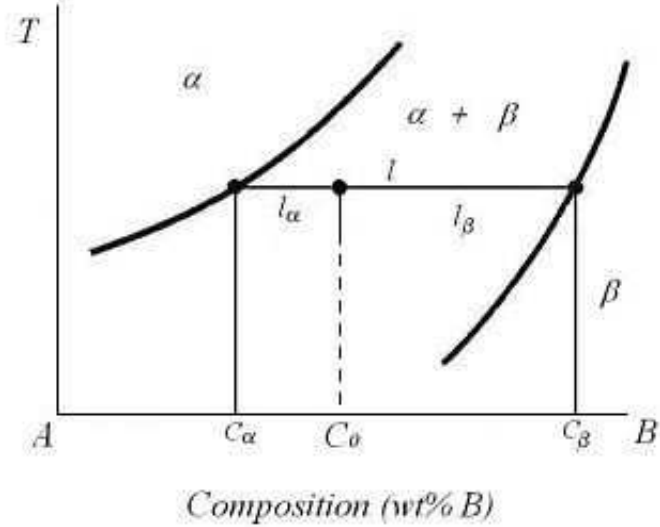}
\end{center}
\caption{The lever rule construction}
\end{figure}

We would like to connect this information with the Landau order parameter $%
\varphi $. To this purpose, we introduce the following notation. Let $%
l_{\beta }=\left\vert C_{\beta }-C_{0}\right\vert $ and $l_{\alpha
}=\left\vert C_{0}-C_{\alpha }\right\vert $ so that\ $l_{\beta }+l_{\alpha
}=\left\vert C_{\beta }-C_{\alpha }\right\vert \equiv l.$ With this
notation, we let the total fraction $w_{\alpha }$ of the $\alpha $ phase be $%
w_{\alpha }=\frac{l_{\beta }}{l}$ and accordingly, $w_{\beta }=\frac{%
l_{\alpha }}{l}$.\bigskip\ \ By construction, $w_{\alpha }+w_{\beta }=1$ so
that the Landau order parameter $\varphi $ can now be defined as $\varphi
=\left\vert w_{\alpha }-w_{\beta }\right\vert $. To check that such a
definition makes sense, it is sufficient to notice that since at criticality 
$C_{\beta }=C_{\alpha }=C_{0},$ we obtain $w_{\alpha }=w_{\beta }=1/2$ in
accord with Section 1. Such a picture assumes existence of a well defined
single critical point as discussed in Ref.[\textbf{6}], page 257. On the
next page of this reference one finds the following statement:
\textquotedblright Strictly speaking, there can be said to be two phases
only when they exist simultaneously and in contact, that is at points lying
on the equilibrium curve. It is clear that the critical point can exist only
for phases such that the difference between them is purely quantitative, for
example a liquid and a gas ...\textquotedblright\ Applying this statement to
Fig.4, one can say that two phases (co)exist as long as $l\neq 0$ and the
critical point corresponds exactly to the case when $l=0$. Next, on the same
page we read \textquotedblright ...such phases as liquid and solid
(crystal), or different crystal modifications of substance, are
qualitatively different, since they have different internal symmetry. It is
clear that we can say only that a particular symmetry...element exists or
does not exist; it can appear or disappear only as a whole, not gradually.
In each state the body will have one symmetry or the other, and so we can
always say to which of the two phases it belongs. Therefore \textit{the
critical point cannot exist for such phases, and the equilibrium curve must} 
\textit{either go to infinity or terminate by intersecting the equilibrium
curves of other phases.\textquotedblright\ }

\bigskip

The background information\ we have just supplied is sufficient for
understanding \ of the CuZn phase diagram depicted in Fig.1. We begin with
the region where Zn concentrations are less than 31.9\% and temperatures $%
T>900^{0}C$ . In this domain the horizontal (temperature) line intersects
two coexistence curves so that locally the picture looks exactly the same as
in Fig.4. Under such conditions we have a phase coexistence between the CuZn
liquid and the Cu Zn solid alloy, whose crystal structure $\alpha $ is that
for Cu, i.e. fcc, Fig. 2a). By lowering the temperature below $900^{0}C$ and
increasing the concentration of Zn we observe the phase coexistence between
two solid phases $\alpha $ and $\beta $ where, according to the Table 1, the 
$\beta $ phase is bcc, Fig.3a). This fact is in accord with results of Table
3 indicating that the free energy of the bcc lattice is \textit{lower} than
that for fcc so that such a structure can exist only at lower temperatures
in accord with empirical observations discussed in the Introduction. Next,
by going to still lower temperatures we obtain the $\beta ^{\prime }$ phase
in coexistence with the $\alpha $ phase. The $\beta ^{\prime }$\ phase is
made of interpenetrating cubic lattices as depicted in Fig.3. Such
observation, again, is in accord with Table 3 (which tells us that the cubic
lattice has still lower free energy than the bcc lattice). The dashed line
in this range of concentrations and temperatures represents \textit{the line}
of the \textit{second} order phase transitions in accord with quotations
from Landau and Lifshitz book, Ref.[\textbf{6}], stated above\footnote{%
The lever rule depicted in Fig.4 can be extended to cover this case by
imagining the dashed line in Fig.1 opening up a little bit initially, thus
forming something like an eye, and then, finally closing up. In this case we
always would get $C_{\beta }=C_{\alpha }=C_{0}$ and $w_{\alpha }=w_{\beta
}=1/2$ in accord with Landau and Lifshitz [\textbf{6}],Chr.14. Under such
conditions the order parameter $\varphi $ is either 0 or 1 in accord with
above cited quotations from Ref.[\textbf{6}]. This causes no problems,
however, in view of Eq.(1.3).}. Exactly the same arguments are applicable to
the next, $\beta +L,\beta -\beta ^{\prime }$ and ($\beta +\gamma )-(\beta
^{\prime }+\gamma ^{\prime })$ portions of the phase diagram so that the
dashed line (extending to concentrations just below 60 \% \ according to
Ref.[\textbf{6}]) is still a line of second order phase transitions where
the Lifshitz theory discussed in the Introduction is applicable. Notice that
if for such concentrations of Zn we raise the temperature, then we would
reach the curved $\beta $ triangle which looks exactly like the curved $%
\alpha $ triangle but is located strictly below in the temperature range.
This is again in complete accord with results of Table 3. Next, notice that
the $\gamma $ phase is made essentially of the body-centered cubic-type
lattice and in this range of concentrations and temperatures \textit{above} $%
700^{0}C$ the situation is analogous to that for the earlier discussed $%
\alpha $ phase. Below $700^{0}C$ \ and for concentrations above 68\% the
diagram exhibits apparent complications as depicted in Fig.5.

%\FRAME{ftbpFU}{4.3223in}{2.9092in}{0pt}{\Qcb{Fragment of Fig.1 exhibiting
%the most complicated phase behavior}}{}{figure55.jpg}{\special{language
%"Scientific Word";type "GRAPHIC";maintain-aspect-ratio TRUE;display
%"USEDEF";valid_file "F";width 4.3223in;height 2.9092in;depth
%0pt;original-width 7.1563in;original-height 4.8023in;cropleft "0";croptop
%"1";cropright "1";cropbottom "0";filename 'figure55.JPG';file-properties
% "NPEU";}}

\begin{figure}[tbp]
\begin{center}
\includegraphics[width=4.60 in]{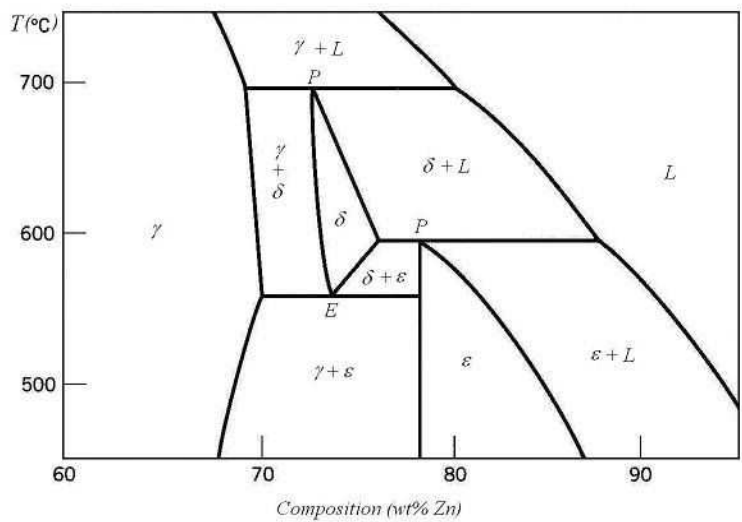}
\end{center}
\caption{Fragment of Fig.1 exibiting the most complicated phase behavior}
\end{figure}

\medskip

Although description of these complications lie beyond the scope of Lifshitz
theory, we would like to discuss them briefly for the sake of completeness
and in view of likely developments in the future.

Since the rest of the phase diagram, Fig.1 (in the range of concentrations
above 80\% and temperatures below $600^{0}C)$, is simpler, we would like to
discuss it first. Provided that we make formal substitutions of the type: $%
\alpha \rightleftharpoons \varepsilon $ and $\beta \rightleftharpoons \eta $%
, this portion of the diagram looks \textit{exactly the same} as the
previously discussed portion including $\alpha +L$, $\alpha ,$ $\alpha
+\beta ,\beta +L$ and $\beta $ phases. This fact should not come as a total
surprise in view of the fact that Cu and Zn have different types of
lattices. According to the Table 1, they are of fcc and hcp type
respectively. Nevertheless, the phase diagram does contain a surprise. It
comes from the apparent contradiction between the results of our Table 3
requiring the free energies of fcc and hcp type crystals to be the same and
obvious asymmetry of the CuZn phase diagram with respect to temperature
axis. Earlier, we noticed that, indeed, the results of computer simulations
support the analytical results of our Table 3. To resolve the apparent
paradox we have to take into account several additional facts. First, we
have to take into account that computer simulations as well as our
calculations summarized in Table 3 have been made with account of the fact
that the packing fractions (Appendix C) of both hcp and fcc lattices are the
same. This is obviously correct for lattices whose vertices contain \textit{%
identical} atoms modelled by the hard spheres and arranged in such a way
that these spheres touch each other. In reality, however, metals are made of
atoms whose packing behavior may or may not be well modelled by the hard
spheres. In particular, the literature data on pure metals, Ref. [\textbf{88}%
], page 260, indicate that only Li, Na and Sr may form standard (Fig.2, $%
c/a\sim 1.633$) hcp lattices. Even for these metals such lattices are not
the most common: for Li and Na the most common form is bcc while for Sr is
fcc. In our case Cu has a stable fcc lattice while Zn has the \textit{%
elongated} (Fig.2, $c/a\sim 1.86)$ fcc lattice. The fact that Li and Na
exist in two modifications: bcc and hcp and that, experimentally, the bcc
lattice is more stable\ than hcp, can be used for explanation of the portion
of the CuZn phase diagram where the transition from $\gamma +L$ to $\gamma
+\delta $ phase is depicted since $\gamma $ is of bcc-type lattice while $%
\delta $ is of hcp-type. This fact is in apparent contradiction with the
order of appearance of phases of lower symmetry with decreasing temperatures
as discussed in the Introduction. Such an order is characteristic only for
the phase transitions of the \textit{second} \textit{order} however (in
accord with Ref.[\textbf{6}]) while a horizontal line between the $\gamma +L$
and $\gamma +\delta $ phases is characteristic for transitions of the 
\textit{first order}\ as demonstrated experimentally in [\textbf{7}].
Moreover, in comparing our results against experimental data one should keep
in mind that the results of Table 3 are for the combination $\beta \mathcal{F%
}$ while thermodynamically we have to look at $\mathcal{F}$ at different
temperatures. When it comes to hard spheres, it is clear that the equality $%
\beta _{1}\mathcal{F}^{fcc}=\beta _{2}\mathcal{F}^{hcp}$ holds if $\beta
_{1}=\beta _{2\text{ }}$ implying $\mathcal{F}^{fcc}=\mathcal{F}^{hcp}$ in
accord with computer simulations [\textbf{86}]. However, for metals\ such as
Cu and Zn, in view of what was said about their crystal structures, it is
sufficient to require $\beta _{1}\mathcal{F}^{fcc}(Cu)=\beta _{2}\mathcal{F}%
^{hcp}(Zn)=\mathcal{N}$ with $\mathcal{N}$ being some constant. This leads
us to the result: 
\begin{equation*}
\mathcal{F}^{fcc}(Cu)=\frac{T_{1}}{T_{2}}\mathcal{F}^{hcp}(Zn).
\end{equation*}%
Since the data presented in Fig.1 imply that $T_{1}/T_{2}>1$ this implies
that $\mathcal{F}^{fcc}(Cu)>\mathcal{F}^{hcp}(Zn)$ . This fact \ formally
explains the source of the apparent temperature asymmetry of the phase
diagram, Fig.1. Clearly, such an explanation is purely formal and should be
replaced by quantum mechanical calculations. This task, however, is beyond
the scope of this work.

In looking at the phase diagram, Fig.1 one should keep in mind \ how such
diagram was obtained experimentally (we mentioned this already). That is,
experimentally one cannot move \textit{horizontally} across the diagram.
This means that the respective free energies also should be compared with
each other only vertically, i.e. temperature changes for the fixed
composition.

Now we are ready to provide our final comments regarding Fig.5, i.e. we
would like to discuss qualitatively the region of concentrations above 60\%
but below 80\% \ and temperatures below $700^{0}C.$ This region contains 3
triple points: two of them, P, at temperature slightly below 700$^{0}$C, and
the third at temperature slightly below 600$^{0}$C. \ These tripple points
are called \textit{peritectics. }They are interesting because in both cases
we have situation when either two phases are transformed into one ($\gamma
+L\rightleftharpoons \delta $ and $\delta +L\rightleftharpoons \varepsilon )$
or, as in the case of point E (called the \textit{eutectic}), one phase $%
\delta $ is transformed into two phases $\gamma $ and $\varepsilon $.
Although the Landau theory can be extended, in principle, to describe
(locally) situations depicted in Fig.5, attempts to describe the entire
phase diagram, Fig.1, quantitatively using the same model \ [\textbf{12}]
thus far have not produced the desired results to our knowledge.

\bigskip

\paragraph{\textbf{Appendix A. Some facts about lattices, especially 3
dimensional.}}

\bigskip\ \ 

\textit{A lattice} $\Lambda $ is a free abelian group given by 
\begin{equation}
\Lambda =\mathbf{Z}v_{1}+...+\mathbf{Z}v_{n}=%
\{a_{1}v_{1}+...+a_{n}v_{n}:a_{i}\in \mathbf{Z},i=1,...,n\}.  \tag{A.1}
\end{equation}%
where $v_{1},...,v_{n}$ is a basis in Euclidean space $\mathbf{E}^{n}$.
Hence each lattice is the universal covering space for some $n-$dimensional
torus $T^{n}.$ The fundamental \textit{parallelotope} $P(\Lambda )$ of the
lattice $\Lambda $ is given by 
\begin{equation}
P(\Lambda )=\{\lambda _{1}v_{1}+...+\lambda _{n}v_{n}:0\leq \lambda _{i}\leq
1,i=1,...,n\}.  \tag{A.2}
\end{equation}%
In solid state physics the fundamental parallelotope is known as \textit{%
primitive cell} [\textbf{13}]. If the crystal lattice is composed of atoms
of several different types, it is required that the atoms of a given type
under basic translations are sent to atoms of the same type. Hence a lattice
as a whole is a \textit{union} of primitive lattices (e.g. see Fig.2) so
that the symmetry of the crystal lattice as a whole \textit{does not} in
general coincide with that for a specific primitive lattice. To distinguish
between primitive lattices the concept of a volume for $P(\Lambda )$ is
useful. It is defined by 
\begin{equation}
volP(\Lambda )=\left\vert \det (v_{1},...,v_{n})\right\vert .  \tag{A.3}
\end{equation}%
If the vector basis is changed by the unimodular transformation, $%
volP(\Lambda )$ remains unchanged$.$ In solid state physics such equivalent
lattices (primitive cells) are called \textit{Bravis} lattices. Since not
all Bravis lattices are connected by the unimodular transformation, in 3
dimensions there are 7 different crystal systems and 14 Bravis lattices
associated with them. These lattices differ from each other by the length
ratios between their edges and by the angles between these edges. The Bravis
lattice of \textit{any} kind has symbol \textbf{P} (for primitive) in solid
state physics. The body-centered and the face-centered lattices are \textit{%
not} primitive and are denoted by \textbf{I} and \textbf{F} respectively.
Let $G_{n}(\Lambda )$ denote a (\textit{crystallographic}) group of
isometries of \textbf{E}$^{n}$ which maps $\Lambda $ to itself. Fedorov,
Schoenflies and Barlow [\textbf{89}] have independently established that for 
$n=3$ there are 230 such groups of isometries. Bieberbach has looked at the
same problem in $n$-dimensions [\textbf{90}]. The 3rd column of Table 1
provides information about the particular catalog number (\#) for the space
group $G_{3}(\Lambda )$. Every group $G_{3}(\Lambda )$ is a composition of
some orthogonal transformation $\alpha $ and translation $T$, i.e. $\forall
g\in G_{3}(\Lambda )$. We should have $g=T\circ \alpha $. The orthogonal
transformation is called \textit{rotation} if $\det \alpha =1$ and $\mathit{%
rotatory}$ $\mathit{reflection}$ if $\det \alpha =-1.$ In general $T\circ
\alpha \neq \alpha \circ T.$ Clearly, $\alpha $ belongs to a point group $%
H_{3}(\Lambda )$ in the sense that for every Bravis lattice $\Lambda _{B}$
it is a stabilizer, i.e. $\alpha \Lambda _{B}=\Lambda _{B}.$ As a group, $%
H_{3}(\Lambda )$ is finite and each of its finitely many elements \ is a
rotation about some axis going trough the (arbitrarily chosen, e.g.
coinciding with one of the lattice sites) origin $O$ by the angle which is
an integer multiple of either $\pi /3$ or $\pi /2\footnote{%
This can be easily understood using two dimensional plane \textbf{R}$^{2}$
as an example. In Section 5 we have mentioned that in case of Euclidean
geometry such plane can be tesselated either by squares or by triangles. The
three dimensional space can then be viewed as $\mathbf{R}^{2}\times \mathbf{%
R.}$}$. Let $C_{n}$ be a rotation which corresponds to the angle $2\pi /n,$
then a rotatory reflection is given by $S_{n}=\sigma \circ C_{n\text{ }}$,
i.e. it is a superposition of a rotation followed by reflection $\sigma $
(rotation by the $\pi $ angle) in the plane perpendicular to the axis of
rotation. In particular, the \textit{inversion} $i=\sigma \circ C_{2},$
while the \textit{reflection itself} is given by $\sigma \circ E=E,$ where $%
E $ is the unit element of the cyclic group $C$. The simplest group after
the cyclic group $C_{n}$ is the group $D_{n}$ consisting of all rotations
transforming regular $n-$sided prism into itself. This group has one $n-$th
order $C_{n}$ axis and $n$ second order axes perpendicular to it thus
containing altogether 2n elements [\textbf{9}]. Next, we need the group $%
\hat{T}$ which consists of all rotations transforming a tetrahedron to
itself. Analogously, the group $\hat{O}$ (the octahedral group) contains all
rotations leaving the cube invariant. To bring all these facts to the level
at which the symbolics of Table 1 can be explained, we need yet a couple
more complications. First, we introduce a group $\hat{C}_{nv}.$ It is a
symmetry group of a regular $n-$gonal pyramid containing one $n-$th order
axis $C_{n}$ coinciding with the height of the pyramid and $n$ vertical
reflection planes that pass through $C_{n}$ axis. Reflections in these
planes are elements of $\hat{C}_{nv}.$ It can be shown [\textbf{9}] that
this group is isomorphic to $\hat{D}_{n}.$ Second, the group $\hat{D}_{nh%
\text{ }}$is the symmetry group of a regular $n-$gonal prism differs from
earlier introduced $\hat{D}_{n}$ by the presence of additional $n$ vertical
reflection planes (just like $\hat{C}_{nv})$ causing extra reflections in
these planes. Third, the group $\hat{T}_{h}$ is related to $\hat{T}$ as $%
S_{n}$ is related to $C_{n}.$ Lastly, the group $\hat{O}_{n}$ is related to $%
\hat{O}$ in the same way as $\hat{T}_{h}$ to $\hat{T}$ .

Since each of 230\ lattice space groups is made the same way as some
composition of the point group symmetry followed \ by\ translational
symmetry, one can arrange these groups either in ascending or descending
level of symmetry. In particular, to explain the data in Table 1 we need
only lattices with point groups associated with cubic and hexagonal
symmetries. The needed information can be arranged for instance in the
following way

\bigskip

\begin{tabular}{|l|l|l|}
\hline
The Bravis lattice & The point symmetry & The crystallographic symbol \\ 
\hline
Cubic ( \textbf{P}, \textbf{I} ) & $O_{h},T_{d},O,\hat{T}_{h},\hat{T}$ & 
m3m, \={4}3m, 432, m3, 23 \\ \hline
Hexagonal (\textbf{P }) & $D_{6h},D_{3h},C_{6v},D_{6},C_{6h},C_{3h},C_{6}$ & 
6/mmm, \={6}m2, 6mm, 622, 6/m, \={6}, 6 \\ \hline
\end{tabular}

\bigskip

Here the elements of point group symmetry are arranged from left to right in
descending order (with the leftmost being the most symmetric). The
crystallographic nomenclature symbols follow the ordering of point symmetry
groups.

All that was said about the \textit{direct} lattice can be said also about
its \textit{dual.} \ To this purpose let $x=(x_{1},...,x_{n})$ and $%
y=(y_{1},...,y_{n})$ be two vectors of $\mathbf{E}^{n}$such that their
scalar product $x\cdot y=x_{1}y_{1}+...+x_{n}y_{n}.$Then, the \textit{dual}
(or \textit{reciprocal}) lattice $\Lambda ^{\ast }$ in $\mathbf{E}^{n}$ is
defined by 
\begin{equation}
\Lambda ^{\ast }=\{x\in \mathbf{E}^{n}:x\cdot y\in \mathbf{Z}\text{ }\forall
y\in \Lambda \}.  \tag{A.4}
\end{equation}%
The \textit{dual basis} $v_{1}^{\ast },...,v_{n}^{\ast }$ is defined in such
way that $v_{i}^{\ast }\cdot v_{j}=\delta _{ij}$ so that the dual lattice
can also be presented as 
\begin{equation}
\Lambda ^{\ast }=\mathbf{Z}v_{1}^{\ast }+...+\mathbf{Z}v_{n}^{\ast } 
\tag{A.5}
\end{equation}%
In solid state physics [\textbf{13}] instead of the requirement $x\cdot y\in 
\mathbf{Z}$ \ in Eq.(A4) the alternative requirement $x\cdot y=2\pi \mathbf{Z%
}$ is used. It is motivated by the Fourier analysis and the Bloch theorem
[11]. It can be easily demonstrated that $vol(\Lambda )\cdot vol(\Lambda
^{\ast })=1$ while in the solid state physics one gets accordingly: $%
vol(\Lambda )\cdot vol(\Lambda ^{\ast })=\left( 2\pi \right) ^{n}.$ A
lattice $\Lambda $ is called \textit{integral} if $\Lambda \subseteq \Lambda
^{\ast }$ . This means that $\forall x,y\in \Lambda $ one has $x\cdot y\in 
\mathbf{Z}$ . If $v_{1},...,v_{n}$ is the basis of such integral lattice,
then the matrix $A_{ij}=v_{i}\cdot v_{j}$ is an integral matrix. In this
case the \textit{discriminant} of lattice $\Lambda $ is defined by 
\begin{equation}
disc(\Lambda )=\det (A).  \tag{A.6}
\end{equation}%
But $\left[ vol(\Lambda )\right] ^{2}=\det A$ and, therefore, $\left[
vol(\Lambda )\right] ^{2}=disc(\Lambda )=\dfrac{vol(\Lambda )}{vol(\Lambda
^{\ast })}.$ Again this result should be amended in the case of solid state
physics where we have instead $disc(\Lambda )=\left( 2\pi \right) ^{n}\dfrac{%
vol(\Lambda )}{vol(\Lambda ^{\ast })}.$ From the definition of the matrix $%
A_{ij}$ it follows that each vector $v_{i}$ has its own decomposition : $%
v_{i}=(v_{i1},...,v_{in}).$ Accordingly, for the integral lattice any vector 
$x$ of such a lattice is made of matrix products of the type: $x=$ $\xi M$ $%
\ $with $\xi $ being a vector with integer coefficients: $\xi =(\xi
_{1},...,\xi _{n})$ and $M$ being a matrix: $M=M_{ij}=\{v_{ij}\}.$ Let $%
M^{T} $ be a transposed matrix. Then, the earlier defined matrix $A$ (called
the Gram matrix [\textbf{52}]) is made of $M$ and $M^{T}$ according to the
rule: 
\begin{equation}
A=MM^{T}  \tag{A.7}
\end{equation}%
Based on this definition, the \textit{norm} $N(x)$ of the vector $x$ can be
defined as: 
\begin{equation}
N_{A}(x)=x\cdot x=\xi A\xi .  \tag{A.8}
\end{equation}

\bigskip

\paragraph{\textbf{Appendix B. Number-theoretic aspects of lattices. }}

\textbf{\bigskip\ }

Based on the information presented in the Appendix A,\ we consider
generalization of CM, discussed in Section 5, to multidimensional lattices.
To this purpose consider the following eigenvalue equation 
\begin{eqnarray}
xv_{1} &=&a_{11}v_{1}+...+a_{1n}v_{n}  \TCItag{B.1} \\
xv_{2} &=&a_{21}v_{1}+...+a_{2n}v_{n}  \notag \\
&&......................  \notag \\
xv_{n} &=&a_{n1}v_{1}+...+a_{nn}v_{n}  \notag
\end{eqnarray}%
where the matrix elements $a_{ij}\in \mathbf{Z.}$ This equation can be
looked upon from different directions. For instance, if we consider it as an
eigenvalue equation, i.e. 
\begin{equation}
\det (x\mathbf{I}-\mathbf{A})=0  \tag{B.2}
\end{equation}%
then, written explicitly, we obtain an $n$-th order algebraic equation 
\begin{equation}
x^{n}+a_{n-1}x^{n-1}+...+a_{0}=0  \tag{B.3}
\end{equation}%
with integer coefficients. By definition, $n$ solutions $\{x_{i}\}$ of such
equation are \textit{algebraic integers }\textbf{[91,92]}. For instance, $%
\pm 1$, $\pm i$ ( the Gaussian integers discussed in Section 5) are
solutions of the equation $x^{4}=1$ while $j=\frac{1}{2}(-1+\sqrt{-3})$
(also discussed in Section 5) is one of the solutions of the equation $%
x^{3}=1.$ In both cases solutions are roots of unity $\zeta (n)_{k}=\exp
(\pm ik\frac{2\pi }{n})$, $k=0,...,n-1($for $n=4$ and $3$ respectively).
Solutions $\zeta (n)_{k}$ by construction are independent of each other and
can be considered as basis vectors (instead of $v_{1},...,v_{n})$ in some
vector space. This analogy can be made more precise. Let $x_{i}$ be one of
the solutions of Eq.(B.3). Substitute it into Eq.(B.3) and rewrite the
result as 
\begin{equation}
-a_{0}=x_{i}a_{1}+...+x_{i}^{n}a_{n}  \tag{B.4}
\end{equation}%
(perhaps, with $a_{n}=1).$ Then, the analogy with Eq.(A.1) becomes complete
if we identify $(x_{i},...,x_{i}^{n})$ with ($v_{i1},...,v_{in}).$ From this
analogy, the norm, the trace and the discriminant can be defined as well.
There are some differences though connected \ with the Galois group
symmetry. Since these topics are not immediately connected with results of
the main text, the interested reader can look up these things in number
theory literature. What is important for us, however, is the fact that only
for quadratic number fields does the norm, Eq.(A.8), coincide with that
defined \ in number theory. For higher order fields this is no longer true [%
\textbf{92}]. Therefore, two dimensional results presented in Section 5 are
not immediately generalizable to higher dimensions.

Fortunately, there is an alternative approach bypassing the difficulty just
described, which we would like to describe now. We begin with defining the%
\textbf{\ }Epstein zeta function $Z(\mathbf{A},s)$ is defined as 
\begin{equation}
Z(\mathbf{A},s)=\frac{1}{2}\sum\limits_{\xi \in \mathbf{Z}^{n}\backslash 0}%
\left[ N_{A}(x)\right] ^{-s},  \tag{B.5}
\end{equation}%
where $N_{A}(x)$ is defined by Eq.(A.8).

Using the known identity 
\begin{equation}
x^{-s}\Gamma (s,x)=\int\limits_{0}^{\infty }dyy^{s-1}\exp (-xy)  \tag{B.6}
\end{equation}%
we can rewrite Eq.(B.5) as 
\begin{equation}
\Lambda (\mathbf{A},s)=\frac{1}{2}\int\limits_{0}^{\infty }dyy^{s-1}(\mathbf{%
\Theta }(A;y)-1)  \tag{B.7}
\end{equation}%
where $\Lambda (\mathbf{A},s)=\pi ^{-s}\Gamma (s)Z(\mathbf{A},s).$ The theta
function 
\begin{equation}
\mathbf{\Theta }(A;y)=\sum\limits_{\xi \in \mathbf{Z}^{n}}\exp (-\pi
N_{A}(x)y)  \tag{B.8}
\end{equation}%
can be represented alternatively as follows 
\begin{equation}
\mathbf{\Theta }(A;y)=\sum\limits_{n=0}^{\infty }a(n)q^{n},  \tag{B.9}
\end{equation}%
where $q=\exp (-\pi ny)$ and $a(n)$ is the number of integer solutions of
the equation $N_{A}(x)=n.$ These results are used in Section 6.

\bigskip

\paragraph{\textbf{Appendix C. Exterema of the positive definite quadratic
forms and the problem of closest packing of spheres.\ }}

\bigskip\ \ 

Although methods discussed in Section 5 are capable of providing
(indirectly) the information about the best (closest) possible packing of
hard discs, they are not immediately generalizable to higher dimensions.
Because of this, we sketch here an alternative approach. To this purpose,
let us consider a positive definite quadratic form 
\begin{equation}
f(\xi _{1},\xi _{2})=a_{11}\xi _{1}^{2}+2a_{12}\xi _{1}\xi _{2}+a_{22}\xi
_{2}^{2}\equiv \xi A\xi  \tag{C.1}
\end{equation}%
whose determinant $D=a_{11}a_{22}-a_{12}^{2}$ . Consider now a subset $%
\mathcal{R}$ of $\xi -$plane such that $f(\xi _{1},\xi _{2})\leq K$ with $K$
being some positive number. Following Cassels, Ref.[\textbf{93}], we would
like to demonstrate that, provided that, $K\geq (4D/3)^{\frac{1}{2}},$ the
subset $\mathcal{R}$ contains some point $(\xi _{1},\xi _{2})$, other than
the origin $(0,0)$, with integer coordinates\footnote{%
In d-dimensions the proof of this fact is known as the Minkowski theorem. \
It plays a major role in geometric number theory and the theory of polytopes
and oriented matroids.}. This result can be understood based on the
following chain of arguments. First, let us assume that $\inf
f(u_{1},u_{2})=a_{11\text{ }}$with $u_{1},u_{2}$ being integers not both
equal to $0$. Clearly, every quadratic form can be rewritten as 
\begin{equation}
f(\xi _{1},\xi _{2})=a_{11}(\xi _{1}+\frac{a_{12}}{a_{11}}\xi _{2})^{2}+%
\frac{D}{a_{11}}\xi _{2}^{2},  \tag{C.2}
\end{equation}%
so that by construction, $f(u_{1},u_{2})\geq a_{11}$. Second, without loss
of generality, we can demand that $f(u_{1},1)\geq a_{11}.$ If, in addition,
we assume that $\left\vert u_{1}+\frac{a_{12}}{a_{11}}\right\vert \leq \frac{%
1}{2}$ then,we obtain the following chain of inequalities 
\begin{equation}
a_{11}\leq f(u_{1},1)\leq \frac{a_{11}}{4}+\frac{D}{a_{11}}  \tag{C.3}
\end{equation}%
leading to 
\begin{equation}
a_{11}^{2}\leq \frac{4D}{3}  \tag{C.4}
\end{equation}%
and, finally, to 
\begin{equation}
\inf f(\xi _{1},\xi _{2})\leq \sqrt{\frac{4D}{3}}.  \tag{C.5}
\end{equation}%
Consider a special case of the quadratic form $f(\xi _{1},\xi _{2})=m\left(
\xi _{1}^{2}+\xi _{1}\xi _{2}+\xi _{2}^{2}\right) .$ It has determinant $D=%
\frac{3m^{2}}{4}$ . So that inequality (C.5) is reduced to the identity $%
:m=m.$ For any other quadratic form the sign $\leq $ in (C.5) should be
replaced by $<$ [\textbf{94}]. The quadratic form $\xi _{1}^{2}+\xi _{1}\xi
_{2}+\xi _{2}^{2}$ corresponds to the hexagonal lattice [\textbf{52,94}]
whose matrix $M$ is given by 
\begin{equation}
M=\left( 
\begin{array}{cc}
1 & 0 \\ 
\frac{1}{2} & \frac{1}{2}\sqrt{3}%
\end{array}%
\right)  \tag{C.6}
\end{equation}%
This matrix gives rise to the matrix $A=MM^{T}$ . Two of such matrices are
equivalent if they are connected by the unimodular transformation made of
matrix $U$ with integer entries whose determinant is one, i.e. $A^{\prime
}=UAU^{T}$. This observation permits us to use as well the matrix [\textbf{52%
}], page 43, 
\begin{equation}
M=\left( 
\begin{array}{cc}
1 & 0 \\ 
\frac{-1}{2} & \frac{1}{2}\sqrt{3}%
\end{array}%
\right)  \tag{C.7}
\end{equation}%
made of two basis vectors (ideals) as discussed in Section 5.

Consider the obtained results from a slightly different angle. Let (in 
\textit{any} dimension) 
\begin{equation}
a=\inf \xi A\xi  \tag{C.8}
\end{equation}%
for $\xi \in \mathbf{Z}^{n}\backslash 0$ . Let each lattice point be filled
with a sphere of radius $r$ and let the radius $r$ be such that spheres at
nearby points touch each other. Then, $2r\leq \sqrt{a}.$ In view of
Eq.(A.6), the volume of the fundamental parallelotope, $vol$($\Lambda )=$ $%
\sqrt{\det A}.$ If the volume $V_{n}$ of the $n$ dimensional sphere is given
by $V_{n}=\frac{\pi ^{\frac{n}{2}}}{\Gamma (1+\frac{n}{2})}r^{n}\equiv
\sigma _{n}r^{n},$ then the most optimal $packing$ $fraction$ $q_{n}$ can be
defined as 
\begin{equation}
\frac{V_{n}}{\sqrt{\det A}}=\frac{\sigma _{n}}{2^{n}}\frac{a^{\frac{n}{2}}}{%
\sqrt{\det A}}\leq \frac{\sigma _{n}}{2^{n}}\gamma _{n}^{\frac{n}{2}}\equiv
q_{n}  \tag{C.9}
\end{equation}%
where the constant $\gamma _{n}^{\frac{n}{2}}$ is defined by the following
inequality 
\begin{equation}
a\leq \gamma _{n}\left( \det A\right) ^{\frac{1}{n}}.  \tag{C.10}
\end{equation}%
According to Eq.(C.5) we obtained in two dimensions : $\gamma _{2}=\frac{2}{%
\sqrt{3}}.$ This produces at once $q_{2}=\frac{2}{\sqrt{3}}\frac{\pi }{2^{2}}%
=\frac{\pi }{2\sqrt{3}}=\frac{\pi }{\sqrt{12}}=0.9069$ in accord with Ref.[%
\textbf{52}], page 110. In the case of 3 dimensions, the analog of the
hexagonal two dimensional lattice is the face centered cubic lattice [%
\textbf{52}] (fcc) whose quadratic form is known to be 
\begin{equation}
f(\xi _{1},\xi _{2},\xi _{3})=\xi _{1}^{2}+\xi _{2}^{2}+\xi _{3}^{2}+\xi
_{1}\xi _{2}+\xi _{2}\xi _{3}+\xi _{3}\xi _{1}.  \tag{C.11}
\end{equation}%
Calculations similar to those leading to inequality (C.5) were made
originally by Gauss\ and can be found in Ref. [\textbf{61,93,94}]. They now
lead to the following inequality 
\begin{equation}
a\leq \left( 2\det A\right) ^{\frac{1}{3}}  \tag{C.12}
\end{equation}%
thus yielding $\gamma _{3}=2^{\frac{1}{3}}$ and $q_{3}=\frac{4\pi }{3}\frac{1%
}{2^{3}}\sqrt{2}=\frac{\pi }{3\sqrt{2}}=0.7405.$ This value is \textit{not}
unique, however, since the hexagonal closed packed (hcp) lattice also
possesses \textit{the same} value for $q_{3}.$

\bigskip

\textit{\ }

\paragraph{\textbf{Appendix D. Representative calculation of determinant
without zeta function regularization\ \ }}

\bigskip\ \ 

The quantum harmonic oscillator is a benchmark example of quantum mechanical
calculations in any course of quantum mechanics. Its eigenvalue spectrum is
known to be: $E_{n}=\hslash \omega (n+\frac{1}{2})$, where $\omega $ is the
classical frequency of the oscillator and $\hslash $ is Planck's constant.
Given this result, the partition function $Z(\beta )$ at temperature $\beta
^{-1}$is obtained in a standard way as 
\begin{equation}
Z(\beta )=\sum\limits_{n=0}^{\infty }\exp (-\beta E_{n})=\frac{\exp (-\frac{%
\beta \hslash \omega }{2})}{1-\exp (-\beta \hslash \omega )}.  \tag{D.1}
\end{equation}%
Using this result the free energy $\mathcal{F}$ is obtained as 
\begin{equation}
\beta \mathcal{F}=-\ln Z(\beta )=\beta \frac{\hslash \omega }{2}+\ln (1-\exp
(-\beta \hslash \omega )).  \tag{D.2}
\end{equation}%
We would like to reproduce this simple result now using Feynman's path
integral approach to quantum and statistical mechanics. To this purpose,
following Feynman [\textbf{95}], we write for the free energy the following
path integral ($\hslash =1)$ 
\begin{equation}
\exp (-\beta \mathcal{F)=}\int\limits_{q(0)=q(\beta )}D[q(\tau )]\exp (-%
\frac{1}{2}\int\limits_{0}^{\beta }d\tau \lbrack \dot{q}^{2}+\omega
^{2}q^{2}])  \tag{D.3}
\end{equation}%
where $\dot{q}=\frac{d}{d\tau }q(\tau ).$ Since for given boundary
conditions for $q(\tau )$ we can write 
\begin{equation}
\int\limits_{0}^{\beta }d\tau \dot{q}^{2}=-\int\limits_{0}^{\beta }d\tau q(%
\frac{d^{2}}{d\tau ^{2}})q  \tag{D.4}
\end{equation}%
this fact allows us to rewrite the path integral, Eq.(D.3), in the following
equivalent form 
\begin{equation}
\exp (-\beta \mathcal{F)=}\int\limits_{q(0)=q(1)}D[q(\tau )]\exp (-\frac{1}{2%
}\int\limits_{0}^{1}d\tau \int\limits_{0}^{1}d\tau ^{^{\prime }}qA(\tau
,\tau ^{\prime })q)  \tag{D.5}
\end{equation}%
where the operator $A(\tau ,\tau ^{\prime })=\left( -\frac{d^{2}}{d\tau ^{2}}%
+\omega ^{2}\beta ^{2}\right) \delta (\tau -\tau ^{\prime }).$ By expanding $%
q(\tau )$ in Fourier series with basis made of eigenfunctions of such
operator the calculation of the path integral is reduced to the calculation
of the Gaussian path integrals of the standard type 
\begin{equation}
\int\limits_{-\infty }^{\infty }dx\exp (-\frac{a_{n}}{2}x^{2})=\sqrt{\frac{%
2\pi }{a_{n}}}  \tag{D.6}
\end{equation}%
where $a_{n}=4\pi ^{2}n^{2}+\omega ^{2}\beta ^{2},n=0,\pm 1,\pm 2,...$ Thus,
we obtain 
\begin{equation}
\det A=\prod\limits_{-\infty }^{\infty }a_{n}  \tag{D.7}
\end{equation}%
and, therefore, the free energy is formally given by 
\begin{equation}
\beta \mathcal{F=}\frac{1}{2}\ln \det A+const  \tag{D.8}
\end{equation}%
where the actual value of $const$ is unimportant to us since it is
temperature- independent and can be dropped since the free energy $\mathcal{F%
}$ is always defined with respect to some reference. To calculate the free
energy we follow the method described in our earlier work [\textbf{82}]. To
this purpose, we let $\omega ^{2}\beta ^{2}=x^{2}$ and consider 
\begin{equation}
\frac{d}{dx^{2}}\ln \det A=\frac{1}{x^{2}}+2\sum\limits_{n=1}^{\infty }\frac{%
1}{4\pi ^{2}n^{2}+x^{2}}.  \tag{D.9}
\end{equation}%
Then, formally we obtain 
\begin{equation}
\ln \det A=\int\limits^{\omega ^{2}\beta ^{2}}dx^{2}\left[ \frac{1}{x^{2}}%
+2\sum\limits_{n=1}^{\infty }\frac{1}{4\pi ^{2}n^{2}+x^{2}}\right] , 
\tag{D.10}
\end{equation}%
where the lower limit of integration will be carefully chosen. By noticing
that 
\begin{equation}
\coth \pi x=\frac{1}{\pi x}+\frac{2x}{\pi }\sum\limits_{n=1}^{\infty }\frac{1%
}{n^{2}+x^{2}},  \tag{D.11}
\end{equation}%
we obtain 
\begin{equation}
\frac{d}{dx^{2}}\ln \det A=\frac{1}{2x}\coth \frac{x}{2}  \tag{D.12}
\end{equation}%
and, accordingly, 
\begin{equation}
\ln \frac{\det A(\omega \beta )}{\det A(0)}=\int\limits_{0}^{\omega \beta
}dx\coth \frac{x}{2}=\omega \beta +2\ln (1-\exp (-\beta \omega ))  \tag{D.13}
\end{equation}%
Finally, taking into account Eq.(D.8) we obtain ($\hslash =1)$ 
\begin{equation}
\beta \mathcal{F=}\frac{\omega \beta }{2}+\ln (1-\exp (-\beta \omega )) 
\tag{D.14}
\end{equation}%
in accord with Eq.(D.2).

\bigskip

\textbf{References}

\bigskip

\ [1] \ J. Maddox, Nature 355 (1998) 201.

\ [2] \ A.R. Oganov, C.W. Glass, J. Chem. Phys. 124, (2006) 244704

\ [3] \ L. D. Landau, I. Sov.Phys. JETP 7 (1937) 19.

\ [4] \ E. M.Lifshitz, Acad.Sci.USSR. J.Phys. 6 (1942) 251.

\ [5] \ V.L. Ginzburg, L.D.Landau, Sov.Phys.JETP 20 (1950) 1064.

\ [6] \ L.D. Landau, E.M. Lifshitz, Statistical Physics.Part 1. Course in

\ \ \ \ \ \ Theoretical Physics. Vol.5., Pergamon Press, Oxford, 1982.

\ [7]\ \ A.P.Miodownik, Cu-Zn . In Phase Diagrams for Binary Copper

\ \ \ \ \ \ Alloys, The Materials Information Society, Materials Park,
OH,1994.

\ [8]\ \ G.E.Schulze, Metallphysik, Academie-Verlag, Berlin, 1967.

\ [9]\ \ G. Y. Lyubarskii, The Application of Group theory in Physics

\ \ \ \ \ \ \ Pergamon Press, Oxford,1960.

[10]\ \ A.G. Khachaturyan, Theory of Structural Transformations in Solids

\ \ \ \ \ \ \ John Wiley \& Sons Inc., New York,1983

[11] \ J-C. Toledano, P. Toledano, The Landau Theory of Phase Transitions

\ \ \ \ \ \ \ World Scientific, Singapore,1987

[12] \ A. Planes, E.Vives, T.Castan, Phys.Rev.B 44 (1991) 6715.

[13] \ M. Ziman, Principles of the Theory of Solids

\ \ \ \ \ \ \ Cambridge U.Press, Cambridge,1972

[14] \ G. Mazenko, Fluctuations, Order and Defects\ 

\ \ \ \ \ \ John Wiley \& Sons Inc., New York, 1983.

[15] \ D. Amit, Field Theory, the Renormalization Group

\ \ \ \ \ \ \ and Critical Phenomena McGraw-Hill Inc., London, 1978

[16] \ C. Nash, D. O'Connor, Ann.Phys. 273 (1999) 72.

[17] \ A.L. Kholodenko, K.F. Freed, J.Chem.Phys.80 (1984) 900.

[18] \ D.I. Uzunov, Theory of Critical Phenomena

\ \ \ \ \ \ \ World Scientific, Singapore,1993.

[19] \ J. Zinn-Justin, Quantum Field Theory and Critical Phenomena

\ \ \ \ \ \ \ Clarendon Press, Oxford, 1989.

[20] \ P. Di Francesco, P. Mathieu, D. Senechal, Conformal Field Theory

\ \ \ \ \ \ \ Springer, Berlin, 1997.

[21] \ H. Yamabe, Osaka Math. J.12 \ (1960) 21.

[22] \ T. Aubin,\ Some Nonlinear Problems in Riemannian Geometry.

\ \ \ \ \ \ \ Springer, Berlin,1998

[23] \ J. Lee, T. Parker, BAMS 17 (1987) 37.

[24] \ R.M. Schoen, LNM 1365 (1989) 120.

[25] \ A. Besse, Einstein Manifolds. Springer-Verlag, Berlin,1987.

[26] \ B. Osgood, \ R. Phillips, P. Sarnak, J.Funct.Anal. 80 (1988)148.

[27] \ M.Goulian, M. Li, PRL 66 (1991) 2051.

[28] \ E. Abdalla, M. Abdalla, D. Dalmazi, A. Zadra, 2D Gravity

\ \ \ \ \ \ \ in Non-Critical Strings, Springer-Verlag, Berlin,1994.

[29] \ S. Chowla, S., Selberg, A. Math. 227 (1967) 86.

[30] \ C. Itzykson, J-B. Zuber, Quantum Field Theory,

\ \ \ \ \ \ \ McGraw-Hill, London,1980.

[31] \ K.Richardson, J.Funct.Anal. 122 (1994) 52.

[32] \ R. Wald, General Relativity, The University

\ \ \ \ \ \ \ of Chicago Press, Chicago,1984.

[33] \ H. Bray, A. Neves, Ann.Math. 159 (2004) 407.

[34] \ J.Escobar, Ann.Math. 136 (1992)1.

[35] \ K.Akutagawa, B. Botvinnik GAFA 13 (2003) 259.

[36] \ P.A.M. Dirac, General Theory of Relativity,

\ \ \ \ \ \ \ Princeton U.Press, Princeton, 1996.

[37] \ A.Polyakov, Phys.Lett.B 103 (1981) 207.

[38] \ O.Alvarez, Nucl.Phys.B 216 (1983)125.

[39] \ W.Weisberger, Comm.Math.Phys. 112 (1987) 633.

[40] \ W. Weisberger, Nucl.Phys. B 284(1987)171.

[41]\ \ B.Osgood, R. Phillips, P. Sarnak, J.Funct. Analysis 80 (1988) 212.

[42] \ B. Osgood,R. Phillips, P.Sarnak, Ann.Math. 129 (1989) 293.

[43] \ B. Hatfield, Quantum Field Theory of Point Particles and Strings,

\ \ \ \ \ \ Addison-Wesley Publ.Co., New York, 1992.

[44] \ J. Cardy, Finite Size Scaling, Nort-Holland, Amsterdam,1988.

[45] \ H. Bl\"{o}te, J. Cardy, M. Nightingale, PRL 56 (1986) 742.

[46] \ I.Affleck, PRL 56 (1986)747.

[47] \ E. Onofri, Comm.Math.Phys. 86 (1882) 321.

[48] \ W. Beckner, Ann.Math. 138 (1993) 213.

[49] \ I. Vardi, SIAM J.Math. Anal.19 (1988) 493.

[50] \ A.Kholodenko, IJMPA 19 (2004)1655.

\ \ \ \ \ \ \ ibid, hep-th/0212189 (extended version)

[51]\ \ D. Husem\"{o}ller, Elliptic Curves, Springer-Verlag, Berlin,1987.

[52] \ J. H.Conway, N.J.A. Sloane, Sphere Packings, Lattices and Groups,

\ \ \ \ \ \ Springer-Verlag, Berlin, 1993.

[53] \ P.Cartier, An introduction to zeta functions,

\ \ \ \ \ \ \ From Number Theory to Physics,

\ \ \ \ \ \ \ Springer-Verlag, Berlin,1992, pp 1-63.

[54] \ A.Terras, Harmonic Analysis on Symmetric Spaces

\ \ \ \ \ \ \ and Applications I, Springer-Verlag, Berlin, 1985.

[55] \ S. Lang, Elliptic Functions,Springer-Verlag, Berlin,1987.

[56] \ D.Zagier, Introduction to modular forms, in From Number Theory to

\ \ \ \ \ \ Physics, Springer-Verlag, Berlin, 1992, pp.238-291.

[57] \ D.Bump, D. Goldfield, A Kronecker limit formula for cubic fields,

\ \ \ \ \ \ \ In :Modular Forms, John Wiley \& Sons, Inc., New York, 1984.

[58] \ A.Weil, Elliptic Functions according to Eisenstein and Kronecker,

\ \ \ \ \ \ \ Springer-Verlag, Berlin, 1976.

[59] \ V.L.Indenbom, Sov.Phys.Krystallography 5 (1960)106.

[60] \ H.McKean, V. Moll, Elliptic Curves,

\ \ \ \ \ \ \ Cambridge University Press, Cambridge, UK, 1999.

[61] \ G.Szpiro, Kepler's Conjecture, John Wiley\&Sons, Inc.,

\ \ \ \ \ \ \ New York, 2003.

[62] \ S.Rosenberg, The Laplacian on a Riemannian Manifold,

\ \ \ \ \ \ Cambridge Univ.Press, Cambridge, UK,1997.

[63] \ T.Parker, S. Rosenberg, J.Diff.Geom. 25 (1987) 535-557.

[64] \ K.Okikiolu, Ann.Math. 153 (2001) 471-531.

[65] \ S.Hawking, Comm.Math.Phys.55 (1977)133-148

[66] \ T.Branson, B. Orsted, Indiana U. Math. Journ. 37 (1988) 83-109.

[67] \ A.Chang,. P.Yang, Ann.Math.142 (1995) 171-212.

[68] \ M.Gromov, H.B. Lawson, IHES\ Publ.58, 295-408 (1983)

[69] \ A.L.Kholodenko, J.Geom. Phys. 35 (2000) 193-238.

[70] \ P.Chui, AMS Proceedings 125 (1997) 723-730.

[71] \ Y.Muto, J.Diff. Geom. 9 (1974) 521-530.

[72] \ N.Koiso, Inv.Math. 73 (1983) 71-106.

[73] \ R.Bruinsma, A. Zangwill, PRL 55 (1985) 214-217.

[74] \ S.Lang, Complex Multiplication. (Springer-Verlag, Berlin, 1983).

[75] \ I.Efrat, J. Number Theory 40(1992) 174-186.

[76] \ P.Sarnak, Some Applications of modular forms,

\ \ \ \ \ \ \ Cambridge U. Press, Cambridge, UK, 1990.

[77] \ D.Bump, LNM 1083 (1984) 1-180.

[78] \ S.Friedberg, AMS Transactions 300 (1987)159.

[79] \ P.Epstein, Math.Ann.56 (1903) 615-644; ibid 63 (1907) 205.

[80] \ A.Terras, AMS Transactions 183 (1973) 477.

[81] \ T. Hiramatsu, G. K\"{o}hler, Coding Theory and Number Theory,

\ \ \ \ \ \ \ Kluver Academic, Boston, 2003.

[82] \ A. L.Kholodenko, A.L. Beyerlein, Phys.Rev. A 34 (1986) 3309.

[83] \ N.W. Ashcroft, N.D. Mermin Solid State Physics, Brooks Cole, 1976.

[84] \ N.J.A. Sloane, B.K. Teo, J.Chem.Phys. 83(12), (1985) 6520

[85] \ J.E. Jones, A.E. Ingham, Proc.Roy.Soc. London Ser.A 107 (1925)

\ \ \ \ \ \ \ 636-653.

[86] \ P.G. Bolhuis, D.Frenkel, S-Ch. Mau, D.Huse, Nature 388 (1997)

\ \ \ \ \ \ 235-235.

[87] \ H. Okamoto, Phase Diagrams for Binary Alloys,

\ \ \ \ \ \ ASM International, Materials Park, OH, 2000.

[88] \ D.M. Adams, Inorganic Solids,

\ \ \ \ \ \ \ John Wiley\&Sons Inc., New York, 1974

[89] \ H.S.M. Coxeter, Regular Polytopes,

\ \ \ \ \ \ Macmillan Co., New York, 1963.

[90] \ L.S. Charlap, Bieberbach Groups and Flat manifolds,

\ \ \ \ \ \ Springer-Verlag, Berlin,1986.

[91] \ G.H. Hardy, E.M. Wright,An Introduction to the

\ \ \ \ \ \ \ Theory of numbers, Clarendon Press, Oxford, 1960.

[92] \ S. Lang, Algebraic Number Theory,

\ \ \ \ \ \ Addison Wesley Inc., New York,1970.

[93] \ J.W.S. Cassels, An Introduction to the Geometry of numbers,

\ \ \ \ \ \ Springer-Verlag, Berlin, 1970.

[94] \ C.L. Siegel, Lectures on Geometry of Numbers,

\ \ \ \ \ \ Springer-Verlag, Berlin,1989.

[95] \ R.P. \ Feynman, A.R. Hibbs, Quantum Mechanics and

\ \ \ \ \ \ Path Integrals, McGraw Hill Inc., New York, 1965.

\textbf{\ }

\ \ \ \ \ \ \ \ \ \ \ \ \ \ \ \ \ \ \ \ \ \ \ \ \ \ \ \ \ \ \ \ \ \ \ \ \ \
\ \ \ \ \ \ \ \ \ \ \ \ \ \ \ \ \ \ \ \ \ \ \ \ \ \ \ \ \ \ \ \ \ \ \ \ \ \
\ \ \ \ \ \ \ \ \ \ \ \ \ \ 

\end{document}